\documentclass[twocolumn,astrosymb,tighten]{aastex631}

\newcommand{\ha}{H$\alpha$}

\def\lessim{\mathrel{\hbox{\rlap{\hbox{\lower4pt\hbox{$\sim$}}}\hbox{$<$}}}}
\def\grtsim{\mathrel{\hbox{\rlap{\hbox{\lower4pt\hbox{$\sim$}}}\hbox{$>$}}}}

\shorttitle{M31 Novae}
\shortauthors{Rector et al.}
\graphicspath{{./}{figures/}{figures/lc/}}

\begin{document}

\title{The Rate and Spatial Distribution of Novae in M31 as Determined by a Twenty-Year Survey}

\author[0000-0001-8164-653X]{Travis A. Rector}
\altaffiliation{National Optical Astronomy Observatory (NOAO), 950 N. Cherry Ave., Tucson, AZ 85719.  Visiting astronomer, Kitt Peak National Observatory at NSF’s NOIRLab, managed by the Association of Universities for Research in Astronomy (AURA) under a cooperative agreement with the National Science Foundation.}
\correspondingauthor{Travis A. Rector}

\affiliation{University of Alaska Anchorage,
Department of Physics \& Astronomy,
Anchorage, AK 99508, USA}
\email{tarector@alaska.edu}

\author[0000-0002-1276-1486]{Allen W. Shafter}
\affiliation{San Diego State University,
Department of Astronomy,
San Diego, CA 92182, USA}

\author[0000-0002-6023-7291]{William A. Burris}
\affiliation{San Diego State University,
Department of Astronomy,
San Diego, CA 92182, USA}

\author{Matthew J. Walentosky}
\affiliation{University of Alaska Anchorage,
Department of Physics \& Astronomy,
Anchorage, AK 99508, USA}

\author{Kendall D. Viafore}
\affiliation{University of Alaska Anchorage,
Department of Physics \& Astronomy,
Anchorage, AK 99508, USA}

\author[0000-0001-6369-1636]{Allison L. Strom}
\affiliation{University of Alaska Anchorage,
Department of Physics \& Astronomy,
Anchorage, AK 99508, USA}

\author{Richard J. Cool}
\affil{Steward Observatory, The University of Arizona, 933 N. Cherry Ave., Tucson, AZ 85721, USA}

\author{Nicole A. Sola}
\affiliation{University of Alaska Anchorage,
Department of Physics \& Astronomy,
Anchorage, AK 99508, USA}

\author{Hannah Crayton}
\affiliation{University of Alaska Anchorage,
Department of Physics \& Astronomy,
Anchorage, AK 99508, USA}


\author[0000-0002-3007-206X]{Catherine A. Pilachowski}
\affil{Astronomy Department, Indiana University Bloomington, Swain West 318, 727 East Third Street, Bloomington, IN 47405-7105, USA}

\author[0000-0001-7970-0277]{George H. Jacoby}
\affil{National Optical Astronomy Observatory (NOAO), 950 N. Cherry Ave., Tucson, AZ 85719, USA}

\author{Danielle L. Corbett}
\affil{Steward Observatory, The University of Arizona, 933 N. Cherry Ave., Tucson, AZ 85721, USA}

\author{Michelle Rene}
\affil{Steward Observatory, The University of Arizona, 933 N. Cherry Ave., Tucson, AZ 85721, USA}

\author{Denise Hernandez}
\affil{Steward Observatory, The University of Arizona, 933 N. Cherry Ave., Tucson, AZ 85721, USA}



\begin{abstract}

A long-term (1995--2016) survey for novae
in the nearby Andromeda galaxy (\object{M31}) was conducted
as part of the Research-Based Science Education initiative.
During the course of the survey 180 nights of observation were completed at Kitt Peak, Arizona.
A total of 262 novae were either discovered or confirmed, 40 of which have not been previously reported. Of these,
203 novae form a spatially-complete sample detected by the KPNO/WIYN~0.9-m telescope within a $20'\times20'$ field centered on the nucleus of M31.
An additional 50 novae are part of a spatially-complete sample detected by the KPNO~4-m telescope within a larger $36'\times36'$ field.
Consistent with previous studies, it is found that the
spatial distribution of novae in both surveys
follows the bulge light of M31 somewhat more closely than the overall
background light of the galaxy.
After correcting for the limiting magnitude and the spatial and
temporal coverage of the surveys, a final nova rate in M31 is found to
be $R=40^{+5}_{-4}$~yr$^{-1}$, which is considerably lower than
recent estimates.
When normalized to the $K$-band luminosity of M31, this value yields a
luminosity-specific nova rate, $\nu_K = 3.3\pm0.4$~yr$^{-1}~[10^{10}~L_{\odot,K}]^{-1}$.
By scaling the M31 nova rate using the relative infrared luminosities of M31 and our Galaxy,
a nova rate of $R_\mathrm{G}=28^{+5}_{-4}$ is found for the Milky Way.


\end{abstract}



\section{Introduction} \label{sec:intro}

Novae are transient sources that arise from a thermonuclear runaway
on the surface of an accreting white dwarf in a close binary system
\citep[see][and references therein]{1995cvs..book.....W,2016PASP..128e1001S}.
Their eruptions are believed to play a significant role in the chemical evolution
of galaxies, particularly with regard to the production of the isotopes of some light elements such as $^7$Li and CNO group nuclei
\citep[see][for a recent review]{2020A&ARv..28....3D}.

The eruptions of novae do not disrupt the progenitor binary, and as a result,
continued accretion onto the white dwarf produces recurrent outbursts on timescales from approximately
a year up to of order $10^5$~yr.  Despite the fact that all novae are recurrent,
only systems with the shortest recurrence
times, and where more than one eruption has been recorded, are
referred to explicitly as Recurrent Novae.

The eruptions
of novae are among the most luminous of any optical transients, with
absolute magnitudes ranging from $M_V\simeq-5$ to $M_V\simeq-10$
for the most luminous systems \citep[e.g.,][]{2009ApJ...690.1148S}.
Their high luminosities make novae
easily visible in nearby galaxies, where they can serve as tracers of the
close binary star content across differing stellar populations.

\begin{figure*}
\includegraphics[angle=0,scale=0.65]{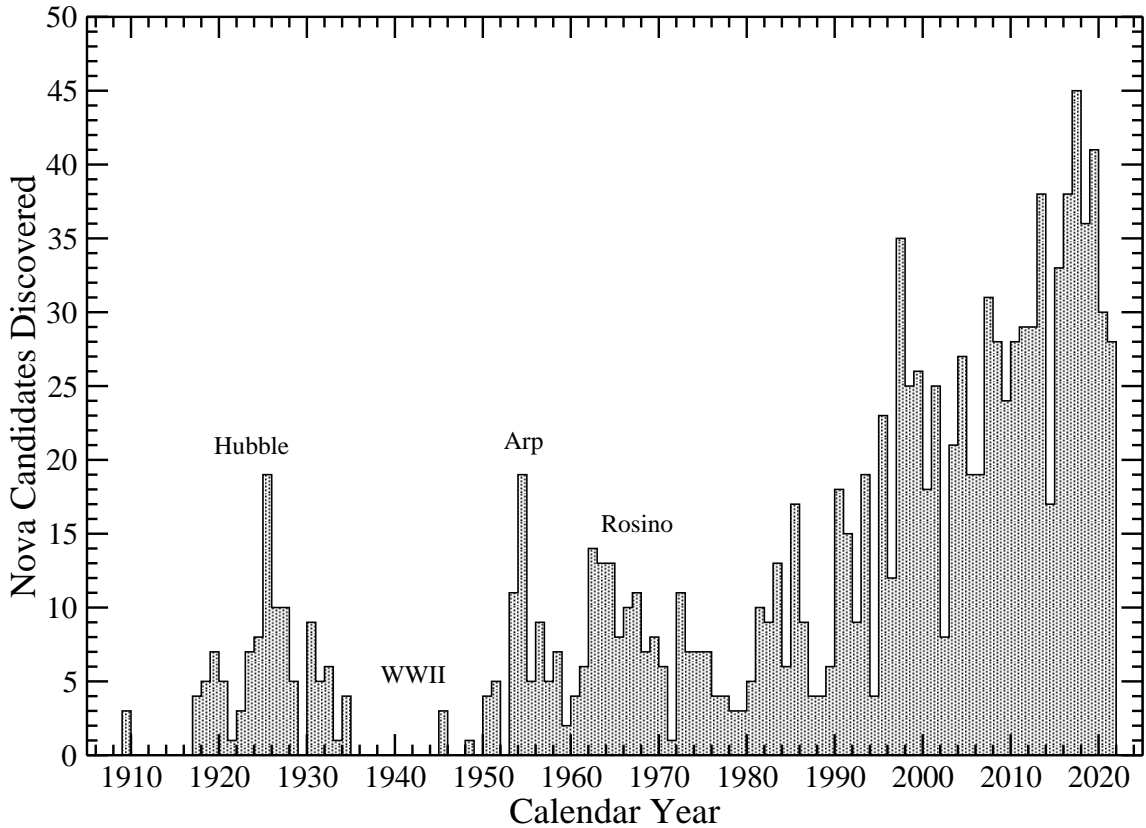}
\caption{The number of nova candidates discovered in M31
in a given calendar year spanning the past 113 years. The average annual number of candidates detected during the seven-year period between 2016 and 2021 was 36. Assuming these candidates are all bona fide novae, this number would appear to place a stringent lower limit to the nova rate in M31.
}
\label{fig:f1}
\end{figure*}

The observed properties of novae, in particular their peak luminosity,
the rate of decline from maximum light, and their rate of recurrence
is primarily a function
of the mass of the white dwarf, its rate of accretion, and the chemical
composition
of the gas from the companion star being pulled onto its surface \citep[e.g.,][]{1982ApJ...253..798N,2005ApJ...628..395T,2014ApJ...793..136K,2016PASP..128e1001S}.
Given that these properties of the progenitor binary
are expected to vary with the underlying
stellar population from which the novae arise, it is reasonable to expect
that the nova rate and light curve properties of novae should
similarly vary between differing stellar populations and Hubble types.
Although novae have now been detected in more than a dozen extragalactic
systems, it is still unclear whether and how the observed properties of novae
differ between these galaxies \citep[e.g., see][and references therein]
{2014ASPC..490...77S,2019enhp.book.....S,2020A&ARv..28....3D}.

More novae have been observed in M31 than all other external galaxies combined.
Discoveries of M31 novae go back more than a century,
with the first novae reported by \citet{1917PASP...29..210R} during
the course of a survey undertaken with the 60-in reflector
on Mount Wilson\footnote{Not including
S~Andromeda (Nova Andromeda 1885),
the supernova that was misidentified as a nova before the distinction
between the two was finally appreciated in the 1930s.}.
Since the first nova was discovered on
1909 September 12 (M31N 1909-09a), more than 1200
additional nova candidates have been identified in M31. Many of these
objects were discovered as part of targeted surveys beginning with
Hubble's classic study \citep{1929ApJ....69..103H}.

After a hiatus
of more than two decades spanning the great Depression and war years,
two additional photographic surveys were initiated. The first
of these was undertaken
(with Hubble's encouragement) by \citet{1956AJ.....61...15A}.
Arp discovered a total of 30 novae during the period between June 1953 and
January 1955. The second, and last of the major photographic
surveys was conducted by Rosino and collaborators with
the 1.22-m and 1.82-m reflectors at the Asiago and Ekar observatories in northern Italy. Their observations
spanned several decades beginning in the mid 1950s
and continuing until 1986 \citep[][]{1964AnAp...27..498R, 1973Ros, 1989AJ.....97...83R}.

By the time the Rosino survey was concluding, modern digital surveys were becoming the
standard \citep[e.g.,][]{1987ApJ...318..520C,2001ApJ...563..749S,2006MNRAS.369..257D}. These surveys,
coupled in recent years with the ever-increasing contributions from
amateur astronomers and the proliferation of automated telescopes
such as the Zwicky Transient Facility \citep[ZTF,][]{2019PASP..131a8002B},
have led to
a substantial increase in the number of nova candidates
discovered in M31 each year (see Figure~\ref{fig:f1}).

Despite the wealth of data, a reliable estimate for the annual rate of novae in M31 has remained elusive, with
estimates over the past century ranging from as few as
26~yr$^{-1}$ \citep{1956AJ.....61...15A} to in excess of 100~yr$^{-1}$ in the case of model simulations
\citep{2016MNRAS.455..668S,2016MNRAS.458.2916C}. These estimates have often been made from surveys with limited temporal coverage and numbers of detected novae. For example, the most recent and widely-accepted estimate of $65^{+16}_{-15}$~yr$^{-1}$ \citep{2006MNRAS.369..257D} is based upon the detection of just 20 novae over the
course of a 4-year survey.

In this paper we describe an intensive, long-term program to study the novae population in M31.
Spanning more than two decades between the years of 1995 and 2016, observations in search of novae in M31 were undertaken as part of the Research-Based Science Education
(RBSE) project conducted at Kitt Peak National Observatory (KPNO).  RBSE is a course-based undergraduate research experience (CURE), wherein students participate in an authentic research project as part of an introductory class \citep{10.1088/2514-3433/ab2b42ch7}. Student gains in understanding the scientific process from participating in RBSE are described in \citet{PhysRevPhysEducRes.14.010151}.
The RBSE observations, which resulted in
the detection of 262 individual novae, provide a rich dataset with which to study the properties of
novae in M31, including their spatial
distribution and overall rate. Here, we present the results of the RBSE survey.

\section{Observations} \label{sec:obs}

For the purposes of this paper, all observations of M31 completed on a single night were combined into one image, representing a single epoch.  A total of 180 epochs  were completed from 1995 September 3 to 2016 October 22. Within the constraints of scheduling and weather, observations were attempted on a monthly basis from 1997 through 2011 during the July-January time period when M31 is accessible. To search for novae that fade rapidly, several campaigns were undertaken in which M31 was observed on a nightly basis for the span of one or more weeks.

All but 15 of the 180 epochs were observed with the KPNO/WIYN 0.9-m telescope\footnote{The WIYN Observatory is a joint facility of the NSF's National Optical-Infrared Astronomy Research Laboratory, Indiana University, the University of Wisconsin-Madison, Pennsylvania State University, the University of Missouri, the University of California-Irvine, and Purdue University.}. Supplemental observations were obtained with the KPNO {\it Mayall} 4-m telescope, the KPNO 2.1-m telescope, the {\it Bok\/} 90-in telescope at Steward Observatory\footnote{Steward Observatory is operated by the University of Arizona.}, and the {\it McGraw-Hill} 1.3-m telescope at the MDM Observatory\footnote{MDM Observatory is operated by a consortium of Dartmouth College, The Ohio State University, Columbia University, the University of Michigan, and Ohio University.}.  Observations from the {\it Local Group Survey} (LGS) \citep[][]{2006AJ....131.2478M} were also used. Multiple cameras were used on the telescopes, with the technical details for each camera and telescope configuration given in Table~\ref{tbl:tel_det}.

Following the pioneering study of \citet{1987ApJ...318..520C}, we chose to
conduct the RBSE survey by imaging in \ha. Shortly after eruption novae develop strong and broad \ha\ emission lines that persist long after the
continuum has faded. Thus \ha\ observations have a distinct advantage in synoptic surveys with sporadic temporal coverage.
In addition, \ha\ observations provide a better contrast against the bright background of the galaxy's bulge where the
nova density is highest.

\begin{figure}
\plotone{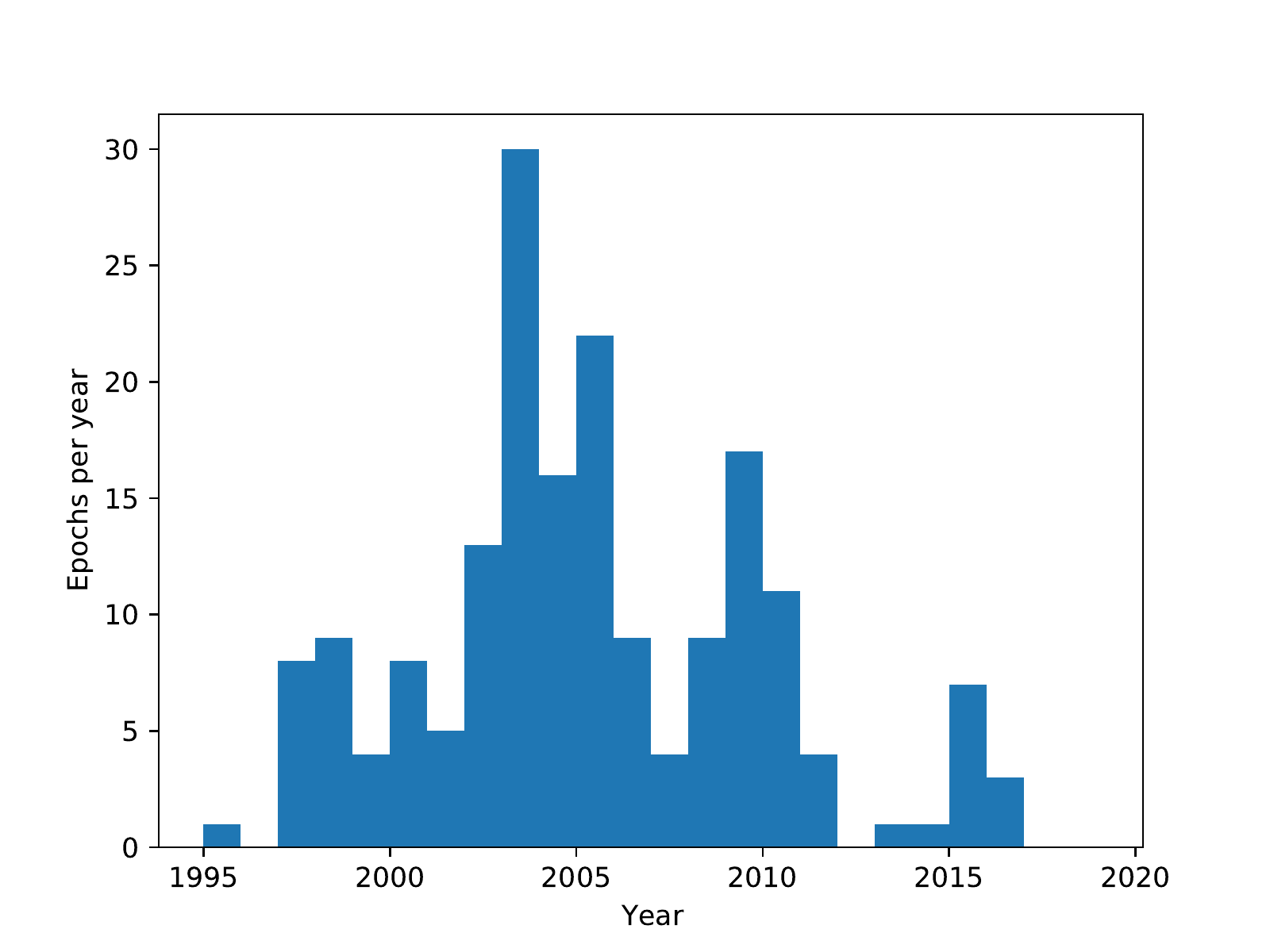}
\caption{The number of epochs per year.
}
\label{fig:date_histo}
\end{figure}

\begin{figure}
\plotone{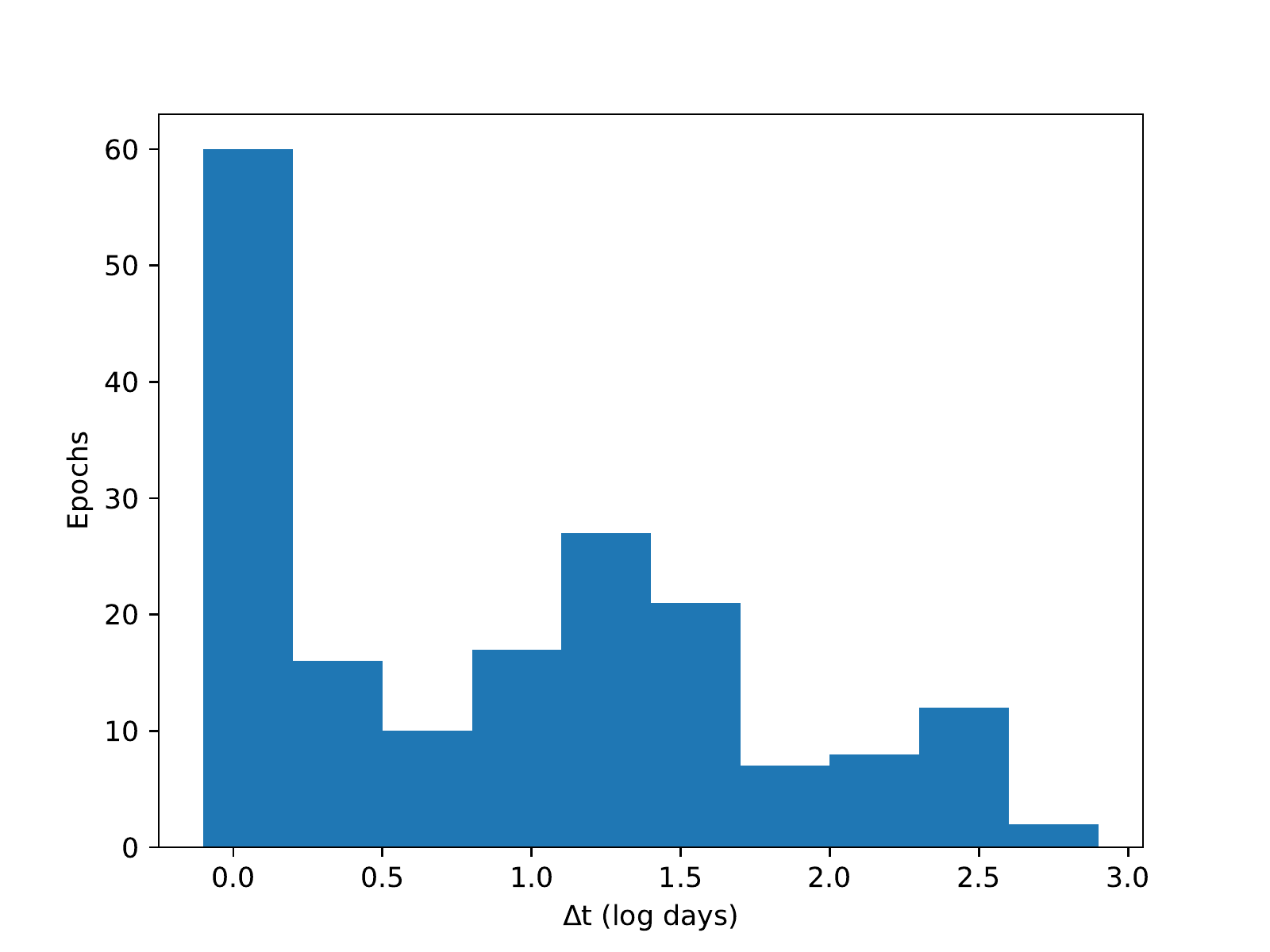}
\caption{The distribution of time between epochs, in units of log$_{10}$ days.
}
\label{fig:delta_histo}
\end{figure}

\begin{figure}
\plotone{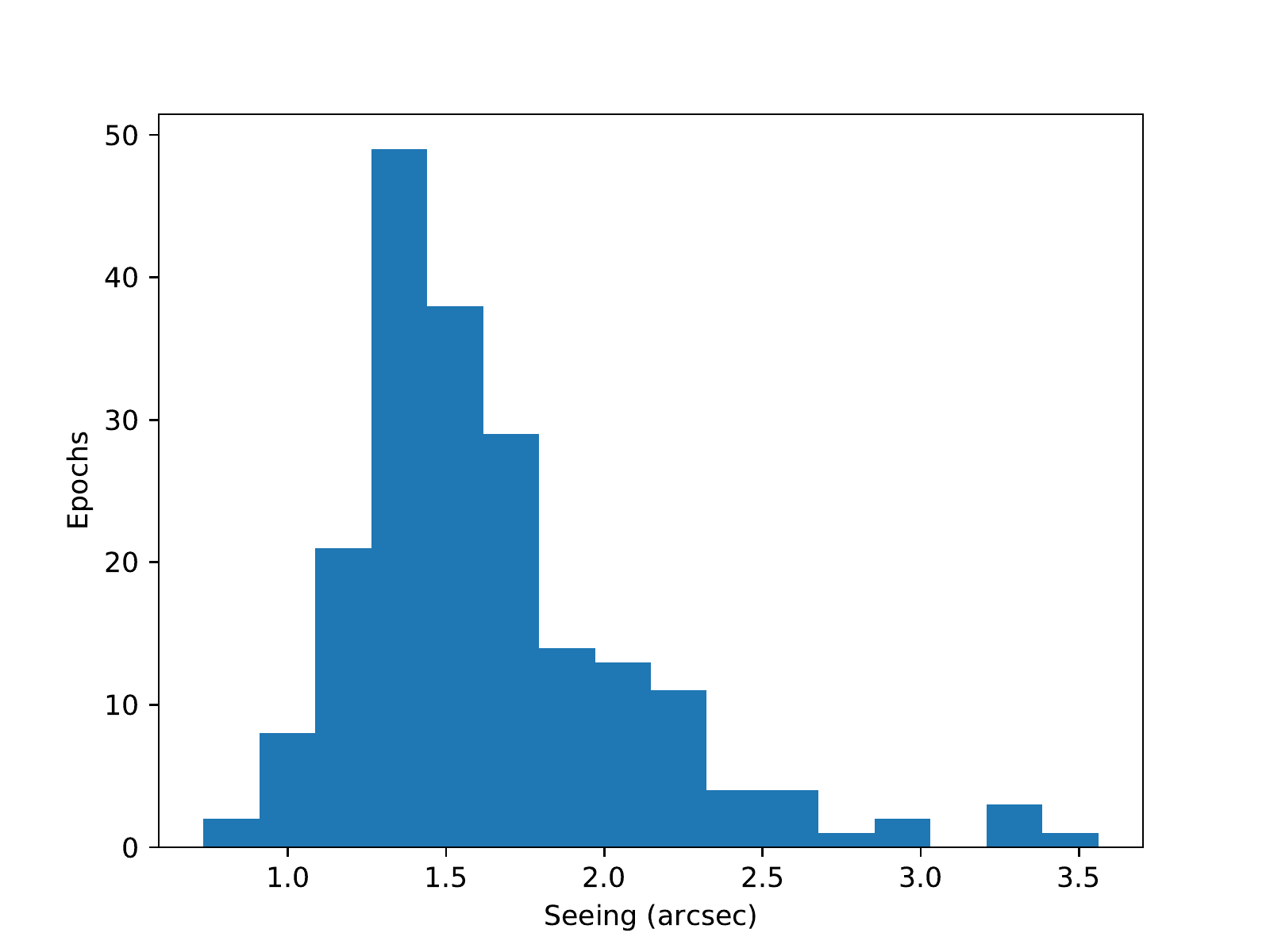}
\caption{The distribution of seeing values.
}
\label{fig:seeing}
\end{figure}

Prior M31 nova surveys in \ha\ \citep[e.g.,][]{1987ApJ...318..520C,2001ApJ...563..749S} have shown that a survey with a limiting absolute magnitude of  $M_{\mathrm{H}\alpha} \sim -7.5$ is sufficiently deep to detect a typical nova several months after eruption.\footnote{
The H$\alpha$ magnitude is defined on the AB system where
$m_{\mathrm{H}\alpha} = 0$ for $f_{\lambda} = 2.53
\times 10^{-9}$~ergs~cm$^{-2}$~s$^{-1}$~\AA$^{-1}$.}
With the 0.9-m telescope it is possible to reach this depth with a total integration time of 30~min.  Multiple observations with small offsets were made between exposures to minimize the effects of bad pixels, cosmic rays, and transient objects (e.g., asteroids and satellite trails).  For single-CCD cameras on the 0.9-m telescope (t2ka, s2kb, and HDI), this was achieved with three 10-min exposures.  For observations with the multi-CCD Mosaic camera, five 6-min exposures were used to facilitate filling in chip gaps.  To avoid saturating the standard stars, integration times were shortened to five 5-min exposures with the 90-in telescope and five 3-min exposures with the 4-m telescope.  The KPNO~2.1-m telescope and the MDM 1.3-m telescope were used only once each.  Because of their smaller fields of view, a mosaic of pointings was taken to produce an effective FOV of $23'\times23'$ for each epoch, matching the 0.9-m/t2ka configuration.  The details of each epoch of observations are given in Table~\ref{tbl:obs_list}. The number of epochs per year is shown in Figure~\ref{fig:date_histo}. The distribution of time between observations is given in Figure~\ref{fig:delta_histo}, while the distribution of seeing values are plotted in Figure~\ref{fig:seeing}.

\begin{deluxetable}{lccr}
\tablecaption{Telescope and Detector Characteristics\label{tbl:tel_det}}
\tablehead{
\colhead{Telescope/} & \colhead{Scale} & \colhead{FOV} & \colhead{Epochs} \\
\colhead{Detector} & \colhead{(\arcsec\ pix$^{-1}$)} & \colhead{(\arcmin\ on a side)} & \colhead{(nights)}
}
\startdata
0.9-m/t2ka & 0.68 & 23 & 20 \\
0.9-m/MOSA & 0.43 & 59 & 41 \\
0.9-m/s2kb & 0.60 & 20 & 96 \\
0.9-m/HDI & 0.43 & 29 & 9 \\
1.3-m/Echelle & 0.50 & 17 & 1\\
2.1-m/t2ka & 0.31 & 10 & 1 \\
Bok/90prime & 0.45 & 70 & 5 \\
4-m/MOSA & 0.26 & 36 & 7 \\
\enddata
\end{deluxetable}

\begin{deluxetable}{llll}
\tablecaption{List of Observations\label{tbl:obs_list}}
\tablehead{
\colhead{Date} & \colhead{Date} & \colhead{Telescope/} & \colhead{Seeing} \\
\colhead{(UT)} & \colhead{(MJD)} & \colhead{Detector} & \colhead{(\arcsec)}
}
\startdata
1995-09-03 & 49963.26 & 0.9-m/t2ka & 1.5 \\
1997-06-18 & 50617.44 & 0.9-m/t2ka & 1.4 \\
1997-07-23 & 50652.32 & 0.9-m/t2ka & 1.6 \\
\enddata
\tablecomments{Only the first three lines of Table~\ref{tbl:obs_list} are shown here.  It is published in its entirety in machine-readable format.}
\end{deluxetable}

\begin{deluxetable}{lllrr}
\tablecaption{Locations of New Novae\label{tbl:astrometry}}
\tablehead{
\colhead{Name} & \colhead{RA} & \colhead{Dec} & \colhead{$\Delta \alpha$\tablenotemark{a}} & \colhead{$\Delta \delta$\tablenotemark{a}} \\
\colhead{(M31N)} & \colhead{(J2000)} & \colhead{(J2000)} & \colhead{(\arcsec)} & \colhead{(\arcsec)}
}
\startdata
1995-09h & 00:43:02.61 & 41:18:41.8 & 274.2 & 154.3 \\
1997-06f & 00:42:18.03 & 41:21:53.5 & -394.5 & 346.0 \\
1997-06g & 00:42:44.78 & 41:16:48.2 & 6.7 & 40.7 \\
1999-01e & 00:42:44.38 & 41:16:16.1 & 0.8 & 8.6 \\
1999-06c & 00:42:38.60 & 41:14:19.2 & -86.0 & -108.3 \\
1999-06d & 00:42:39.32 & 41:16:11.3 & -75.2 & 3.8 \\
1999-07c & 00:42:44.04 & 41:17:05.6 & -4.3 & 58.1 \\
1999-12f & 00:43:06.37 & 41:30:14.0 & 330.6 & 846.5 \\
1999-12g & 00:41:37.15 & 41:05:51.2 & -1007.7 & -616.3 \\
2001-11b & 00:41:49.49 & 41:04:04.0 & -822.6 & -723.5 \\
2002-06a & 00:42:22.52 & 41:09:48.5 & -327.2 & -379.0 \\
2002-06b & 00:41:56.16 & 41:08:31.3 & -722.5 & -456.2 \\
2002-06c & 00:42:50.91 & 41:24:13.6 & 98.7 & 486.1 \\
2002-08c & 00:42:56.23 & 41:22:21.9 & 178.5 & 374.4 \\
2002-09a & 00:44:33.24 & 41:36:07.5 & 1633.7 & 1200.0 \\
2002-09b & 00:41:58.12 & 41:02:13.4 & -693.1 & -834.1 \\
2002-11a & 00:41:57.83 & 41:08:46.3 & -697.5 & -441.2 \\
2003-06e & 00:42:43.93 & 41:15:33.9 & -6.0 & -33.6 \\
2003-08d & 00:44:36.50 & 41:33:50.4 & 1682.5 & 1062.9 \\
2003-08e & 00:43:55.98 & 41:36:33.2 & 1074.8 & 1225.7 \\
2003-10d & 00:43:08.85 & 41:03:43.4 & 367.8 & -744.1 \\
2004-01c & 00:42:46.21 & 41:05:38.2 & 28.2 & -629.3 \\
2004-06d & 00:42:44.75 & 41:16:35.5 & 6.3 & 28.0 \\
2004-06e & 00:42:10.33 & 41:09:50.6 & -510.0 & -376.9 \\
2004-06f & 00:42:44.79 & 41:16:35.4 & 6.9 & 27.9 \\
2005-11a & 00:42:46.52 & 41:15:33.9 & 32.8 & -33.6 \\
2006-11d & 00:43:05.61 & 41:01:29.8 & 319.2 & -877.7 \\
2006-11e & 00:42:06.41 & 40:56:58.4 & -568.8 & -1149.1 \\
2006-11f & 00:41:56.88 & 40:58:43.9 & -711.7 & -1043.6 \\
2007-06c & 00:42:38.44 & 41:17:22.2 & -88.3 & 74.7 \\
2008-12c & 00:42:43.95 & 41:16:59.1 & -5.7 & 51.6 \\
2008-12d & 00:42:33.17 & 41:15:56.4 & -167.4 & -11.1 \\
2011-09c & 00:42:51.32 & 41:20:41.6 & 104.8 & 274.1 \\
2011-09d & 00:44:05.96 & 41:26:50.0 & 1224.4 & 642.5 \\
2014-08a & 00:43:47.72 & 41:33:13.2 & 950.8 & 1025.7 \\
2014-08b & 00:43:56.63 & 41:15:16.8 & 1084.5 & -50.7 \\
2015-11d & 00:42:43.89 & 41:16:01.9 & -6.6 & -5.6 \\
2016-10e & 00:42:47.60 & 41:10:24.8 & 49.0 & -342.7 \\
2016-10f & 00:42:34.37 & 41:15:52.6 & -149.4 & -14.9 \\
2016-10g & 00:42:23.67 & 41:12:05.6 & -309.9 & -241.9 \\
\enddata
\tablecomments{Only the 40 previously unreported novae are shown here.  The astrometry for all 262 novae detected in this survey is published in its entirety in machine-readable format.}
\tablenotetext{a}{Offset from the nucleus of M31.}
\end{deluxetable}

The data were reduced with IRAF in the standard manner.  The world coordinate system (WCS) was determined via stars from the USNO-B1.0 catalog \citep{2003AJ....125..984M} with a global solution RMS of better than $0\farcs6$ in all cases. This is assumed as the accuracy for all measured positions.  To remove geometric distortions and to allow for the images to be aligned, all of the images were projected to a common WCS and plate scale of $0\farcs43$ pixel$^{-1}$, the plate scale of 0.9-m/MOSA.  This scale was chosen so as to not degrade the stellar PSFs for most of the images.  Data obtained with 4-m/MOSA had a smaller inherent plate scale and often better seeing, occasionally creating undersampled PSFs when rescaled.  The original data were checked whenever the authenticity of a nova was in question. Because 0.9-m/MOSA also had the largest field of view ($59\arcmin\times 59\arcmin$), data obtained with other camera and telescope configurations were scaled, aligned, and inset into larger blank images of this size so that epochs could be easily compared.

\subsection{Nova Search Strategy}

While automated searches for novae in M31 have made significant gains \citep[e.g.,][]{2017A&A...599A..48S}, searching for novae in M31 is an ideal project for students because it is a relatively simple task that is well suited to human physiology, as our brains have evolved to detect subtle changes in complex environments.  Over the two decades of the survey, each epoch was examined by hundreds of students as a blind study.  Final confirmation of candidates was completed by the PIs.

Images were searched for novae with the following strategy: The 42 images taken with 0.9-m/MOSA were used to create a median image. Each epoch of observations was then divided by this median image, producing a ``difference" image wherein stars and galactic structure largely canceled and novae appeared more prominently.  This allowed the images to be searched in a uniform manner despite the tremendous differences in brightness between the core and outer parts of M31.  Once a nova candidate was identified, $100\times100$~pixel ``cutouts" of that region were extracted from all 180 epochs to determine not only the veracity of the detection but its long-term behavior.  Long-period variables could then be excluded.  The positions of all nova candidates were compared with their locations on a deep 4-m image to see if any stellar objects were coincident.  

\subsection{Photometry}

Aperture photometry was performed using the IRAF~APPHOT package.  To improve the photometry of novae within the bulge, the background light of M31 was fitted and subtracted with the IRAF~IMSURFIT task. Tests indicated that removal of the background did not degrade the accuracy of photometric measurements for standard stars.   

The 74 \ha\ standard stars given in \citet{1987ApJ...318..520C} were used to calibrate each epoch.  Standards that deviated more than 3$\sigma$ were clipped, typically 1 to 5 stars per epoch.  For the epochs obtained with the 1.3-m and 2.1-m telescopes, only 10-20 standards were in the FOV.  For most epochs the standard star calibration is internally consistent to $\sim$0.05~mag.  

The photometric calibration obtained from the \citet{1987ApJ...318..520C} standards was compared to that determined by \citet{2007AJ....134.2474M} for the LGS fields.  The Ciardullo et al. observations were obtained with a 75\AA\ FWHM \ha\ filter, whereas the Massey et al. data, as well as ours, were obtained with 50\AA\ FWHM \ha\ filters.  To precisely compare the two we would need to know the full transmission profile for all of the filters; however, the two calibrations are consistent if the 75\AA-wide filter allows about 30\% more light to pass.  For consistency with other work we use the Ciardullo et al. values.

Aperture photometry was completed for all nova candidates detected in our survey.  For each nova a $5\sigma$ detection upper limit was also calculated at the location of the nova in the epoch directly before first detection and in the epoch after it was last detected.  Photometric measurements, as well as detection upper limits, are given in Table~\ref{tbl:photo}.

\begin{deluxetable}{llll}
\tablecaption{Photometry for all Novae\label{tbl:photo}}
\tablehead{
\colhead{Name} & \colhead{Date} & \colhead{Mag} & \colhead{Err} \\
\colhead{(M31N)} & \colhead{(MJD)} & \colhead{(\ha)} 
}
\startdata
1994-09a & 49963.26 & 18.68 & $\pm 0.07$ \\
1994-09a & 50617.44 & $>21.0$ \\
1995-07a & 49963.26 & 16.70 & $\pm 0.03$ \\
\enddata
\tablecomments{Only the first three lines of Table~\ref{tbl:photo} are shown here.  It is published in its entirety in machine-readable format.}
\end{deluxetable}

\section{Nova Rate in our Surveyed Regions} \label{sec:novrat}

Of our observations only data obtained with two telescopes, the KPNO/WIYN~0.9-m and the KPNO~4-m, provide sufficient depth and temporal coverage to allow the spatial distribution and nova
rate of M31 to be reliably determined. A spatially complete sample was defined of 203 novae detected by the 0.9-m telescope within a $20'\times20'$ field centered on the nucleus of M31. An additional 50 novae are part of a separate spatially complete sample discovered with the 4-m telescope within a $36'\times36'$ field centered on the nucleus.  These surveys, which we will henceforth refer to as survey~A and survey~B respectively, have been analyzed separately and therefore provide two independent estimates of M31's nova rate.

For both surveys, a determination of the nova rate in M31 depends on the intrinsic properties
of the M31 nova population (i.e., their peak magnitudes, rates of decline, and
whether these properties vary as a function of spatial position within the galaxy) and on the characteristics of the surveys themselves (i.e., the dates of observation, the survey's effective limiting magnitude, and the number of novae discovered during the course of the survey).
Of these parameters, only the dates of observation and the number of novae
discovered are known with certainty. The photometric properties of the M31 nova population (what
we will refer to as the instantaneous nova luminosity function), and the effective limiting magnitude
of the survey
(i.e., the completeness of the survey to a given apparent magnitude) are quantities that must either be assumed or estimated.
The former is independent of the specifics of the survey, and will be the same for both surveys, while the latter must be determined for each survey separately.
We explore each of these important considerations further below.

\subsection{The M31 Nova Luminosity Function}

The intrinsic photometric (light curve) properties of the M31 nova population are not known a priori,
and must be estimated using existing nova \ha\ light curves. Our best option to characterize the M31 nova luminosity function is to
take the observed light curves of novae from
this paper, augmented with similar observations from previous surveys \citep[][]{1987ApJ...318..520C, 2001ApJ...563..749S, 2004AJ....127..816N}.
A potential drawback to this approach is that the
existing observations of M31 nova light curves may suffer from observational
selection biases. For example, a population of unusually faint and
rapidly evolving novae, if it exists, could be under-represented in our observed light curve sample. Fortunately, we see
no evidence for such a putative population of faint and fast novae in our extensive data set.
Finally, given that there is no compelling evidence to the contrary,
we have assumed that
the nova properties do not vary significantly with spatial position
within M31.

Not all of the novae discovered in our surveys have sufficient temporal coverage to allow reliable light curve
parameters (i.e., peak magnitudes and fade rates) to be determined. In particular, the discovery magnitude
will usually underestimate the peak brightness of the nova unless maximum light is well covered by the observations,
and this is typically not the case. In order to mitigate the discrepancy between the observed peak and the true peak
magnitude, we have restricted our light curve sample to those novae where either maximum light can be constrained by
observations made on the rise to peak, or in cases where a previous observation (or null detection) was available
up to a month before the observed maximum. In the latter case we have estimated peak magnitude by extrapolating
the best fit of the declining branch of the light curve backwards to the midpoint between the time of the observed
maximum and that of the previous measurement or non-detection.

\begin{figure}
\label{fig:lightcurves}
\plotone{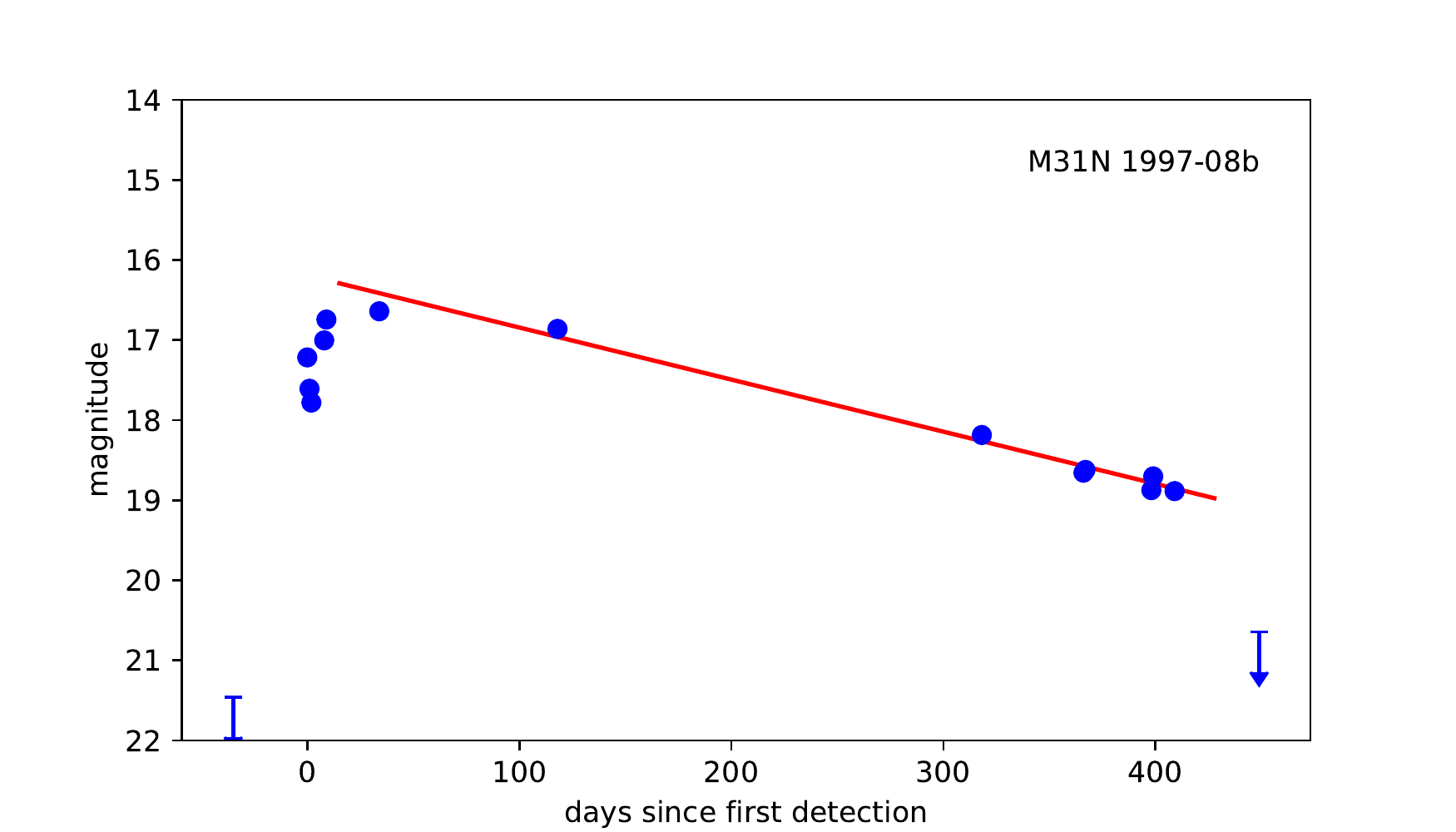}
\caption{The light curve for nova M31N~1997-08b, one of 30 novae used in the Monte Carlo simulations.  The complete figure set of 30 plots is available in the online journal.  The vertical axis shows the \ha\ magnitude.  The horizontal axis shows the number of days since first detection.  Upper-limits on detection are also plotted for prior observations within 31 days of the first detection, as well as for observations obtained within 100 days of the last detection.  The red line shows the linear best fit during the decline.  In some figures a red plus sign marks the estimated location of peak magnitude, as extrapolated from the rise and decay slopes.}
\end{figure}

\begin{figure}
\includegraphics[angle=0,scale=0.32]{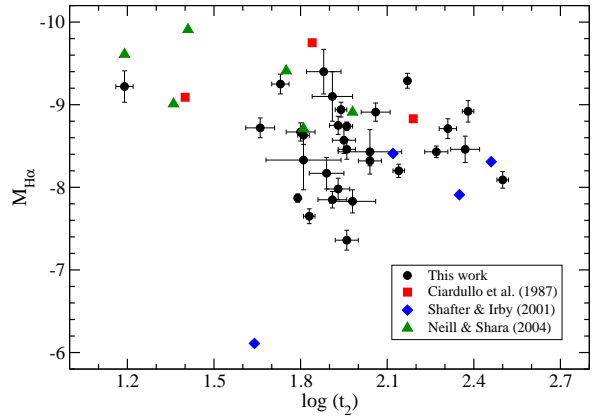}
\caption{The peak \ha\ absolute magnitudes for the novae used in our Monte Carlo analysis plotted as a function of the fade rate as measured by log($t_2$). There is no evidence for an \ha\ MMRD relation as first demonstrated by
\citet{1987ApJ...318..520C}.
}
\label{fig:mmrd}
\end{figure}

After applying these selection criteria to our overall nova
sample, we were left with 30 novae with sufficiently well-sampled light curves suitable for
use in our M31 nova rate simulations (see Figure~\ref{fig:lightcurves} for their light curves and Table~\ref{tab:lcparam} for their calculated parameters). In addition to these novae,
Table~\ref{tab:lcparam} also lists 13 novae from the literature with measured peak \ha\ magnitudes and fade rates.
In Figure~\ref{fig:mmrd} we show the absolute \ha\ magnitudes for all 43 novae plotted
as a function of the fade rate, as parameterized by the log of the time in days it takes for a nova to fade by two magnitudes. As expected based on the work
of \citet{1987ApJ...318..520C}, there is no compelling evidence for a maximum magnitude, rate-of-decline (MMRD) relationship for the \ha\ light curves.

\begin{deluxetable*}{lccccccrrcc}
\tablecolumns{11}
\tablecaption{Light Curve Parameters\label{tab:lcparam}}
\tablehead{\colhead{Nova} & \colhead{} & \colhead{} & \colhead{} & \colhead{} & \colhead{$f_{\mathrm{H}\alpha}$} & \colhead{$\sigma_{f_{\mathrm{H}\alpha}}$} & \colhead{$t_2(\mathrm{H}\alpha)$} &
\colhead{$\sigma_{t_2}$} & \colhead{} & \colhead{} \\
\colhead{(M31N)} & \colhead{$m_{\mathrm{H}\alpha}$} & \colhead{$\sigma_m$} & \colhead{$M_{\mathrm{H}\alpha}$} & \colhead{$\sigma_M$} & \colhead{(mag~d$^{-1}$)} & \colhead{(mag~d$^{-1}$)} & \colhead{(d)} & \colhead{(d)} & \colhead{log($t_2$)} & \colhead{Ref\tablenotemark{a}}
}
\startdata
1997-08b & 16.42 & 0.09 & $ -8.09 $ & 0.10 &0.0063 &0.0003 & 318.0 & 15.3 & 2.50 & 1 \cr
1998-07d & 15.81 & 0.11 & $ -8.71 $ & 0.12 &0.0098 &0.0007 & 204.8 & 14.0 & 2.31 & 1 \cr
1998-08a & 16.05 & 0.16 & $ -8.46 $ & 0.16 &0.0086 &0.0009 & 232.7 & 24.6 & 2.37 & 1 \cr
1998-08b & 15.41 & 0.29 & $ -9.10 $ & 0.30 &0.0245 &0.0038 &  81.6 & 12.6 & 1.91 & 1 \cr
2000-10a & 16.64 & 0.02 & $ -7.87 $ & 0.05 &0.0321 &0.0003 &  62.2 &  0.6 & 1.79 & 1 \cr
2002-08a & 15.60 & 0.10 & $ -8.91 $ & 0.11 &0.0174 &0.0018 & 115.1 & 12.0 & 2.06 & 1 \cr
2003-06b & 16.05 & 0.11 & $ -8.46 $ & 0.12 &0.0220 &0.0015 &  91.0 &  6.1 & 1.96 & 1 \cr
2003-06c & 15.12 & 0.26 & $ -9.40 $ & 0.27 &0.0262 &0.0034 &  76.2 &  9.8 & 1.88 & 1 \cr
2003-06d & 15.95 & 0.16 & $ -8.57 $ & 0.16 &0.0224 &0.0021 &  89.3 &  8.4 & 1.95 & 1 \cr
2003-07b & 16.86 & 0.08 & $ -7.65 $ & 0.09 &0.0295 &0.0016 &  67.8 &  3.8 & 1.83 & 1 \cr
2003-08a & 16.69 & 0.13 & $ -7.83 $ & 0.14 &0.0210 &0.0041 &  95.5 & 18.5 & 1.98 & 1 \cr
2003-08b & 15.58 & 0.07 & $ -8.94 $ & 0.09 &0.0230 &0.0009 &  87.0 &  3.2 & 1.94 & 1 \cr
2003-09a & 15.88 & 0.09 & $ -8.63 $ & 0.11 &0.0311 &0.0014 &  64.4 &  2.8 & 1.81 & 1 \cr
2003-09b & 16.08 & 0.26 & $ -8.43 $ & 0.27 &0.0184 &0.0046 & 108.6 & 27.2 & 2.04 & 1 \cr
2003-10c & 15.77 & 0.01 & $ -8.74 $ & 0.05 &0.0219 &0.0008 &  91.2 &  3.2 & 1.96 & 1 \cr
2003-11a & 16.53 & 0.12 & $ -7.98 $ & 0.13 &0.0235 &0.0021 &  85.1 &  7.6 & 1.93 & 1 \cr
2003-11b & 15.26 & 0.11 & $ -9.25 $ & 0.12 &0.0368 &0.0029 &  54.3 &  4.2 & 1.73 & 1 \cr
2003-12a & 16.31 & 0.06 & $ -8.20 $ & 0.08 &0.0144 &0.0006 & 139.1 &  5.3 & 2.14 & 1 \cr
2003-12b & 16.20 & 0.04 & $ -8.32 $ & 0.06 &0.0182 &0.0016 & 109.7 &  9.4 & 2.04 & 1 \cr
2003-12c & 15.30 & 0.18 & $ -9.22 $ & 0.19 &0.1282 &0.0080 &  15.6 &  1.0 & 1.19 & 1 \cr
2004-01a & 15.23 & 0.07 & $ -9.29 $ & 0.09 &0.0134 &0.0003 & 148.8 &  3.8 & 2.17 & 1 \cr
2004-10a & 15.79 & 0.11 & $ -8.72 $ & 0.12 &0.0441 &0.0054 &  45.3 &  5.5 & 1.66 & 1 \cr
2005-10b & 15.77 & 0.10 & $ -8.75 $ & 0.11 &0.0238 &0.0018 &  84.2 &  6.2 & 1.93 & 1 \cr
2009-08e & 16.08 & 0.05 & $ -8.43 $ & 0.07 &0.0108 &0.0010 & 184.5 & 16.5 & 2.27 & 1 \cr
2009-10b & 15.60 & 0.12 & $ -8.92 $ & 0.13 &0.0084 &0.0004 & 237.8 & 12.2 & 2.38 & 1 \cr
2009-11b & 16.66 & 0.09 & $ -7.85 $ & 0.10 &0.0247 &0.0027 &  81.1 &  8.9 & 1.91 & 1 \cr
2009-11e & 15.84 & 0.10 & $ -8.67 $ & 0.11 &0.0314 &0.0037 &  63.7 &  7.5 & 1.80 & 1 \cr
2010-06a & 17.15 & 0.11 & $ -7.36 $ & 0.12 &0.0218 &0.0022 &  91.7 &  9.2 & 1.96 & 1 \cr
2010-10a & 16.35 & 0.18 & $ -8.17 $ & 0.19 &0.0258 &0.0035 &  77.6 & 10.5 & 1.89 & 1 \cr
2010-10d & 16.19 & 0.36 & $ -8.33 $ & 0.36 &0.0308 &0.0094 &  64.8 & 19.7 & 1.81 & 1 \cr
1982-09e  & 15.42 & \dots & $  -9.09 $ & \dots & 0.0800 & \dots &  25.0 & \dots & 1.40 & 2 \cr
1985-10b  & 14.76 & \dots & $  -9.75 $ & \dots & 0.0290 & \dots &  69.0 & \dots & 1.84 & 2 \cr
1986-09a  & 15.68 & \dots & $  -8.83 $ & \dots & 0.0130 & \dots & 153.8 & \dots & 2.19 & 2 \cr
1992-12a  & 16.20 & \dots & $  -8.31 $ & \dots & 0.0070 & \dots & 285.7 & \dots & 2.46 & 3 \cr
1995-08d  & 18.40 & \dots & $  -6.11 $ & \dots & 0.0460 & \dots &  43.5 & \dots & 1.64 & 3 \cr
1995-08e  & 16.10 & \dots & $  -8.41 $ & \dots & 0.0150 & \dots & 133.3 & \dots & 2.12 & 3 \cr
1995-11d  & 16.60 & \dots & $  -7.91 $ & \dots & 0.0090 & \dots & 222.2 & \dots & 2.35 & 3 \cr
2003-01b\tablenotemark{b} & 15.60\tablenotemark{c} & \dots & $  -8.91 $ & \dots & 0.0210 & \dots &  95.2 & \dots & 1.98 & 4 \cr
2003-02a\tablenotemark{b} & 14.90\tablenotemark{c} & \dots & $  -9.61 $ & \dots & 0.1290 & \dots &  15.5 & \dots & 1.19 & 4 \cr
2003-02b\tablenotemark{b} & 15.80\tablenotemark{c} & \dots & $  -8.71 $ & \dots & 0.0310 & \dots &  64.5 & \dots & 1.81 & 4 \cr
2003-03a\tablenotemark{b} & 14.60\tablenotemark{c} & \dots & $  -9.91 $ & \dots & 0.0780 & \dots &  25.6 & \dots & 1.41 & 4 \cr
2003-05d\tablenotemark{b} & 15.10\tablenotemark{c} & \dots & $  -9.41 $ & \dots & 0.0360 & \dots &  55.6 & \dots & 1.75 & 4 \cr
2003-06a\tablenotemark{b} & 15.50\tablenotemark{c} & \dots & $  -9.01 $ & \dots & 0.0870 & \dots &  23.0 & \dots & 1.36 & 4 
\enddata
\tablenotetext{a}{(1) This work; (2) \citet{1990ApJ...356..472C}; (3) \citet{2001ApJ...563..749S}; (4) \citet{2004AJ....127..816N}
\tablenotetext{b}{M81N}
\tablenotetext{c}{Corrected for \ha\ bandpass (30\AA\ vs 75\AA) and $\Delta m$, adopting $\mu_\mathrm{o}(\mathrm{M81})=27.8$ \citep[][]{2001ApJ...553...47F}.}
}
\end{deluxetable*}

\figsetstart
\figsetnum{5}
\figsettitle{Light curves for novae used in the Monte Carlo simulations}

\figsetgrpstart
\figsetgrpnum{5.1}
\figsetgrptitle{Light curve for nova M31N~1997-08b}
\figsetplot{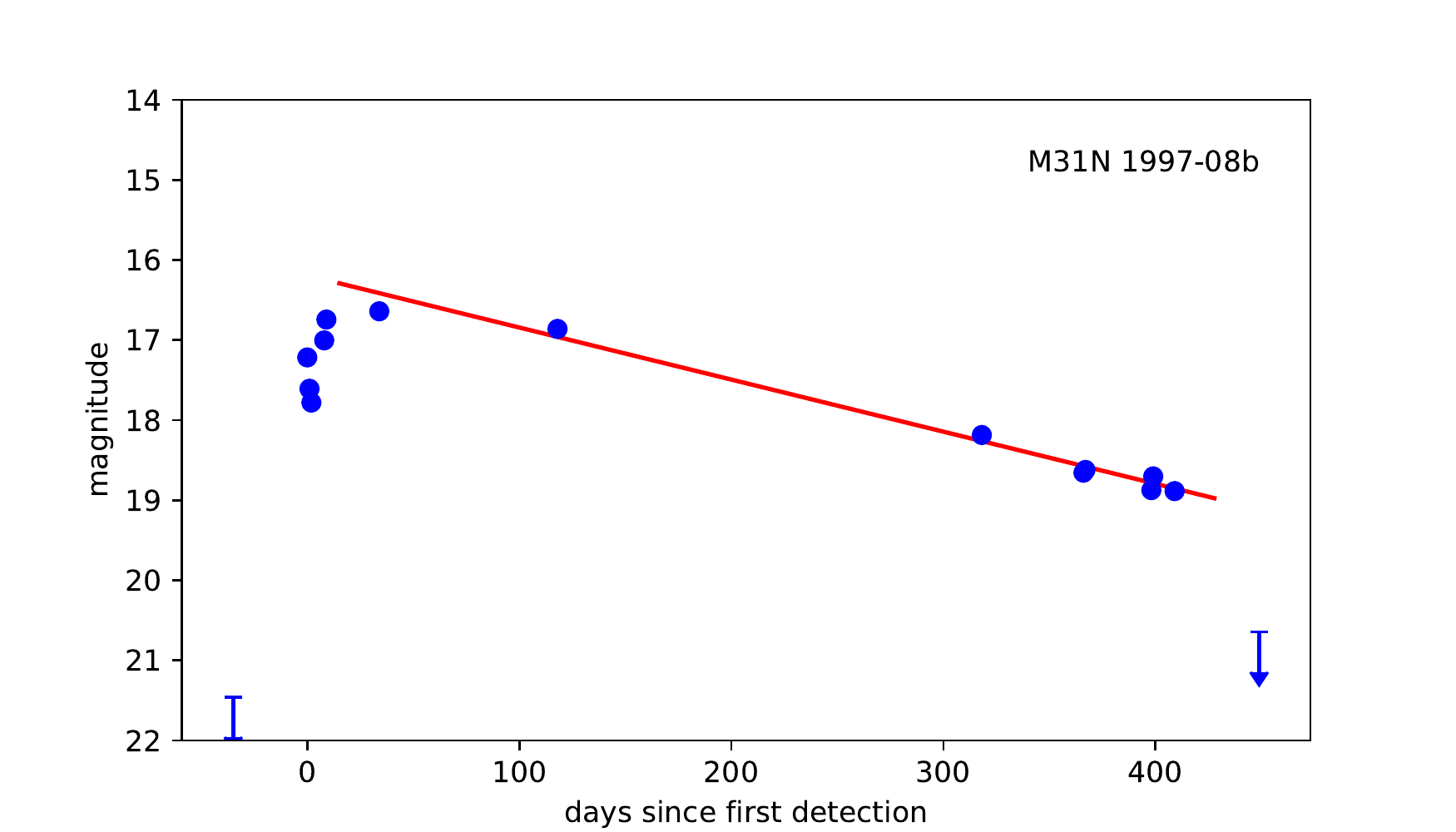}
\figsetgrpnote{The vertical axis shows the magnitude.  The horizontal axis shows the number of days since first detection.}
\figsetgrpend

\figsetgrpstart
\figsetgrpnum{5.2}
\figsetgrptitle{Light curve for nova M31N~1998-07d}
\figsetplot{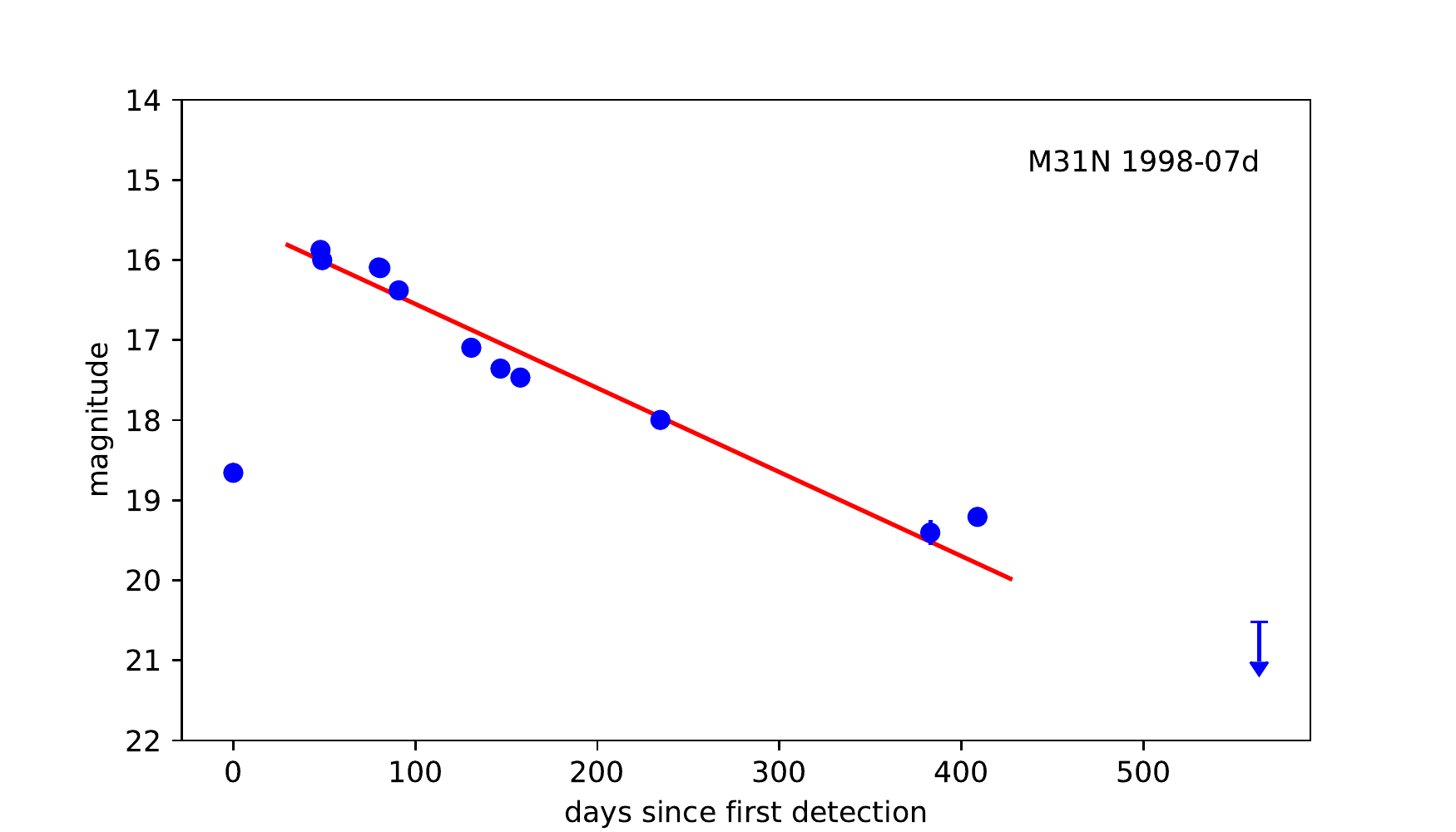}
\figsetgrpnote{The vertical axis shows the \ha\ magnitude.  The horizontal axis shows the number of days since first detection.}
\figsetgrpend

\figsetgrpstart
\figsetgrpnum{5.3}
\figsetgrptitle{Light curve for nova M31N~1998-08a}
\figsetplot{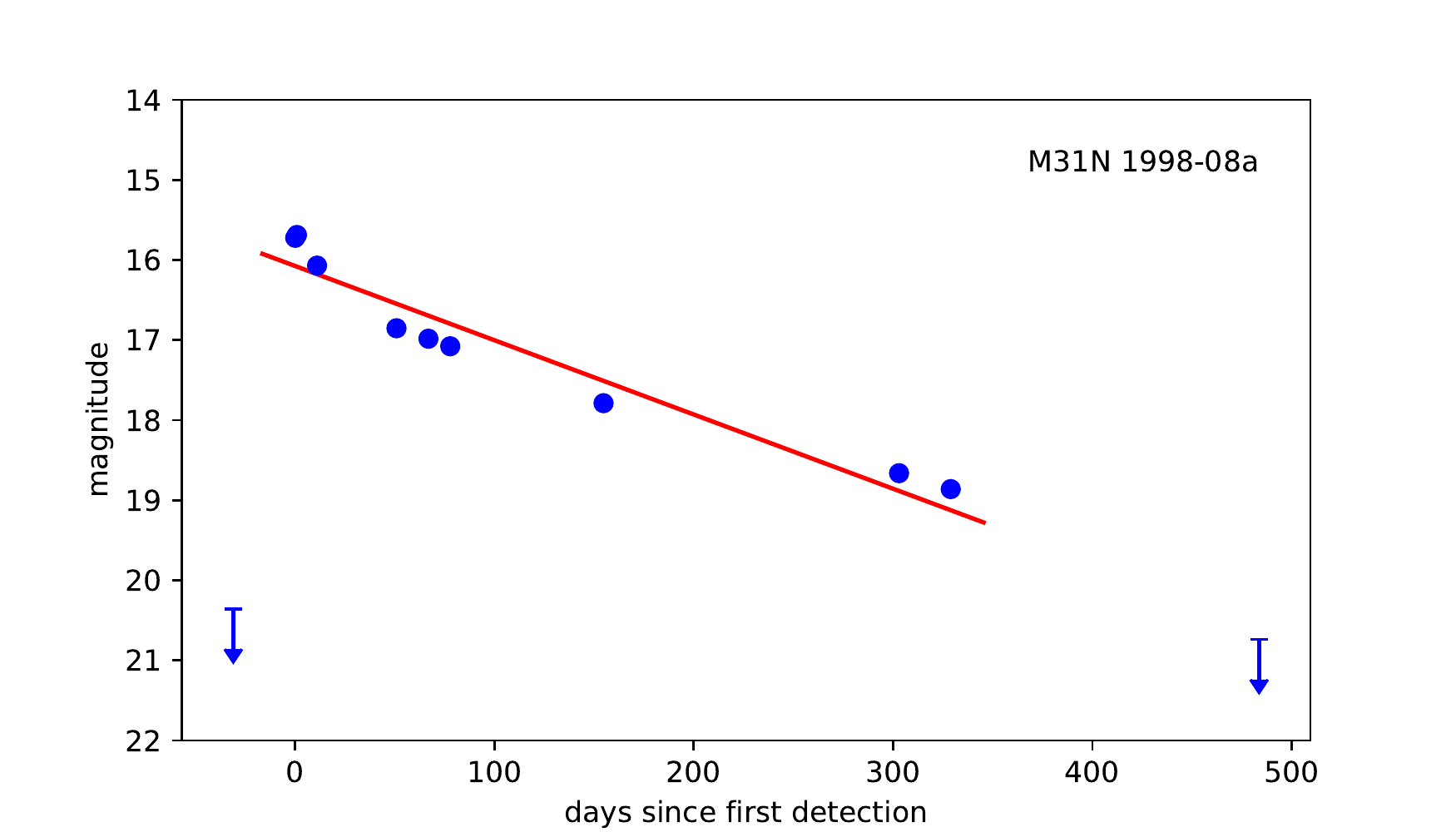}
\figsetgrpnote{The vertical axis shows the \ha\ magnitude.  The horizontal axis shows the number of days since first detection.}
\figsetgrpend

\figsetgrpstart
\figsetgrpnum{5.4}
\figsetgrptitle{Light curve for nova M31N~1998-08b}
\figsetplot{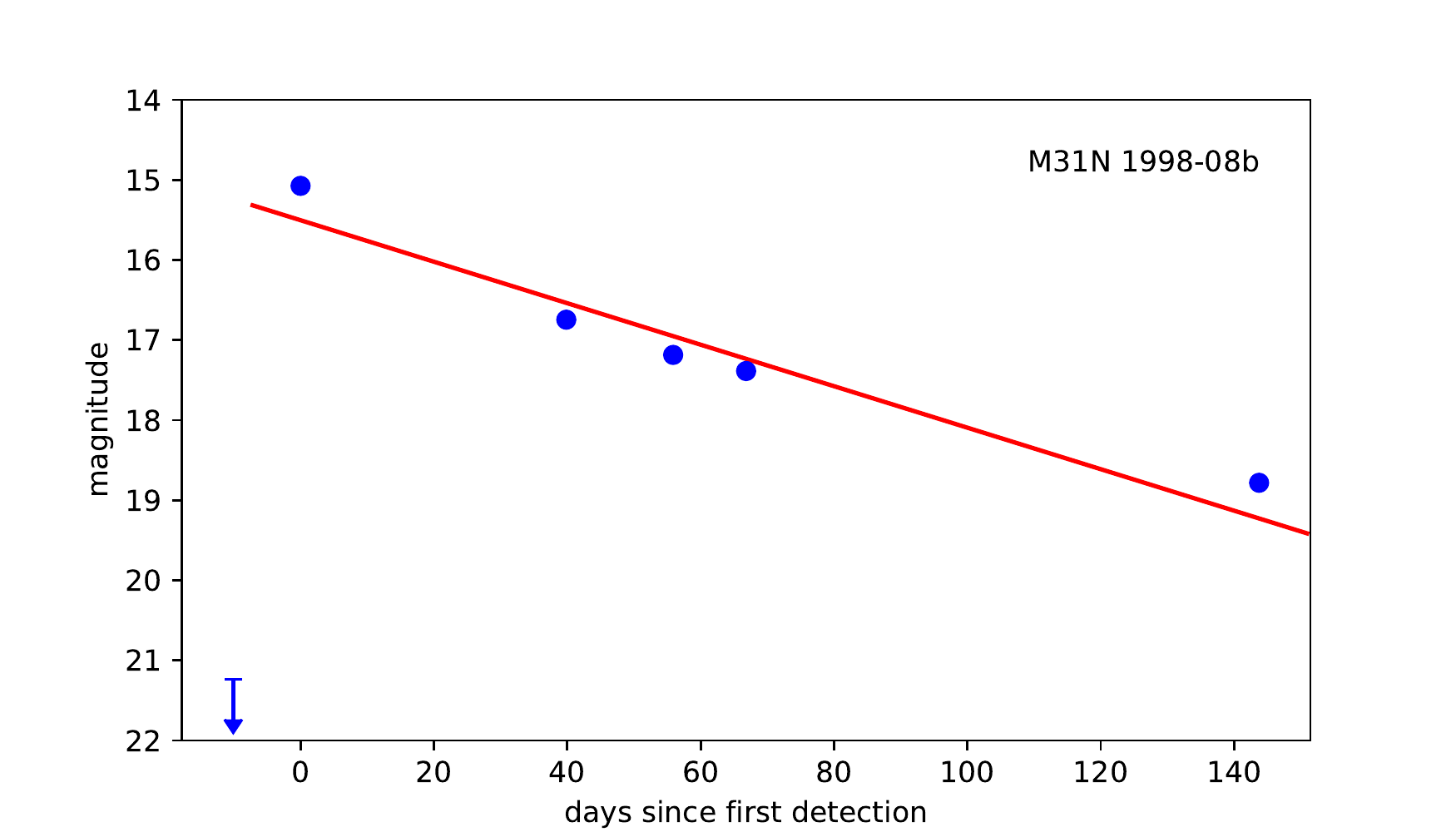}
\figsetgrpnote{The vertical axis shows the \ha\ magnitude.  The horizontal axis shows the number of days since first detection.}
\figsetgrpend

\figsetgrpstart
\figsetgrpnum{5.5}
\figsetgrptitle{Light curve for nova M31N~2000-10a}
\figsetplot{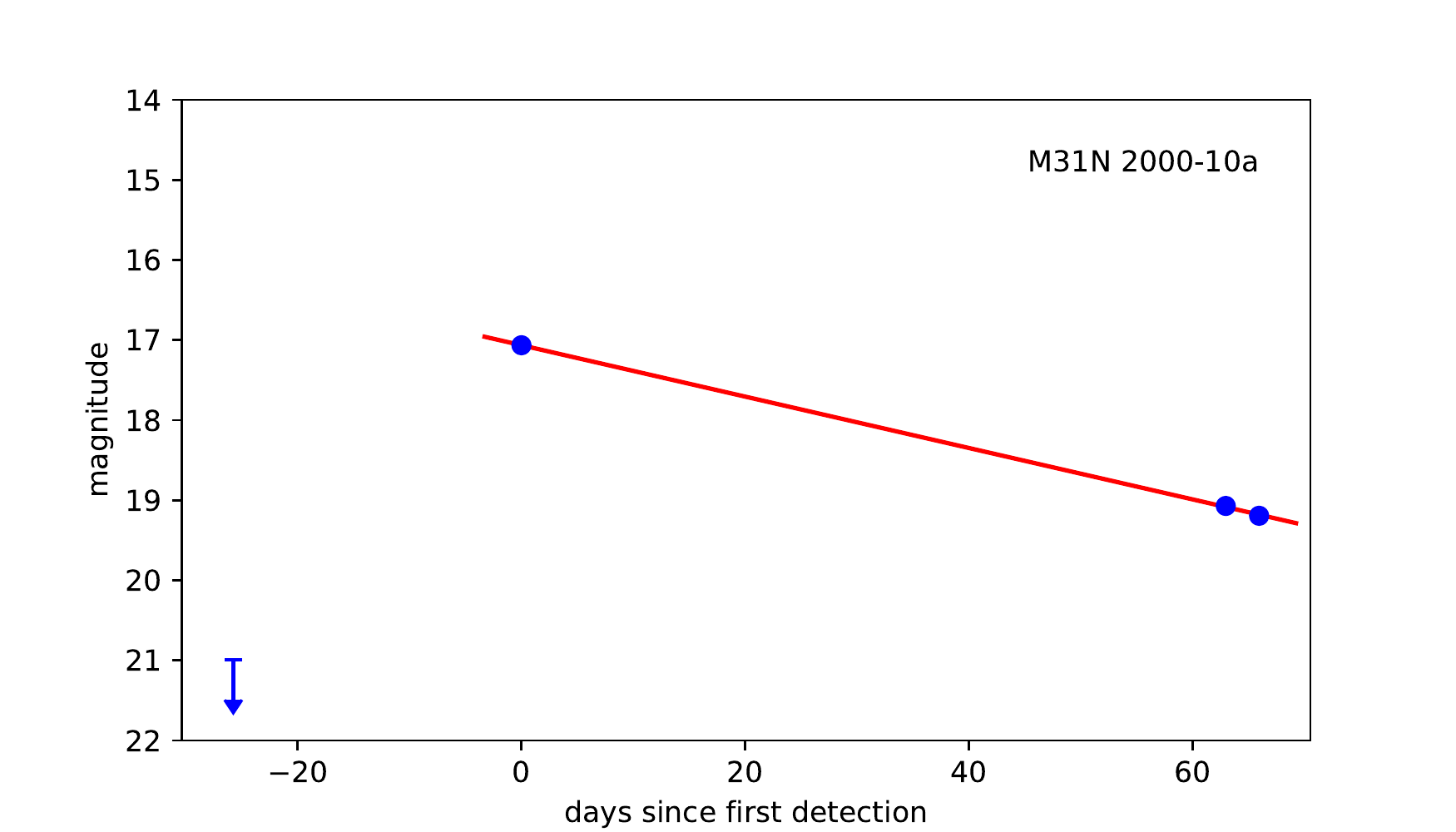}
\figsetgrpnote{The vertical axis shows the \ha\ magnitude.  The horizontal axis shows the number of days since first detection.}
\figsetgrpend

\figsetgrpstart
\figsetgrpnum{5.6}
\figsetgrptitle{Light curve for nova M31N~2002-08a}
\figsetplot{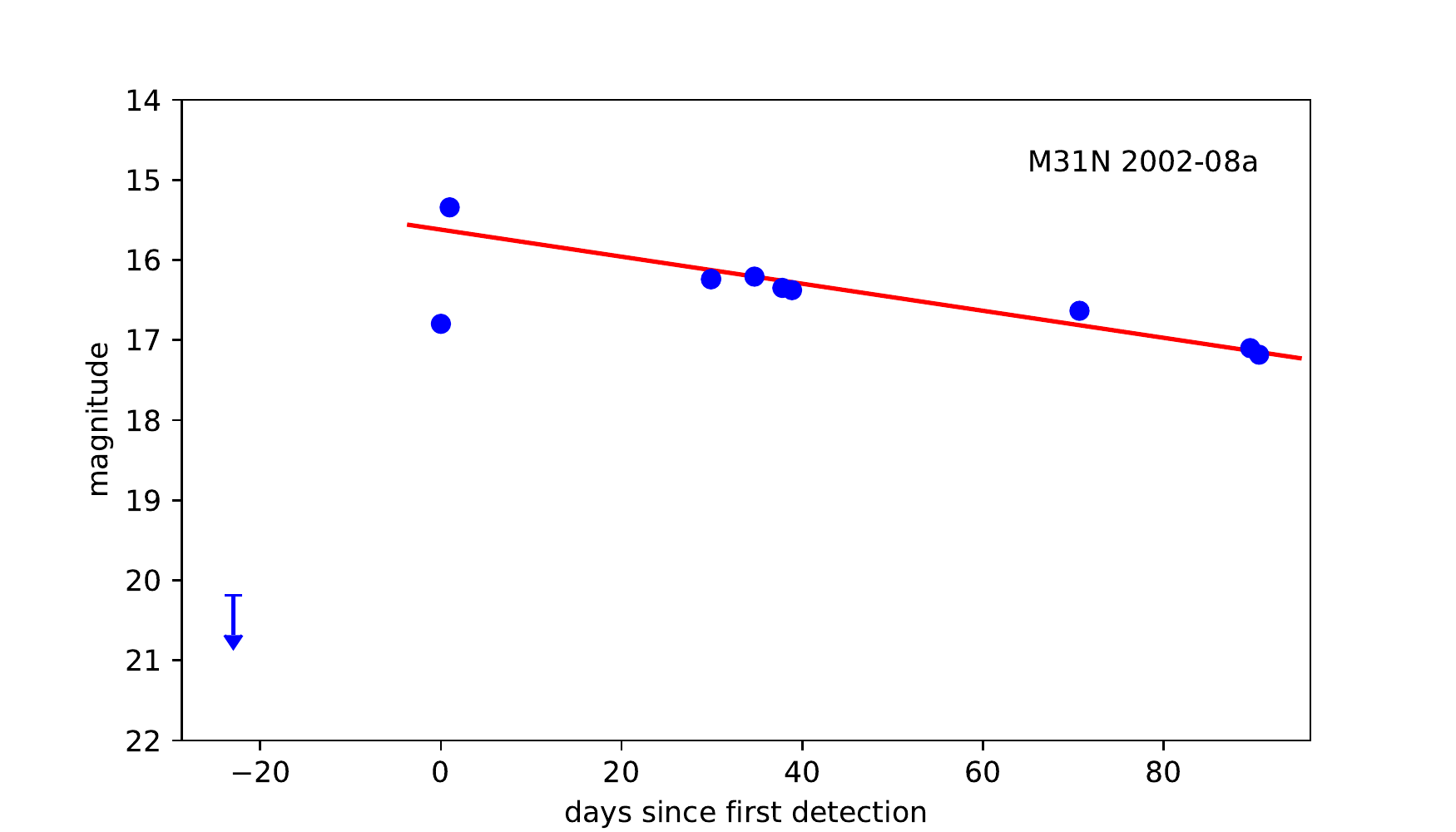}
\figsetgrpnote{The vertical axis shows the \ha\ magnitude.  The horizontal axis shows the number of days since first detection.}
\figsetgrpend

\figsetgrpstart
\figsetgrpnum{5.7}
\figsetgrptitle{Light curve for nova M31N~2003-06b}
\figsetplot{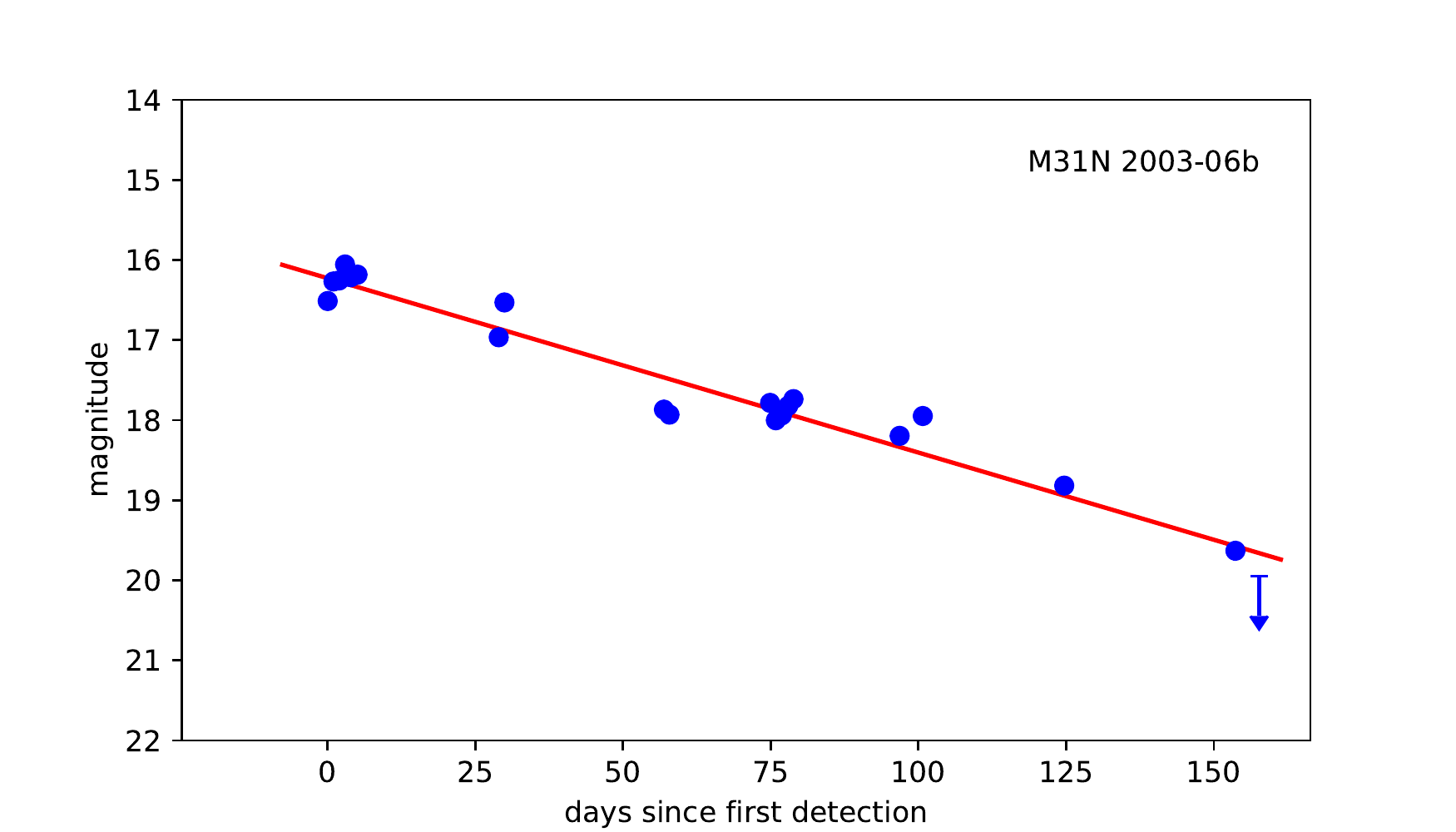}
\figsetgrpnote{The vertical axis shows the \ha\ magnitude.  The horizontal axis shows the number of days since first detection.}
\figsetgrpend

\figsetgrpstart
\figsetgrpnum{5.8}
\figsetgrptitle{Light curve for nova M31N~2003-06c}
\figsetplot{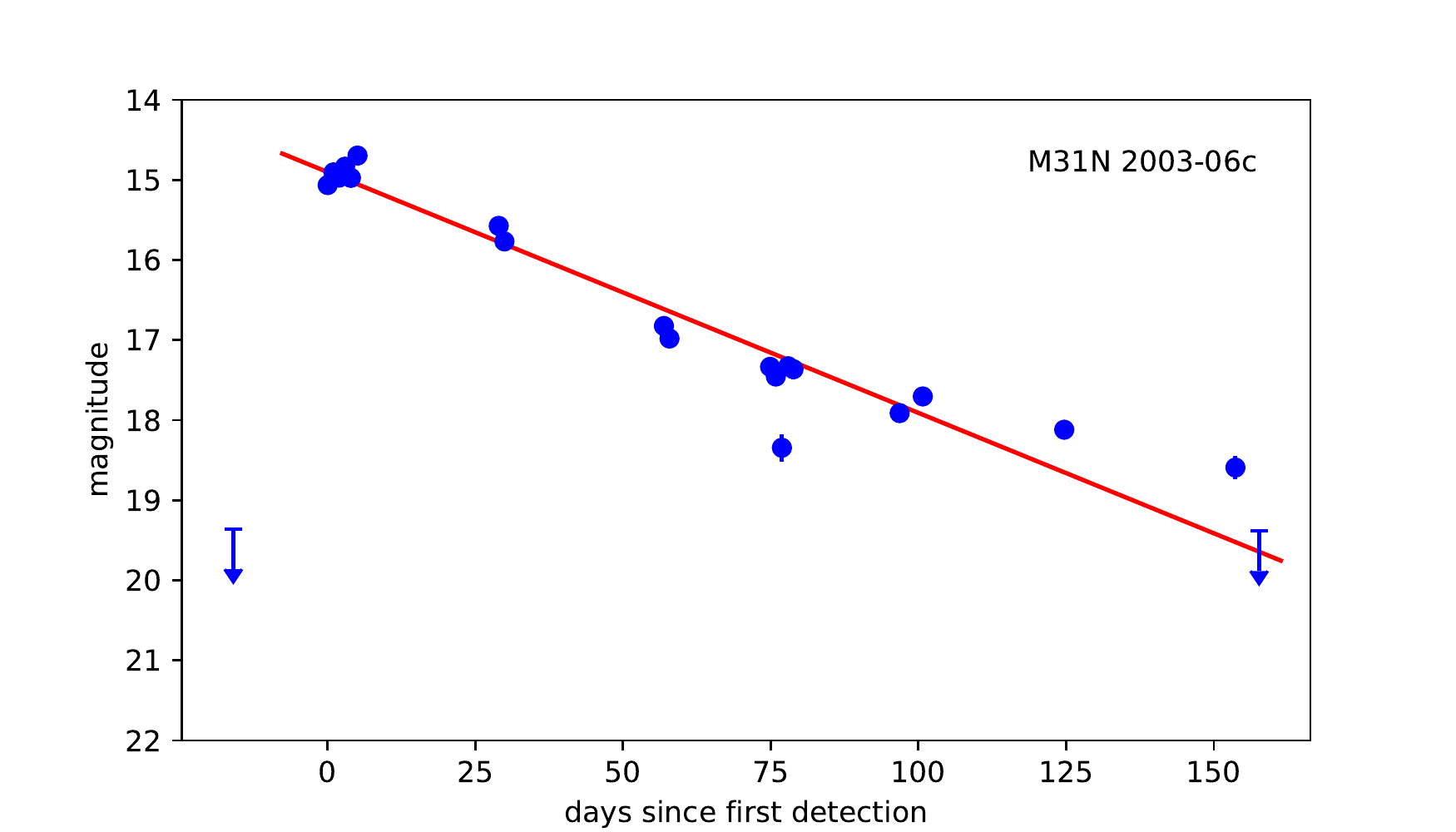}
\figsetgrpnote{The vertical axis shows the \ha\ magnitude.  The horizontal axis shows the number of days since first detection.}
\figsetgrpend

\figsetgrpstart
\figsetgrpnum{5.9}
\figsetgrptitle{Light curve for nova M31N~2003-06d}
\figsetplot{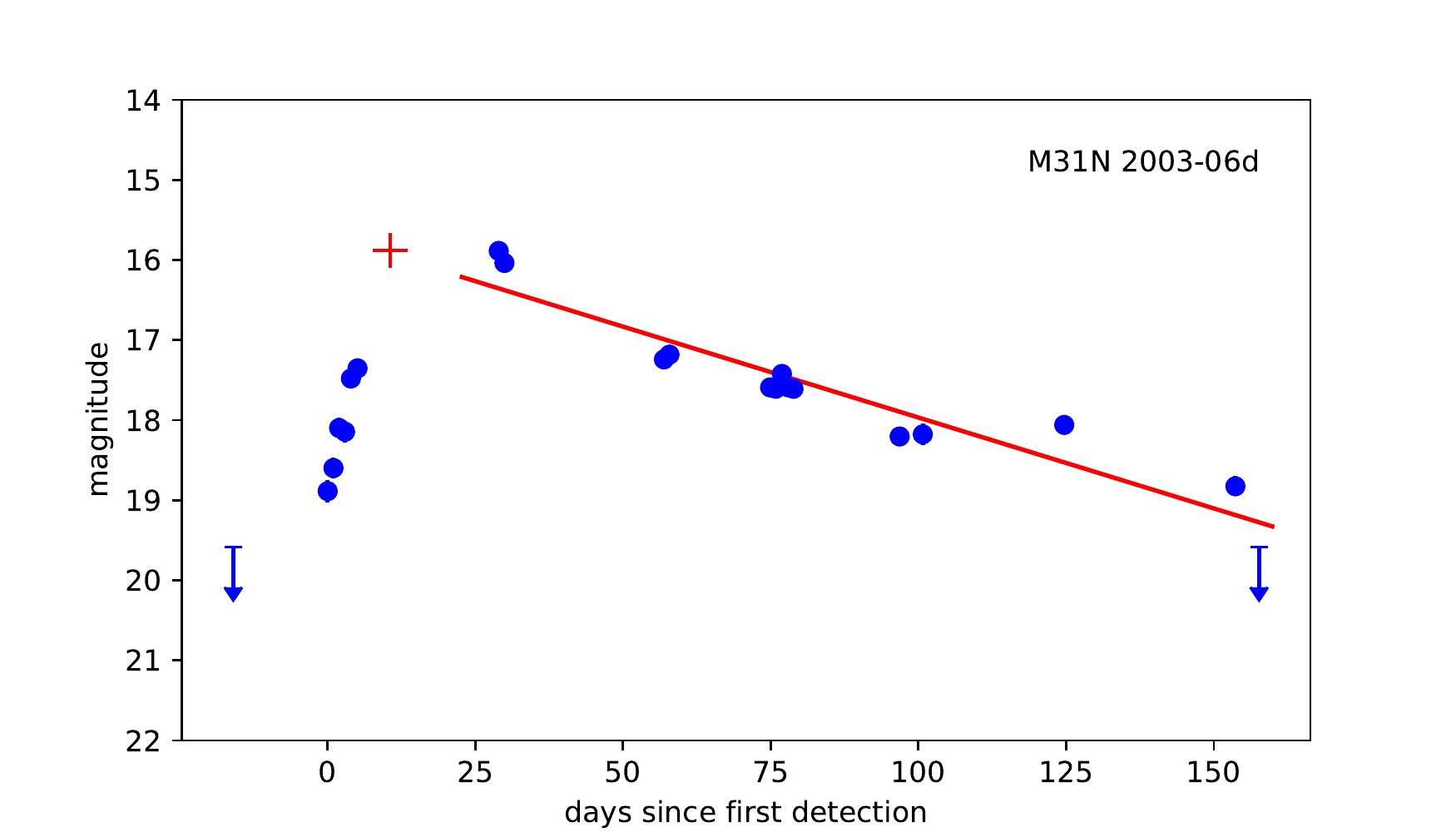}
\figsetgrpnote{The vertical axis shows the \ha\ magnitude.  The horizontal axis shows the number of days since first detection.  The red plus sign marks the estimated location of peak magnitude, as extrapolated from the rise and decay slopes.}
\figsetgrpend

\figsetgrpstart
\figsetgrpnum{5.10}
\figsetgrptitle{Light curve for nova M31N~2003-07b}
\figsetplot{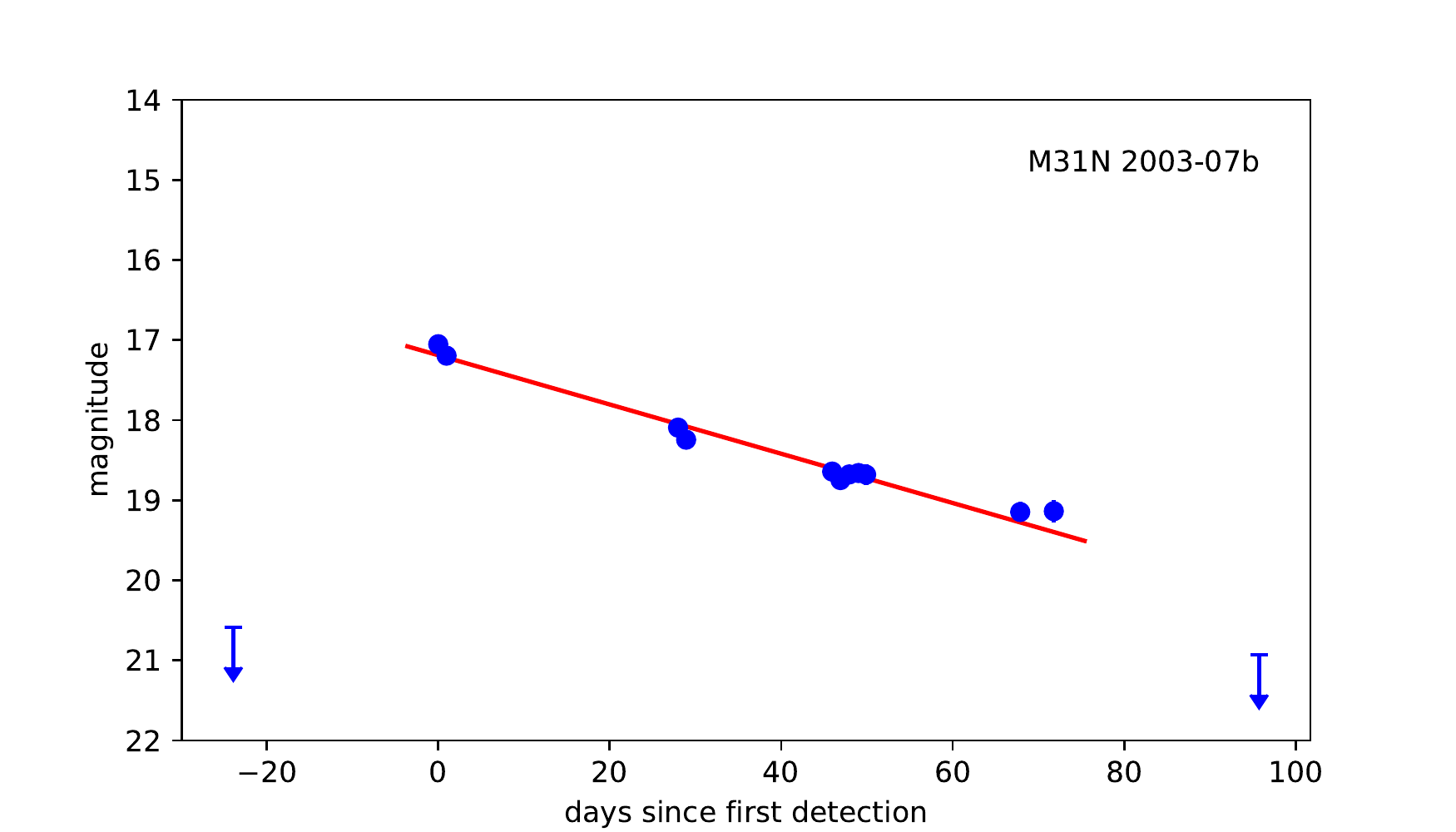}
\figsetgrpnote{The vertical axis shows the \ha\ magnitude.  The horizontal axis shows the number of days since first detection.}
\figsetgrpend

\figsetgrpstart
\figsetgrpnum{5.11}
\figsetgrptitle{Light curve for nova M31N~2003-08a}
\figsetplot{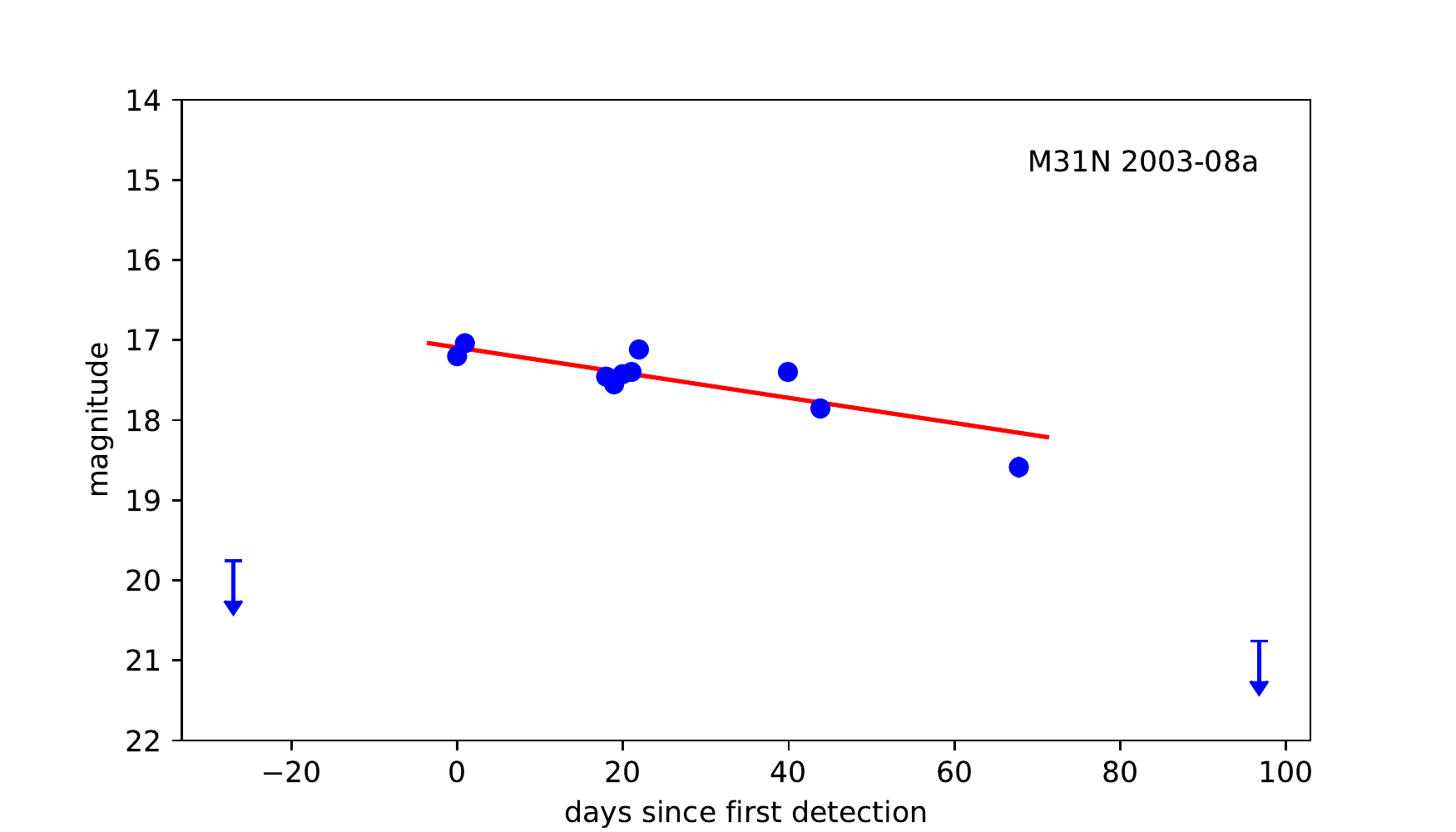}
\figsetgrpnote{The vertical axis shows the \ha\ magnitude.  The horizontal axis shows the number of days since first detection.}
\figsetgrpend

\figsetgrpstart
\figsetgrpnum{5.12}
\figsetgrptitle{Light curve for nova M31N~2003-08b}
\figsetplot{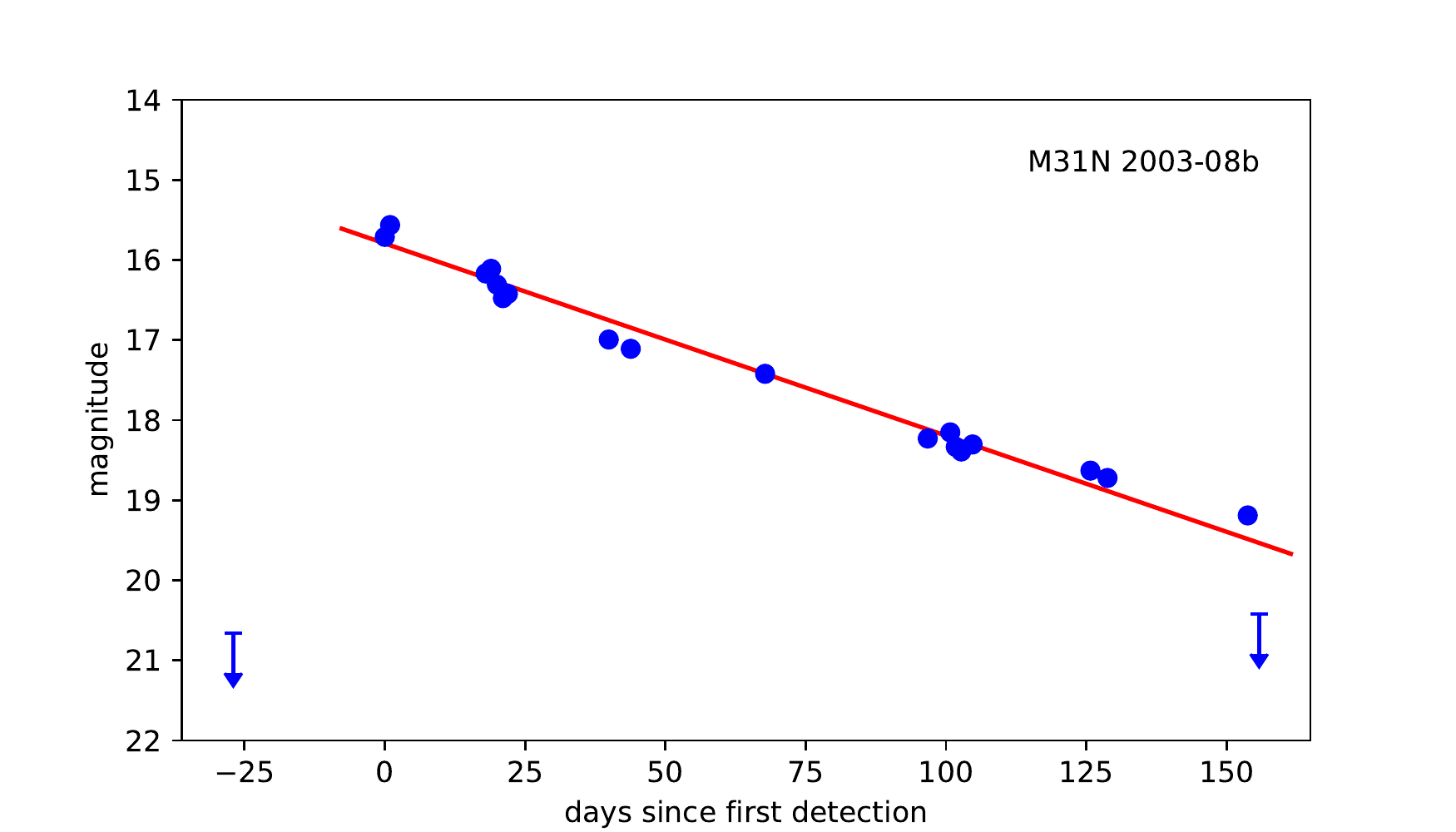}
\figsetgrpnote{The vertical axis shows the \ha\ magnitude.  The horizontal axis shows the number of days since first detection.}
\figsetgrpend

\figsetgrpstart
\figsetgrpnum{5.13}
\figsetgrptitle{Light curve for nova M31N~2003-09a}
\figsetplot{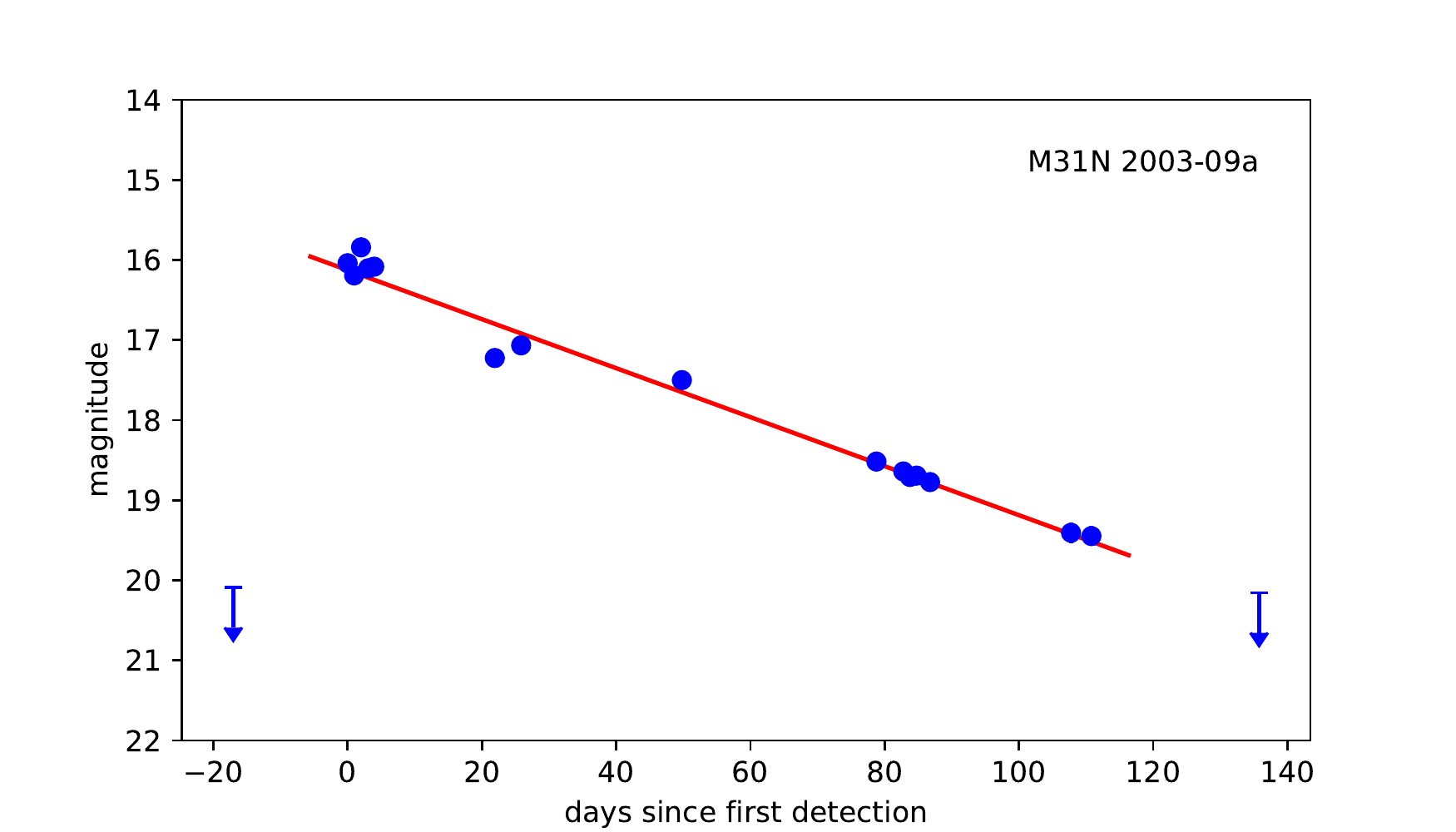}
\figsetgrpnote{The vertical axis shows the \ha\ magnitude.  The horizontal axis shows the number of days since first detection.}
\figsetgrpend

\figsetgrpstart
\figsetgrpnum{5.14}
\figsetgrptitle{Light curve for nova M31N~2003-09b}
\figsetplot{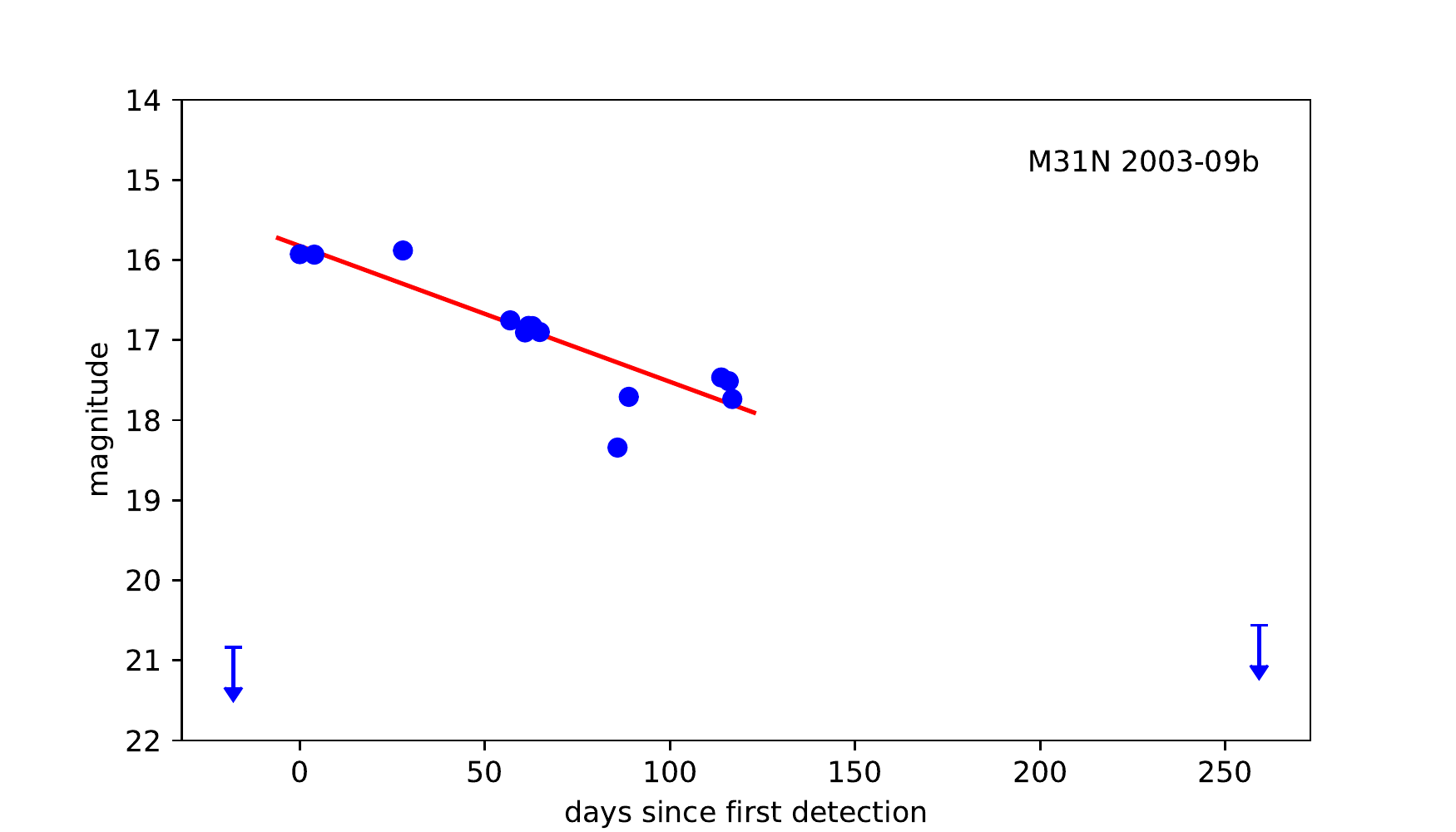}
\figsetgrpnote{The vertical axis shows the \ha\ magnitude.  The horizontal axis shows the number of days since first detection.}
\figsetgrpend

\figsetgrpstart
\figsetgrpnum{5.15}
\figsetgrptitle{Light curve for nova M31N~2003-10c}
\figsetplot{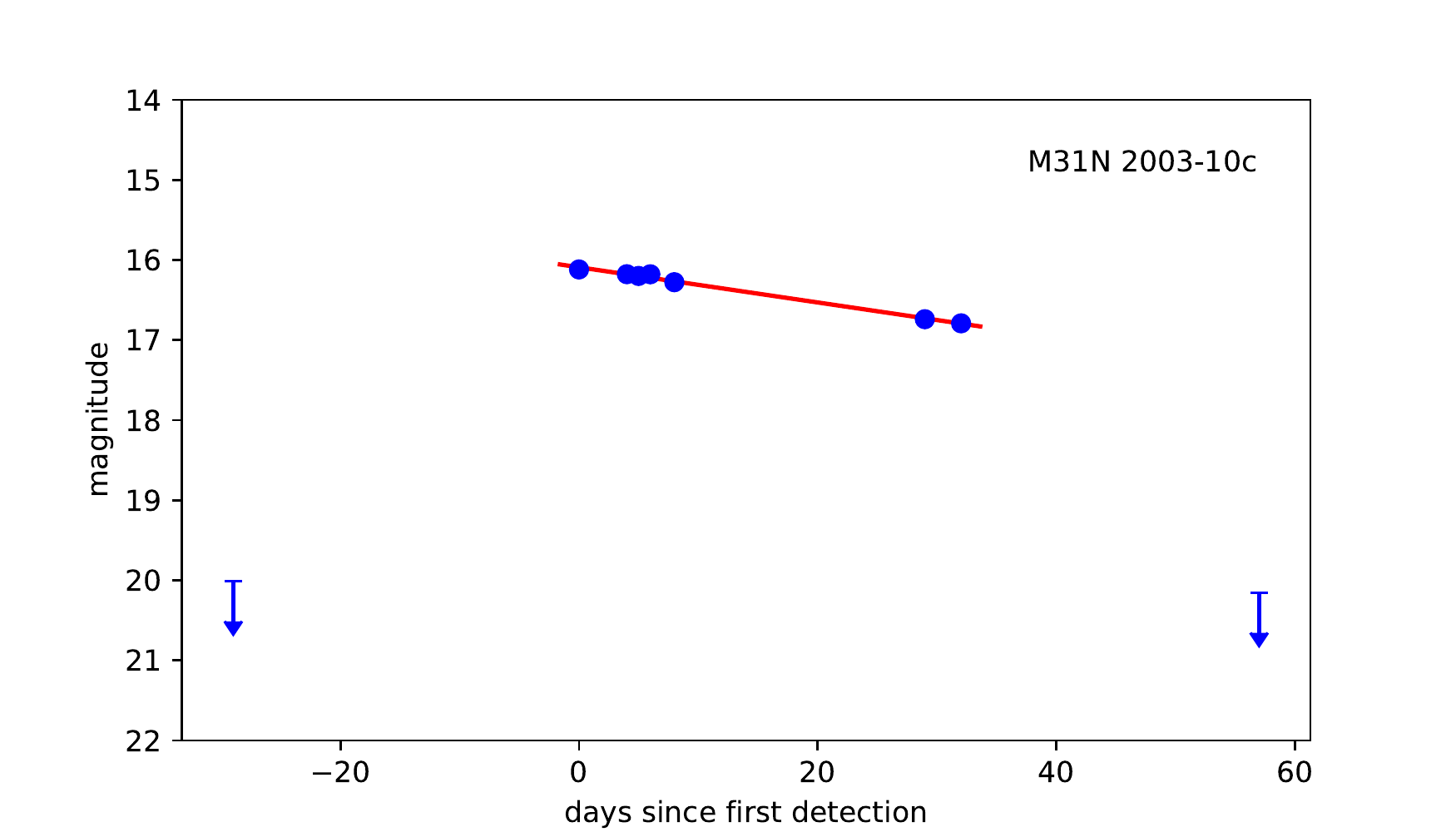}
\figsetgrpnote{The vertical axis shows the \ha\ magnitude.  The horizontal axis shows the number of days since first detection.}
\figsetgrpend

\figsetgrpstart
\figsetgrpnum{5.16}
\figsetgrptitle{Light curve for nova M31N~2003-11a}
\figsetplot{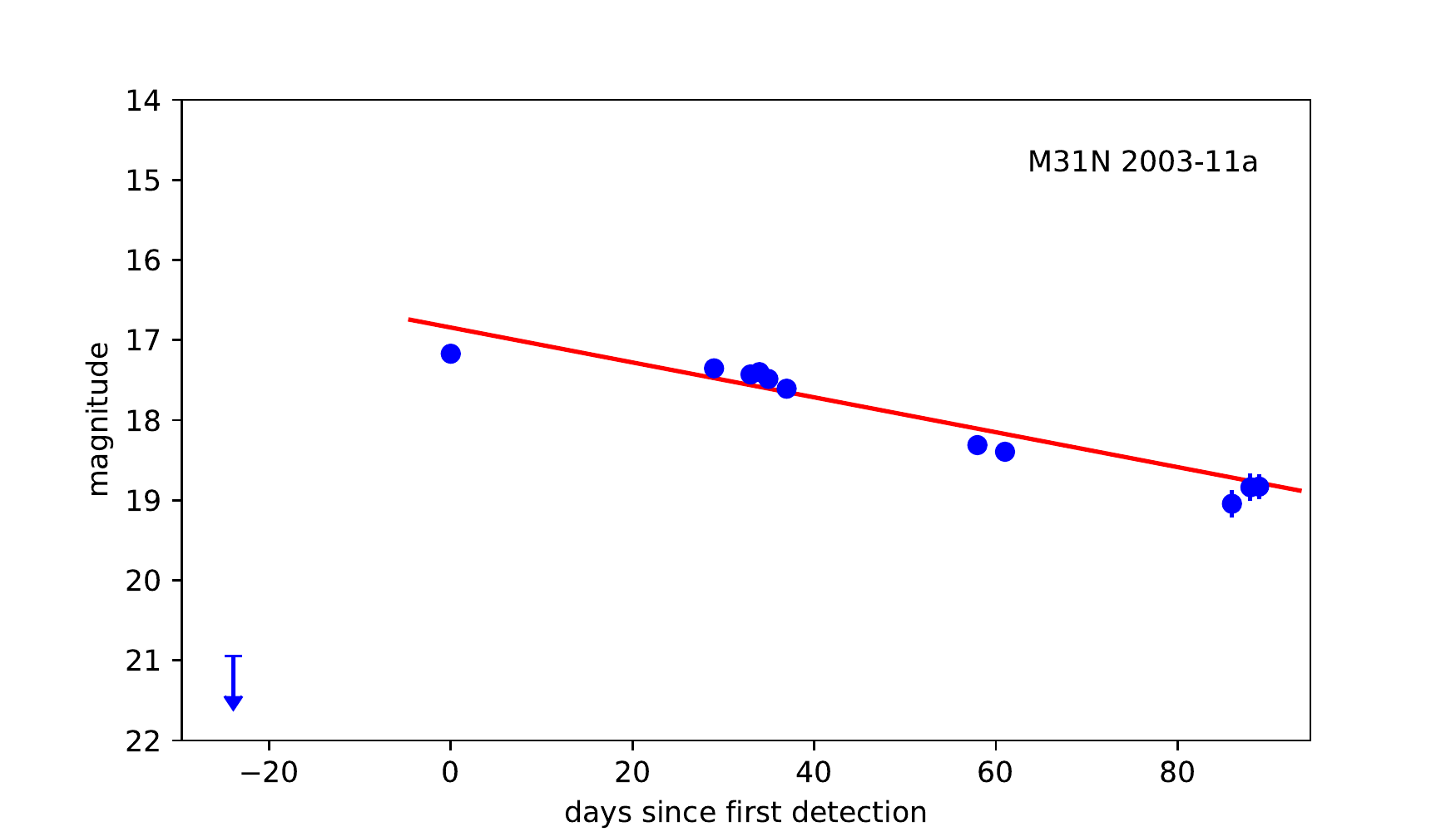}
\figsetgrpnote{The vertical axis shows the \ha\ magnitude.  The horizontal axis shows the number of days since first detection.}
\figsetgrpend

\figsetgrpstart
\figsetgrpnum{5.17}
\figsetgrptitle{Light curve for nova M31N~2003-11b}
\figsetplot{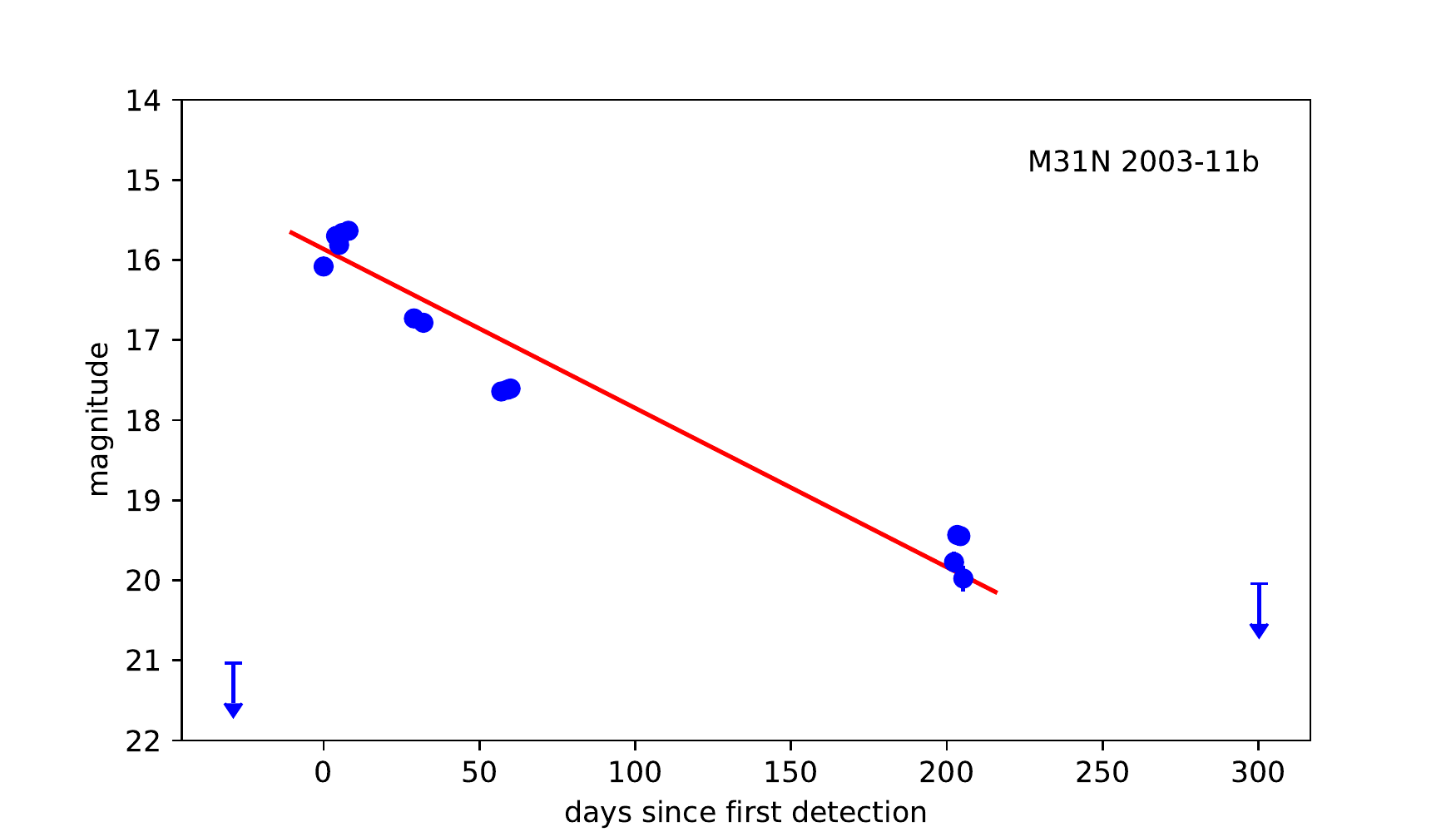}
\figsetgrpnote{The vertical axis shows the \ha\ magnitude.  The horizontal axis shows the number of days since first detection.}
\figsetgrpend

\figsetgrpstart
\figsetgrpnum{5.18}
\figsetgrptitle{Light curve for nova M31N~2003-12a}
\figsetplot{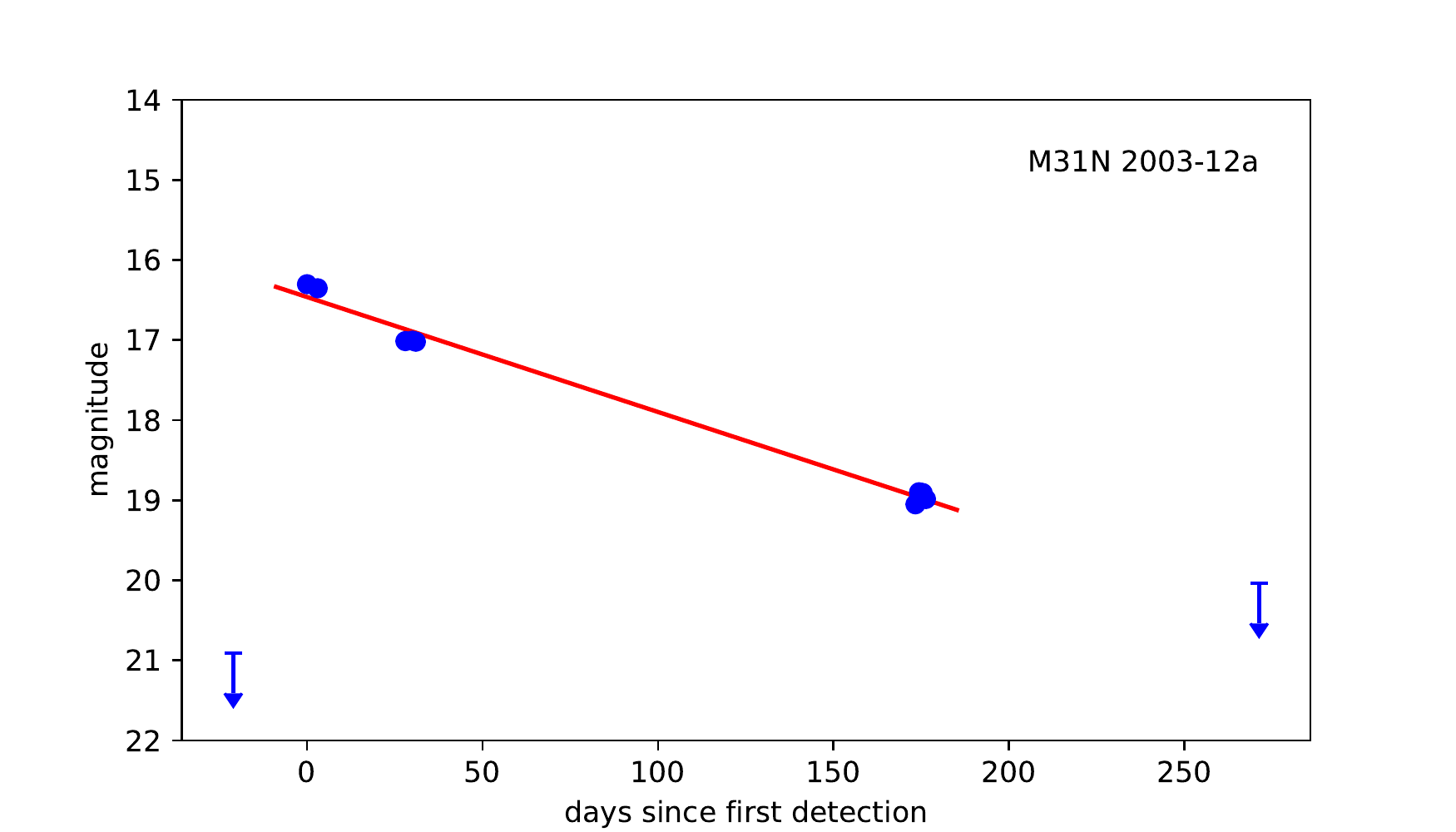}
\figsetgrpnote{The vertical axis shows the \ha\ magnitude.  The horizontal axis shows the number of days since first detection.}
\figsetgrpend

\figsetgrpstart
\figsetgrpnum{5.19}
\figsetgrptitle{Light curve for nova M31N~2003-12b}
\figsetplot{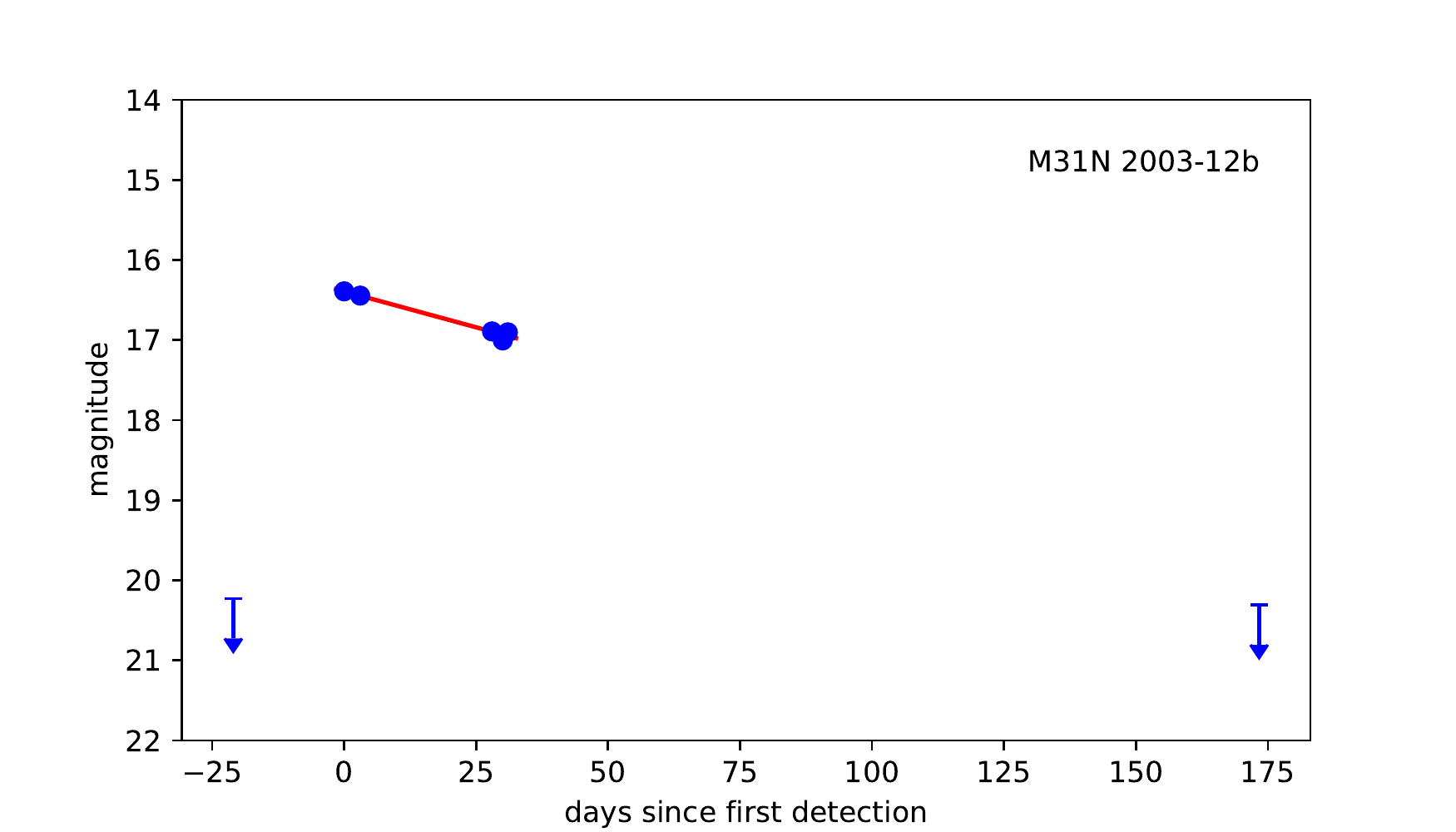}
\figsetgrpnote{The vertical axis shows the \ha\ magnitude.  The horizontal axis shows the number of days since first detection.}
\figsetgrpend

\figsetgrpstart
\figsetgrpnum{5.20}
\figsetgrptitle{Light curve for nova M31N~2003-12c}
\figsetplot{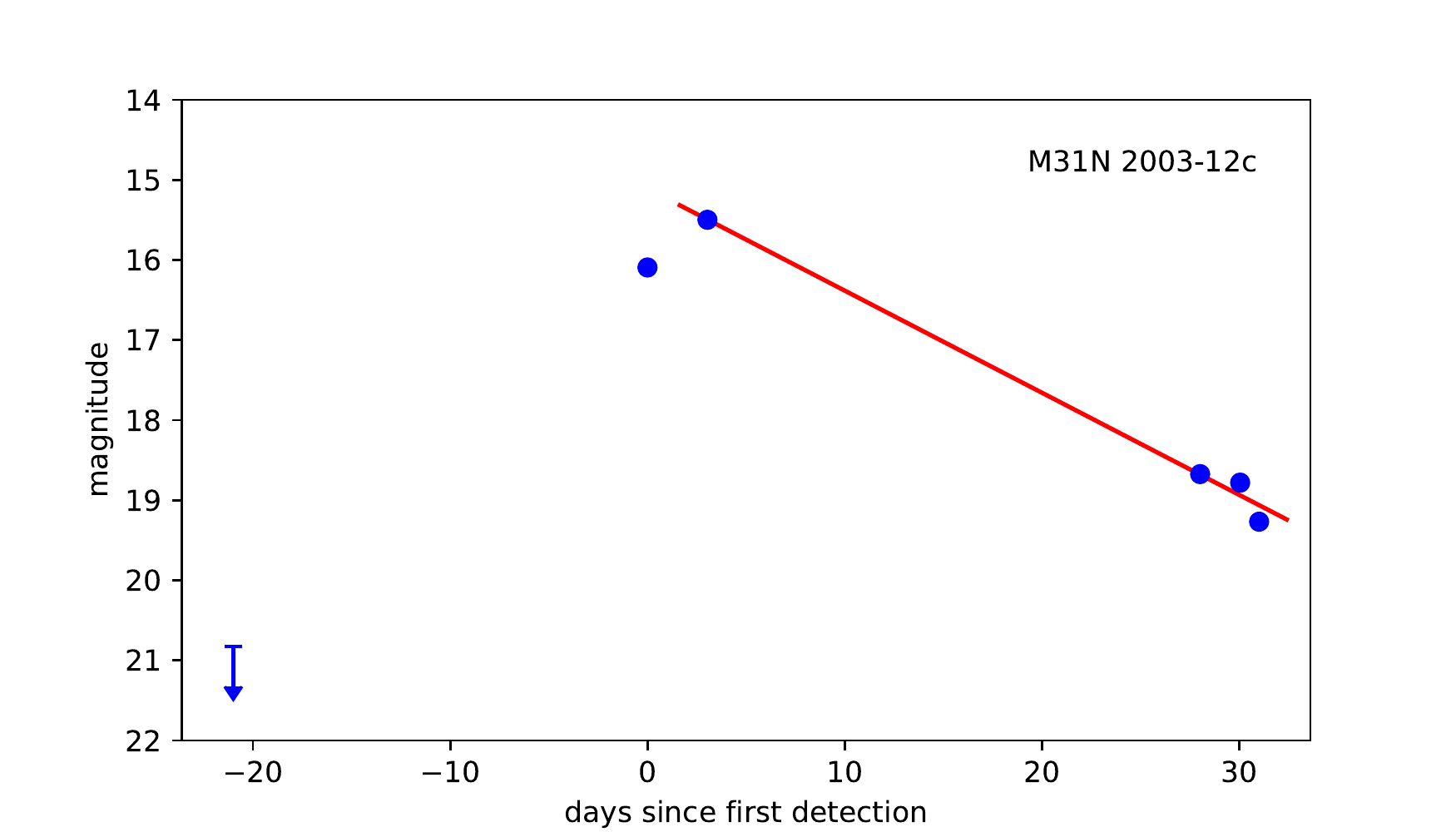}
\figsetgrpnote{The vertical axis shows the \ha\ magnitude.  The horizontal axis shows the number of days since first detection.}
\figsetgrpend

\figsetgrpstart
\figsetgrpnum{5.21}
\figsetgrptitle{Light curve for nova M31N~2004-01a}
\figsetplot{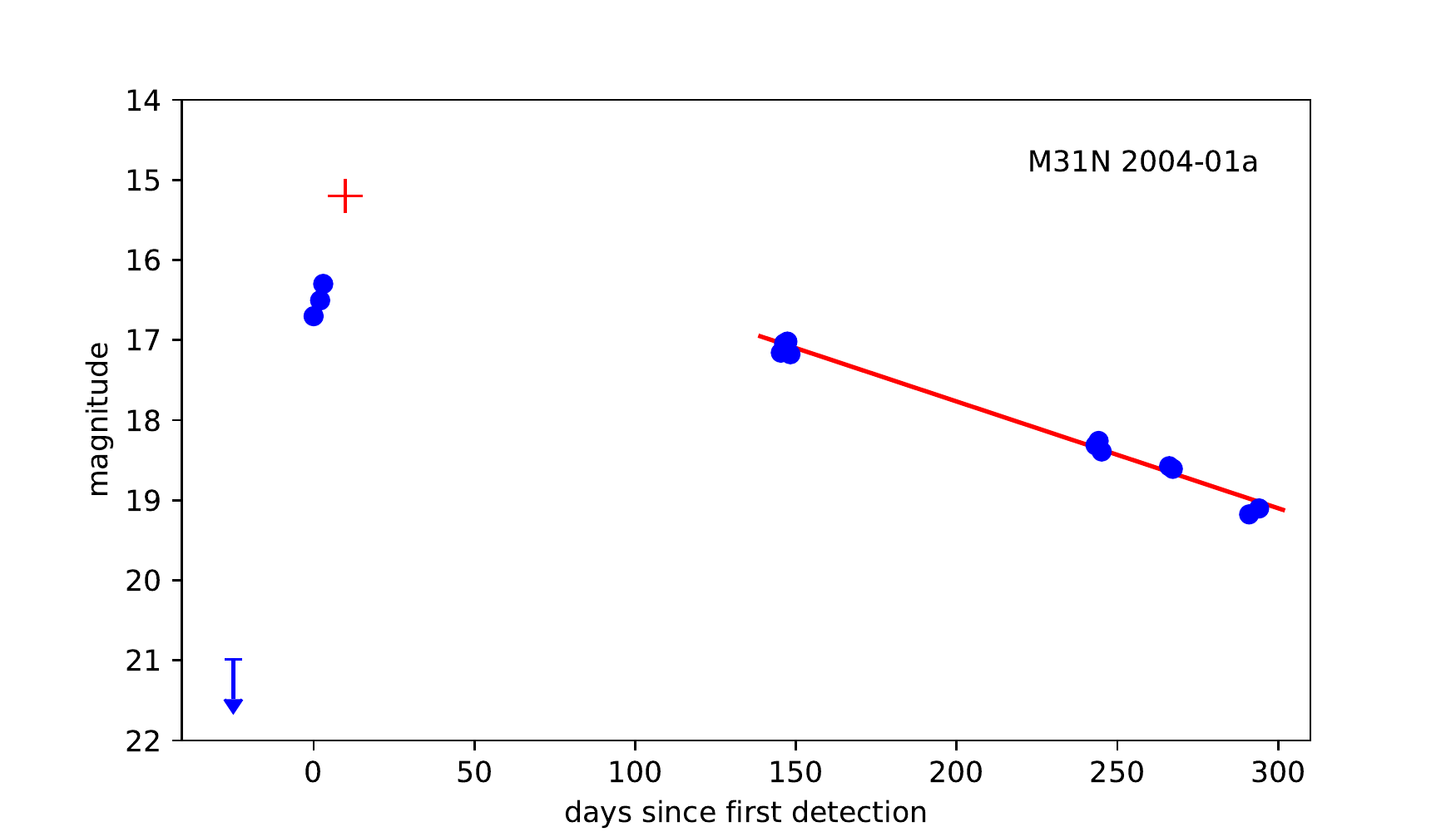}
\figsetgrpnote{The vertical axis shows the \ha\ magnitude.  The horizontal axis shows the number of days since first detection. The red plus sign marks the estimated location of peak magnitude, as extrapolated from the rise and decay slopes.}
\figsetgrpend

\figsetgrpstart
\figsetgrpnum{5.22}
\figsetgrptitle{Light curve for nova M31N~2004-10a}
\figsetplot{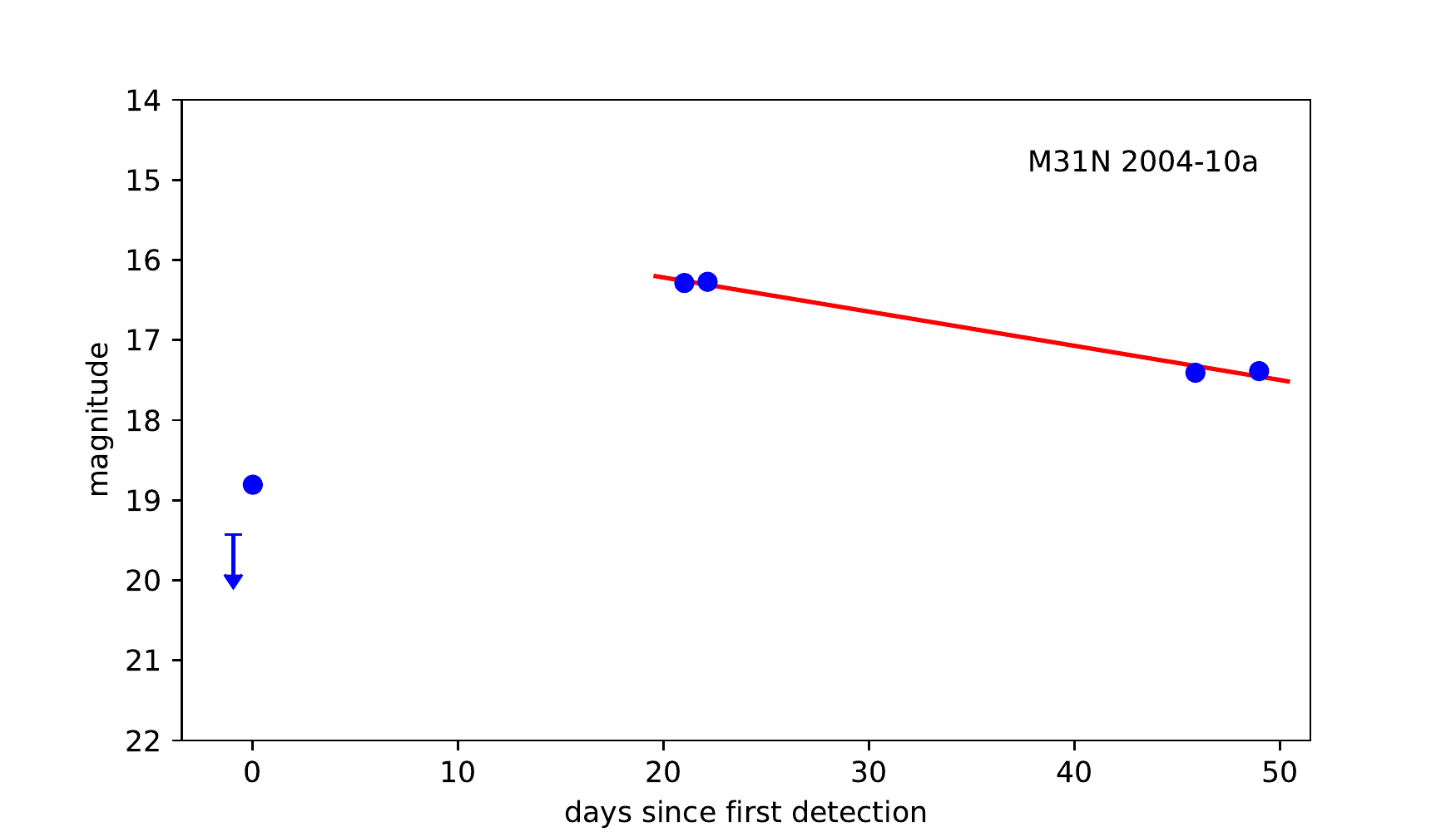}
\figsetgrpnote{The vertical axis shows the \ha\ magnitude.  The horizontal axis shows the number of days since first detection.}
\figsetgrpend

\figsetgrpstart
\figsetgrpnum{5.23}
\figsetgrptitle{Light curve for nova M31N~2005-10b}
\figsetplot{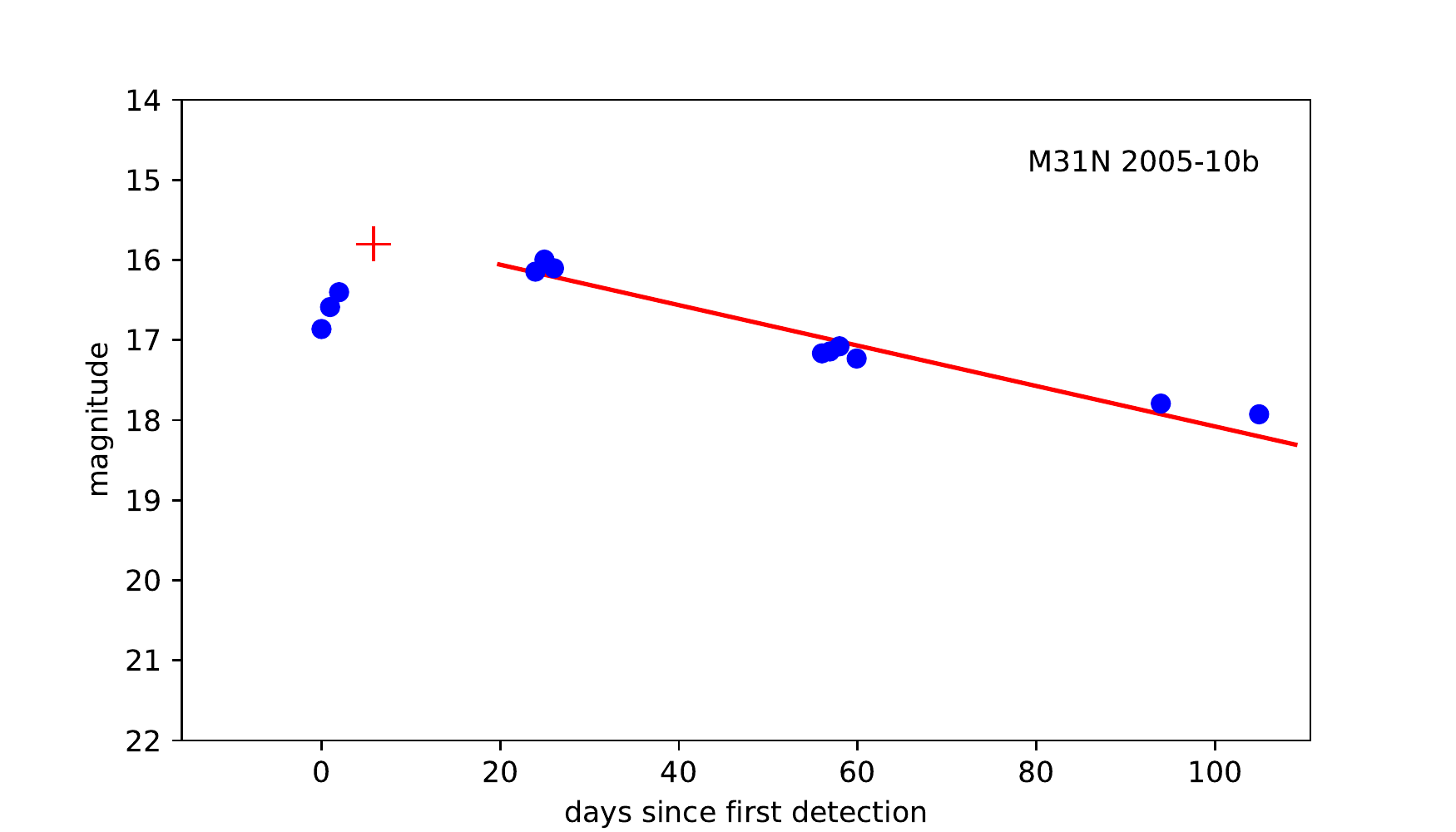}
\figsetgrpnote{The vertical axis shows the \ha\ magnitude.  The horizontal axis shows the number of days since first detection. The red plus sign marks the estimated location of peak magnitude, as extrapolated from the rise and decay slopes.}
\figsetgrpend

\figsetgrpstart
\figsetgrpnum{5.24}
\figsetgrptitle{Light curve for nova M31N~2009-08e}
\figsetplot{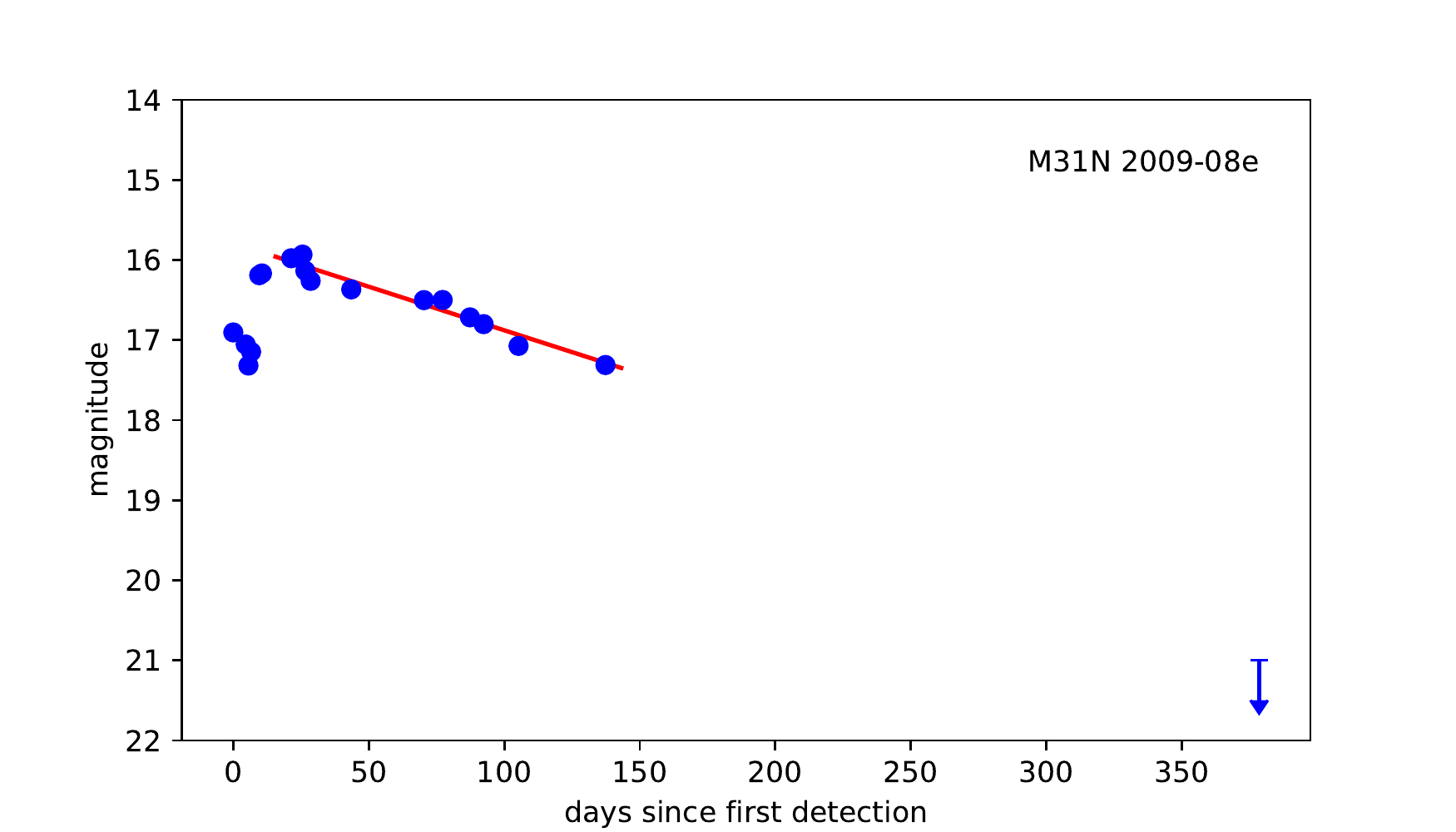}
\figsetgrpnote{The vertical axis shows the \ha\ magnitude.  The horizontal axis shows the number of days since first detection.}
\figsetgrpend

\figsetgrpstart
\figsetgrpnum{5.25}
\figsetgrptitle{Light curve for nova M31N~2009-10b}
\figsetplot{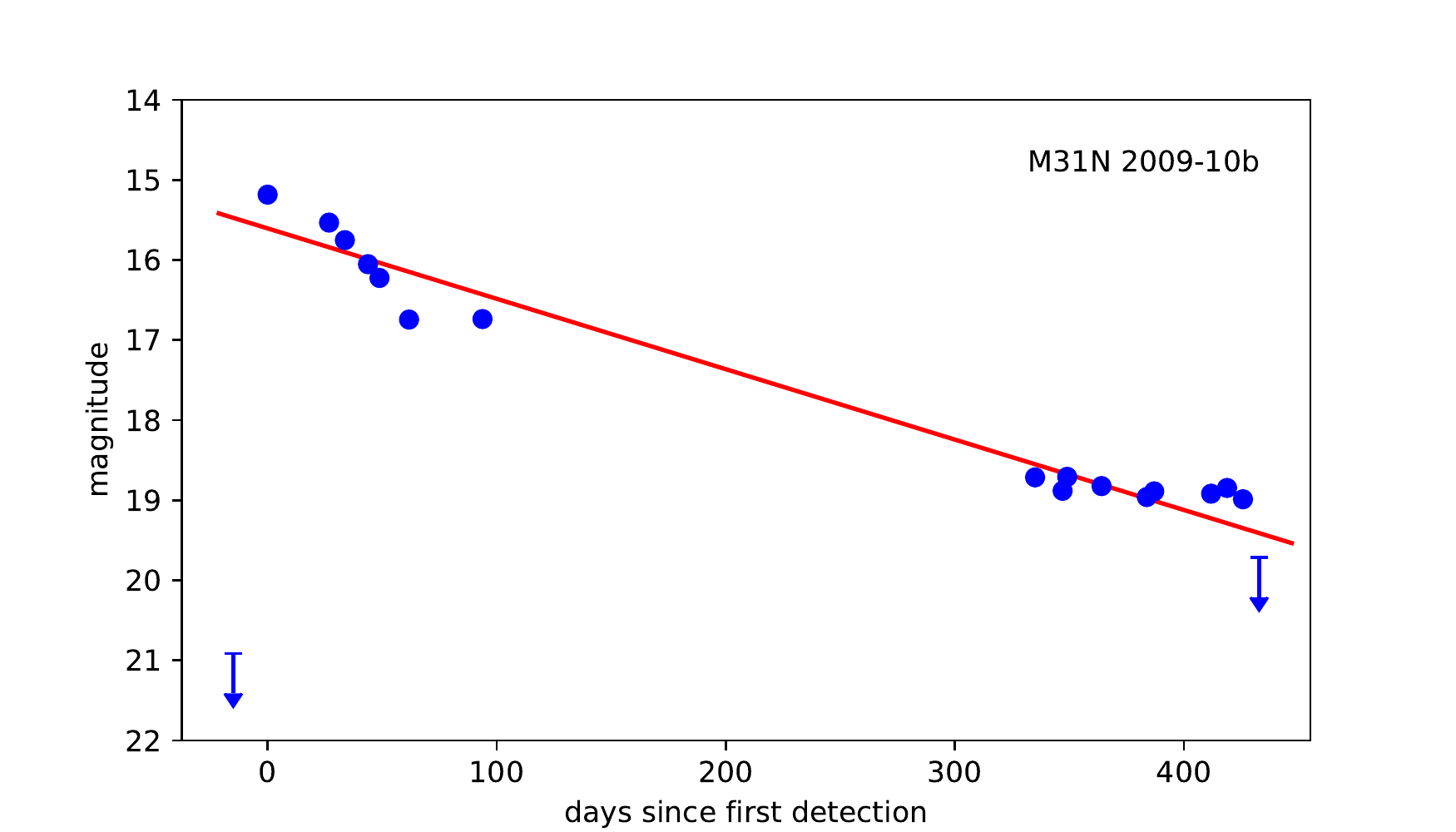}
\figsetgrpnote{The vertical axis shows the \ha\ magnitude.  The horizontal axis shows the number of days since first detection.}
\figsetgrpend

\figsetgrpstart
\figsetgrpnum{5.26}
\figsetgrptitle{Light curve for nova M31N~2009-11b}
\figsetplot{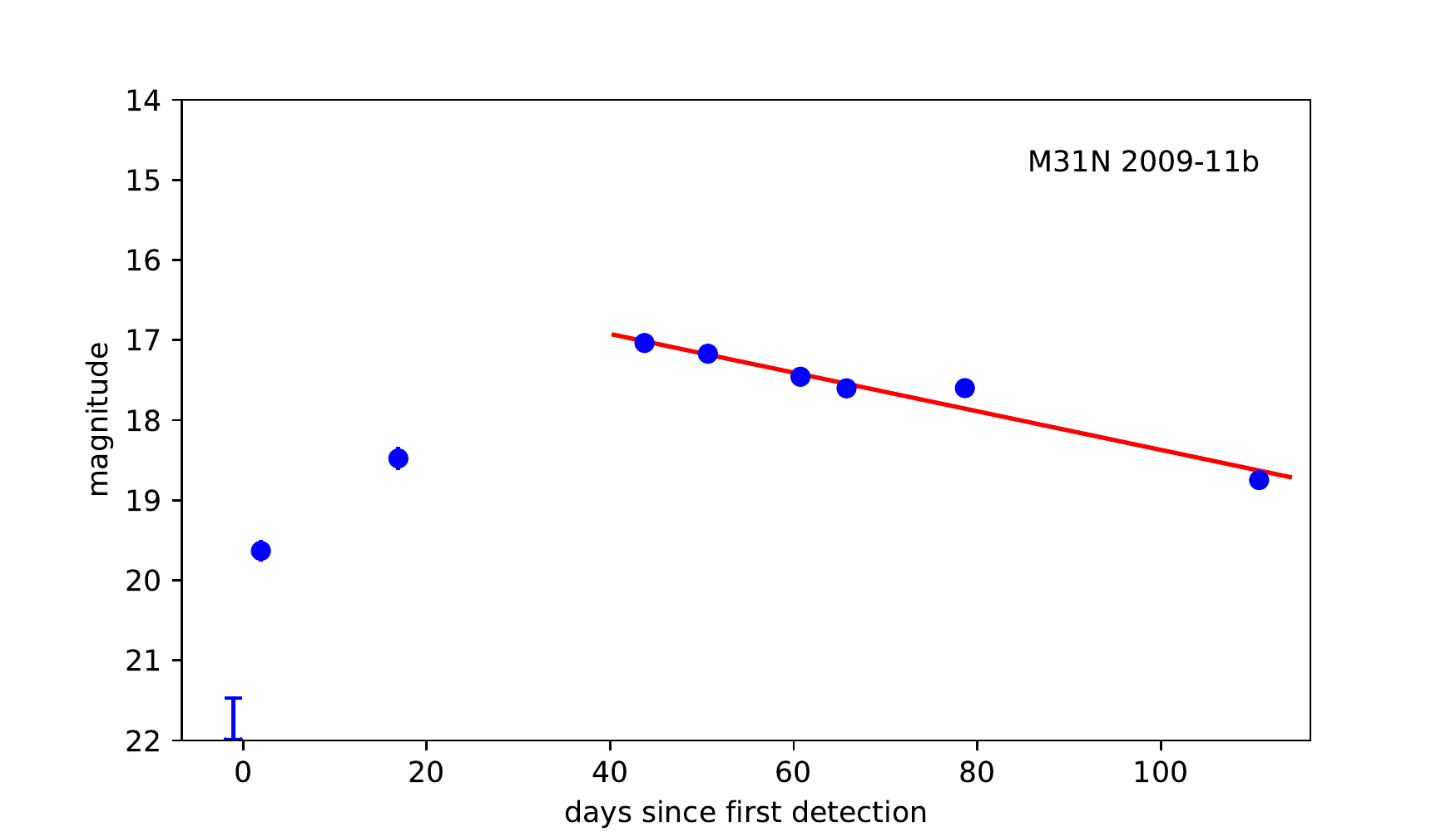}
\figsetgrpnote{The vertical axis shows the \ha\ magnitude.  The horizontal axis shows the number of days since first detection.}
\figsetgrpend

\figsetgrpstart
\figsetgrpnum{5.27}
\figsetgrptitle{Light curve for nova M31N~2009-11e}
\figsetplot{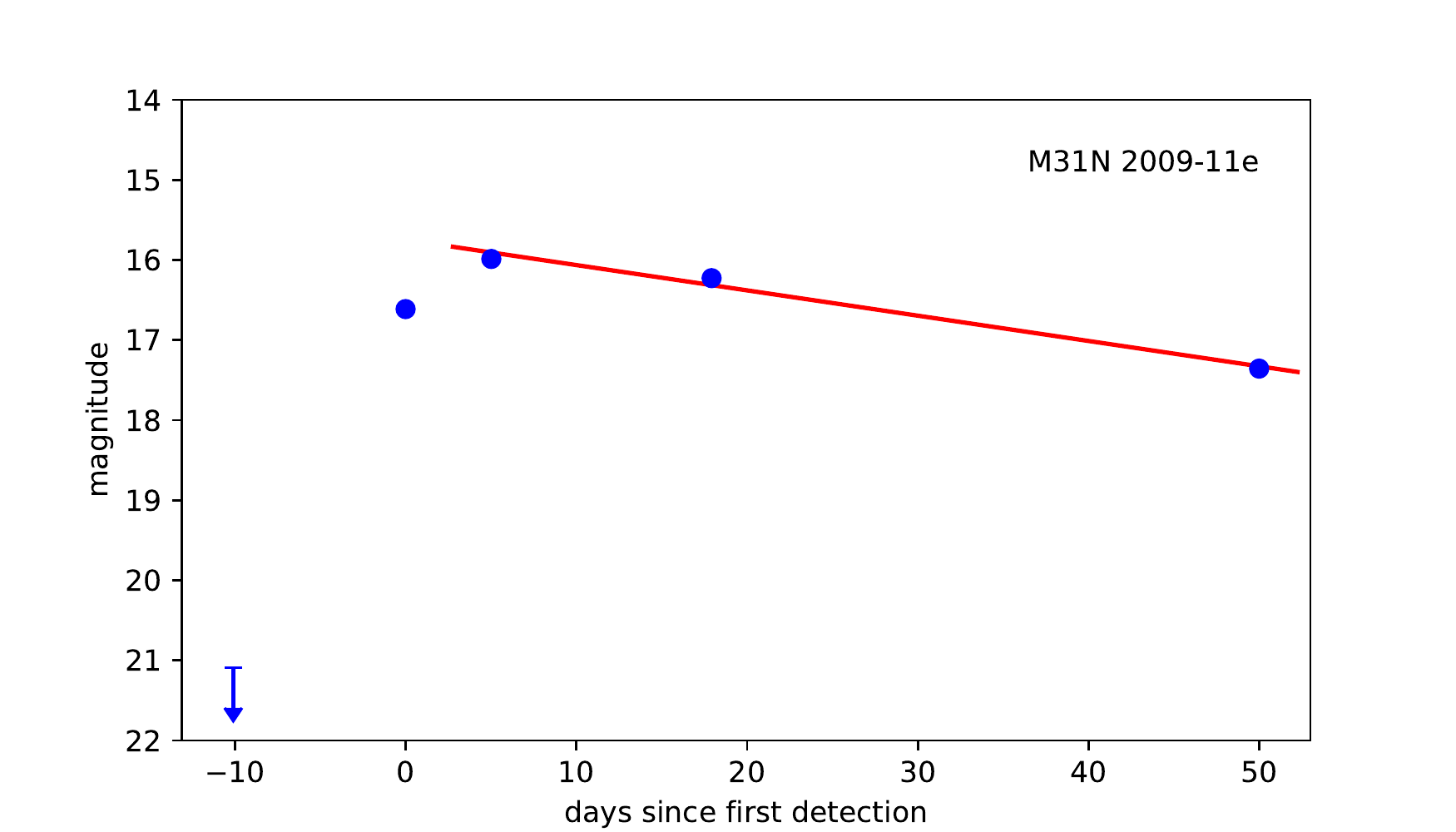}
\figsetgrpnote{The vertical axis shows the \ha\ magnitude.  The horizontal axis shows the number of days since first detection.}
\figsetgrpend

\figsetgrpstart
\figsetgrpnum{5.28}
\figsetgrptitle{Light curve for nova M31N~2010-06a}
\figsetplot{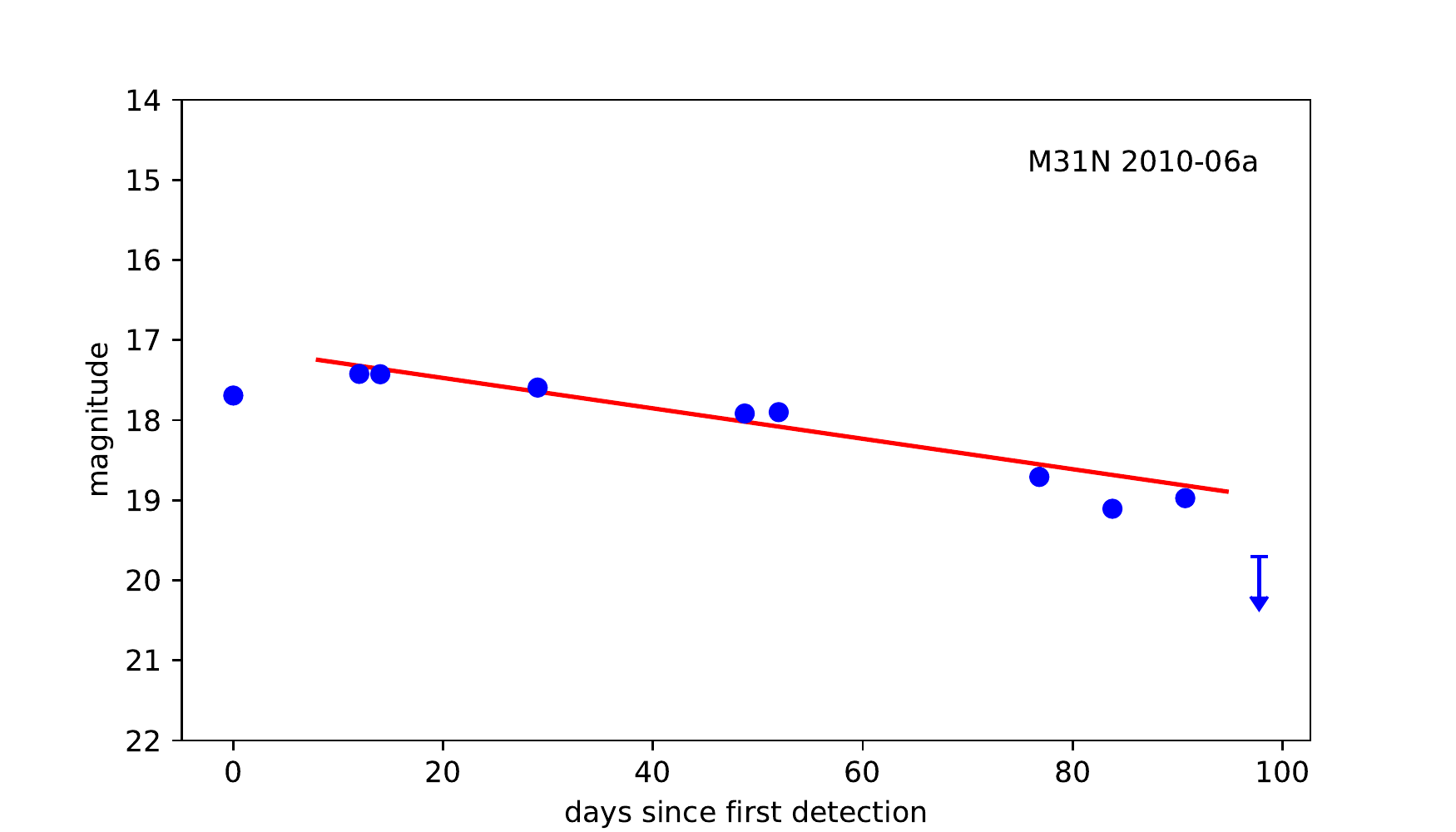}
\figsetgrpnote{The vertical axis shows the \ha\ magnitude.  The horizontal axis shows the number of days since first detection.}
\figsetgrpend

\figsetgrpstart
\figsetgrpnum{5.29}
\figsetgrptitle{Light curve for nova M31N~2010-10a}
\figsetplot{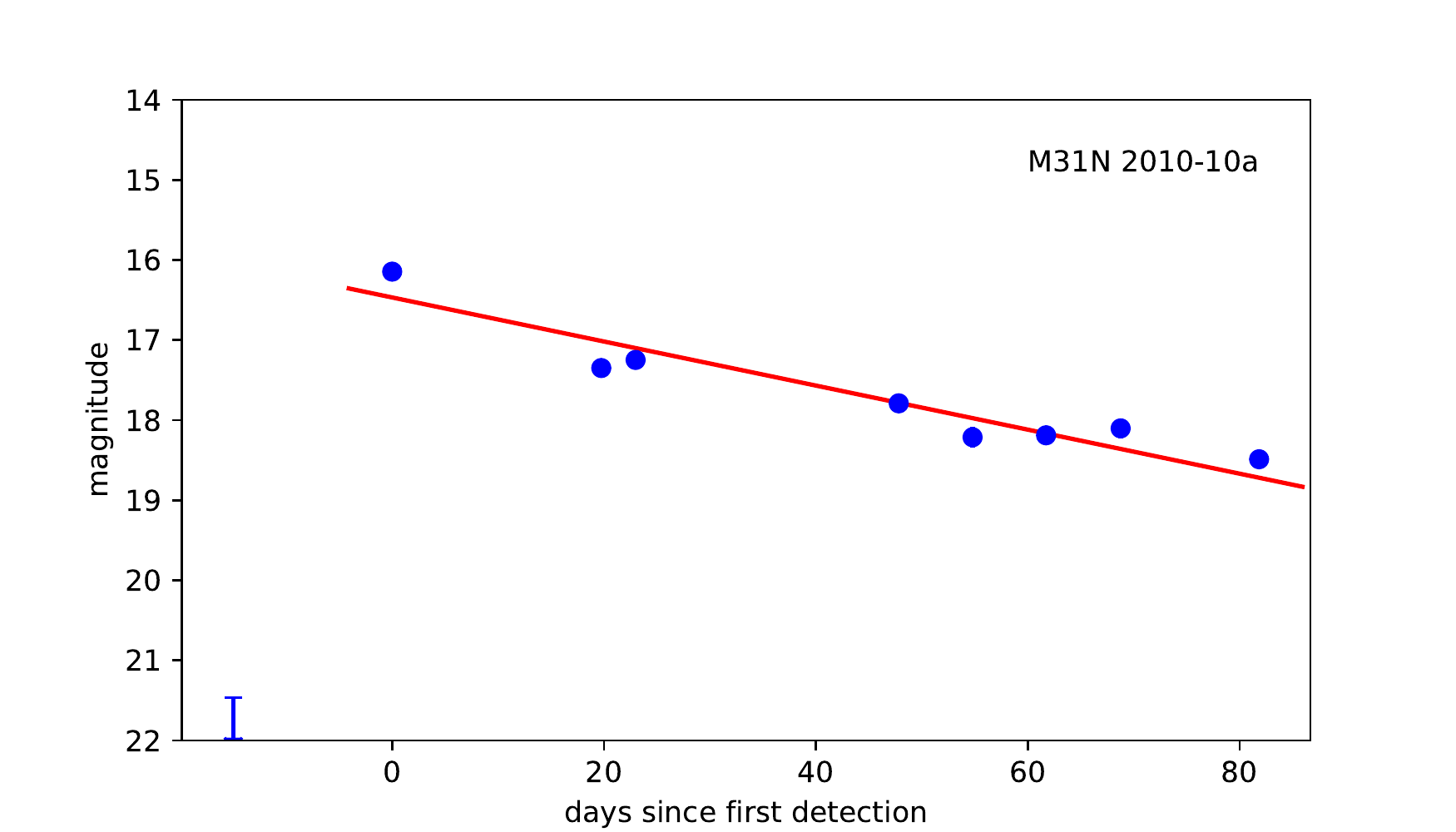}
\figsetgrpnote{The vertical axis shows the \ha\ magnitude.  The horizontal axis shows the number of days since first detection.}
\figsetgrpend

\figsetgrpstart
\figsetgrpnum{5.30}
\figsetgrptitle{Light curve for nova M31N~2010-10d}
\figsetplot{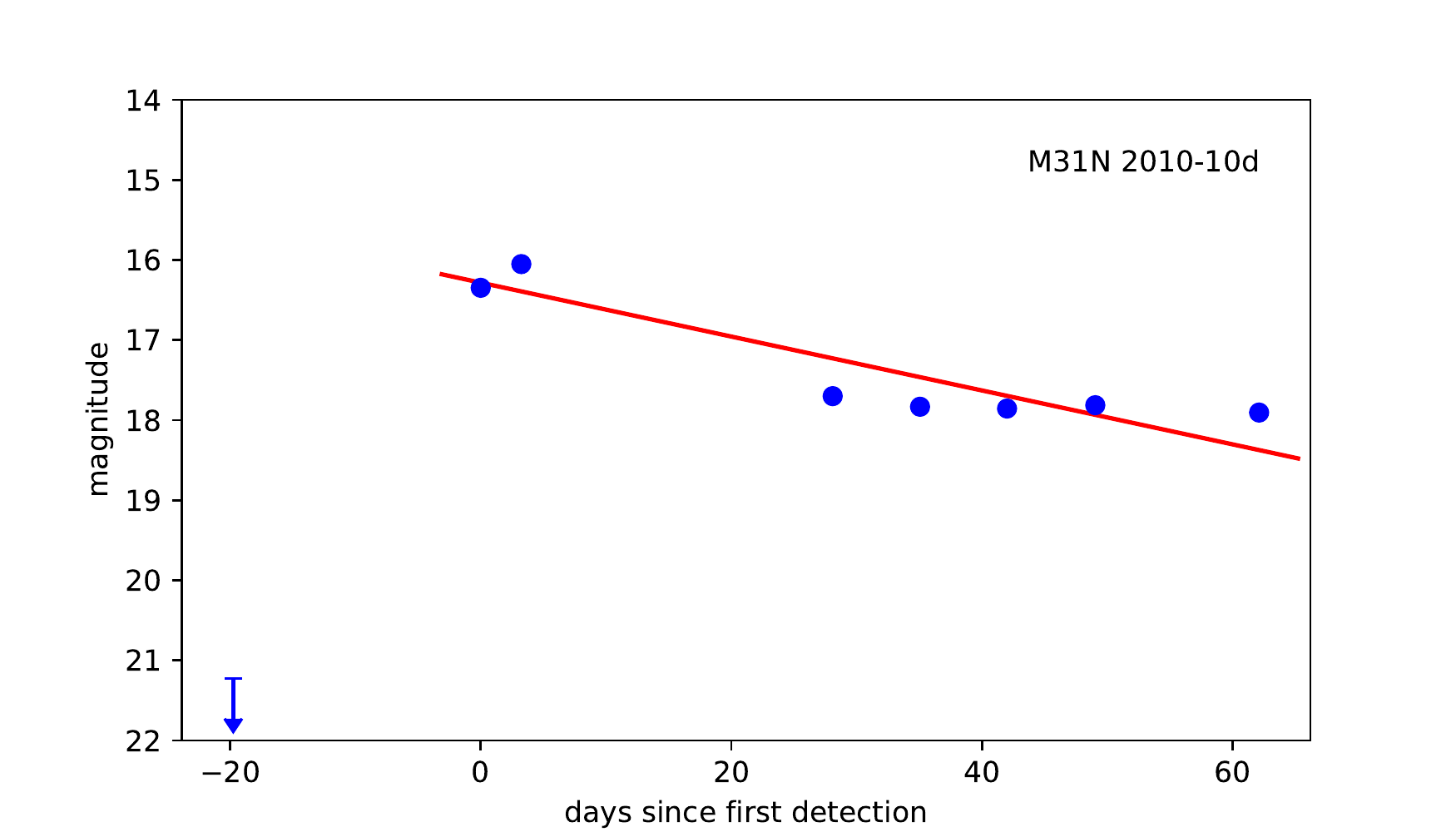}
\figsetgrpnote{The vertical axis shows the \ha\ magnitude.  The horizontal axis shows the number of days since first detection.}
\figsetgrpend

\figsetend

\subsection{The Effective Limiting Magnitude of the Survey}

A determination of the completeness of survey A (0.9-m) and survey B (4-m) is
central to a determination of the galaxy's nova rate.
In addition to the characteristics of the surveys themselves (e.g., the telescope apertures, the integration times, and the plate scales), the
overall completeness to a given apparent magnitude $C(m)$
will be strongly dependent on the brightness of, and variation in, the background light of the host galaxy. Since the galaxy background light
is non-uniform across the survey images,
the simplest and most effective approach to estimating
the completeness of extragalactic nova surveys
is to conduct artificial star (nova) tests. Thus, following our earlier work \citep[e.g., see][and references therein]{2021ApJ...923..239S}, we have
conducted artificial nova tests
on representative images from both survey~A and survey~B (hereafter the fiducial images).

In performing the simulations, one must decide how to treat the spatial distribution of the artificial novae. Here, as in previous studies, we have chosen to proceed
under the assumption that the spatial
distribution of the artificial novae follows the background light of the galaxy. Such an assumption might be questionable if the nova rate was a strong function of stellar population, for example. However, since our central field is dominated by light from a single population (M31's bulge), we consider our approach to be justified.

\begin{figure}
\includegraphics[angle=0,scale=0.32]{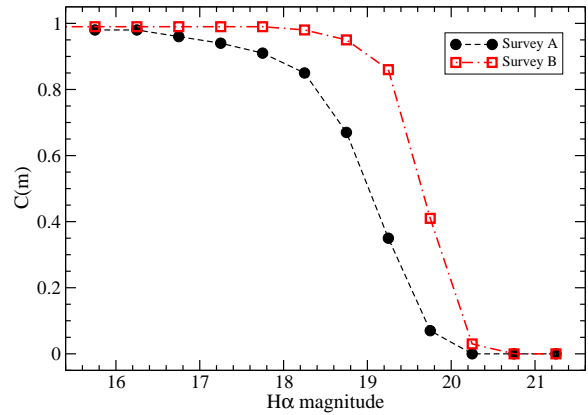}
\caption{The completeness of surveys A and B as a function of apparent \ha\ magnitude as determined from the artificial nova tests.
}
\label{fig:cm}
\end{figure}

Artificial novae with PSFs that matched the stars in our fiducial images
were generated using tasks in the IRAF DAOPHOT package.
The fiducial
images were then seeded with 100 artificial novae with random apparent magnitudes
lying within each of eight 0.5 mag wide bins using the routine \texttt{addstar}.
For each of the magnitude bins, the artificial novae were distributed
at random locations within each image, but with a spatial density that was constrained
to follow the surface brightness of M31.
Artificial novae were then recovered using the same procedures
that were employed in identifying novae in our survey images.

The fraction of artificial novae recovered in our searches
in each magnitude bin yielded the completeness functions $C(m)$
shown in Figure~\ref{fig:cm}. In cases where the different
epochs of observation were obtained under different conditions
(e.g., different telescopes, integration times, etc.),
this completeness function can be generalized
to any epoch, $i$, of observation
by applying a shift,
$\Delta m_i$ ($= m_{\mathrm{lim},0} - m_{\mathrm{lim},i}$).
Thus, for any epoch, $i$, we have $C_i(m) = C(m+\Delta m_i)$.
In the case of our M31 nova surveys, the limiting magnitudes of
the individual images were sufficiently similar that we were able to use
the same completeness function for all epochs from a given survey.

\subsection{Numerical Simulations}

Armed with an estimate of the instantaneous nova luminosity function and the completeness
functions $C(m)$ for surveys~A and B, we are now in a position to estimate the
nova rate $R$ in our two surveyed regions of M31.
Following the procedure adopted in our previous extragalactic nova
studies \citep[e.g.,][]{2021ApJ...923..239S,2012ApJ...760...13F,2010ApJ...720.1155G,2008ApJ...686.1261C,2004ApJ...612..867W},
we have performed numerical experiments
to estimate the intrinsic nova rate $R$ that will most likely produce the number of M31 novae
that we have detected during the course of our surveys.

We begin by producing trial novae
erupting at random times throughout the time span covered by a given survey.
Each simulated nova was given the parameters of one of the 30 novae selected at random. Again it is worth noting that
these parameters are based on novae discovered in our survey,
supplemented by light curves measured in previous M31 nova studies \citep{1990ApJ...356..472C,2001ApJ...563..749S}. They are therefore subject to observational selection biases
and may not necessarily represent the full range of nova properties.
Ideally, we would like to use light curve parameters specific to the
full population of M31 novae, but such an unbiased sample does not exist.
Instead, we used the M31 light curve parameters from our observations
given in Table~\ref{tab:lcparam}, augmented with additional \ha\ light curve
parameters from the nova light curves observed by \citet{1990ApJ...356..472C}, \citet{2001ApJ...563..749S}, and \citet{2004AJ....127..816N}.

The number of novae expected to be discovered during the course of
a given survey as a function of the assumed nova rate N$_{\mathrm obs}(R)$ can then be
computed by convolving the simulated apparent magnitude distribution
for a given epoch of observation $n_i(m,R)$
with the completeness function $C(m)$ and then summing
over all epochs of observation:

\begin{equation}
N_\mathrm{obs}(R) = \sum_{i}\sum_{m}{C(m)~n_i(m,R)}.
\label{eqn:NovaNum}
\end{equation}

\begin{figure}
\includegraphics[angle=0,scale=0.32]{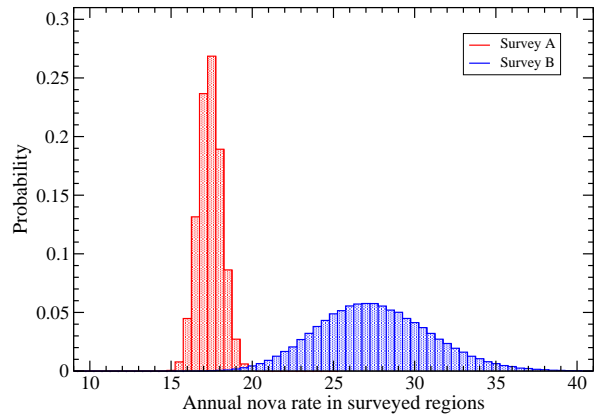}
\caption{The results of the numerical nova rate simulation
for the $20'\times20'$ and the $36'\times36'$ surveyed regions of surveys~A and~B. The most probable nova rates in the surveyed regions are $R_{\mathrm A}=17.5^{+0.9}_{-0.4}$ and $R_{\mathrm B}=27.0^{+3.8}_{-3.0}$ per year, respectively.
}
\label{fig:montecarlo}
\end{figure}

The best estimate of the intrinsic nova rate $R$ in the regions covered by each survey, and their associated uncertainties
is determined by comparing the
number of novae found in a given survey $n_\mathrm{obs}$
with the number of novae predicted by equation (1).
We explored trial nova rates ranging from $R=1$ to 50 novae per year,
repeating the numerical simulation a total of 10$^{5}$ times for each trial
value of $R$. The number
of matches $M(R)$ between the predicted number of observable novae
$N_\mathrm{obs}(R)$ and the actual number of novae discovered in each of our surveys,
$n_\mathrm{obs}$ = 203 for survey~A and $n_\mathrm{obs}$ = 50 in survey~B, was recorded for each trial value of $R$.
The values of $M(R)$ were
then normalized by the total number of matches for all $R$
to give the probability distribution functions, $P(R)=M(R)/\sum_{R}{M(R)}$
shown in
Figure~\ref{fig:montecarlo}. The most probable nova rate in the
portion of the galaxy covered by our survey images
is $R_{\mathrm A}=17.5^{+0.9}_{-0.4}$~yr$^{-1}$ and $R_\mathrm{B}=27.0^{+3.8}_{-3.0}$~yr$^{-1}$ for surveys~A and~B, respectively. The 1$\sigma$ error range for the probability distributions
were computed by assuming they can be approximated as bi-Gaussian functions.

\begin{figure*}
\includegraphics[angle=0,scale=0.70]{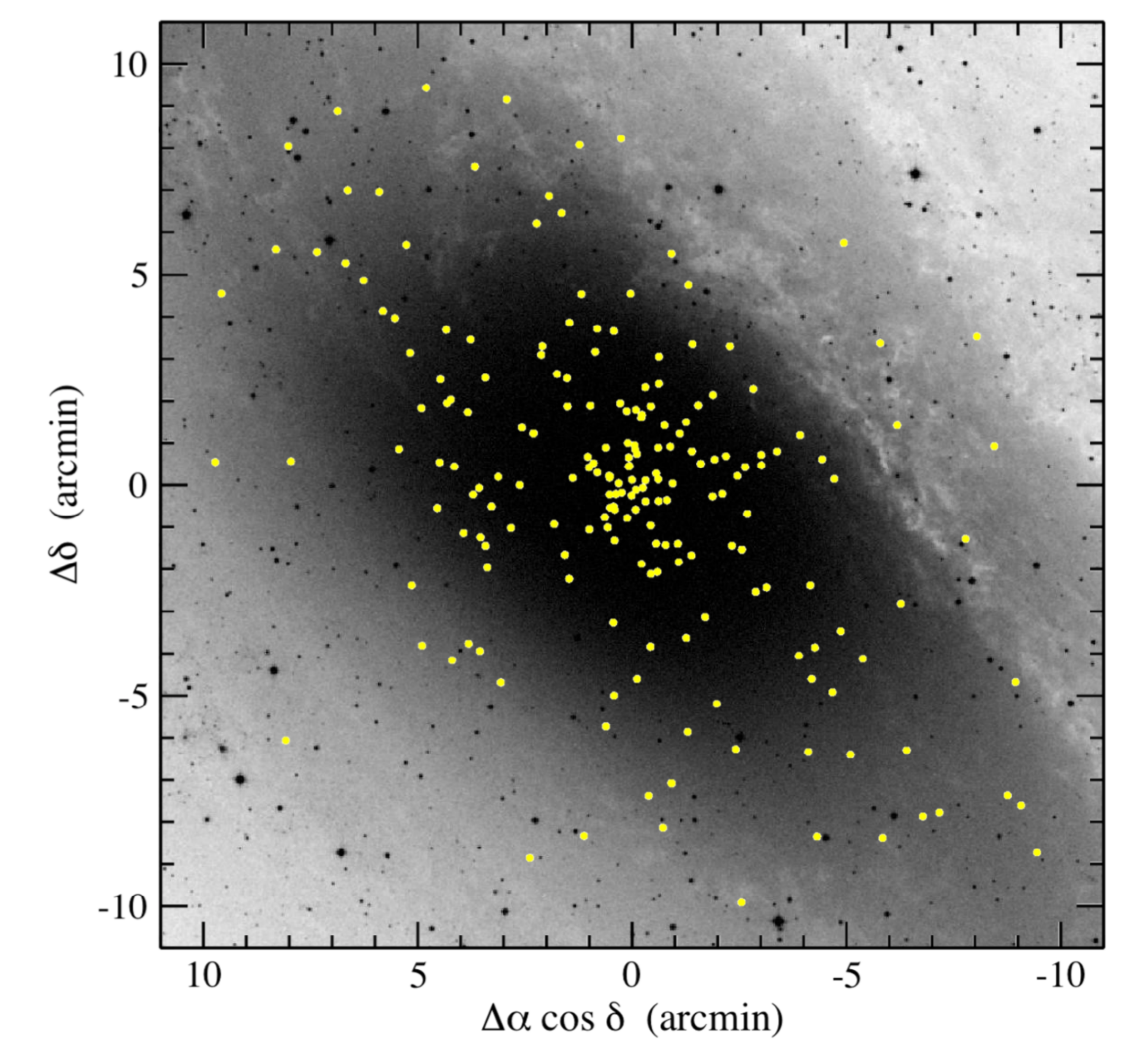}
\caption{The spatial distribution of the 203 novae found within our $20'\times20'$ field of survey~A, superimposed
on a DSS2 red image of M31.
\label{fig:0.9m_spatial}
}
\end{figure*}

\begin{figure*}
\includegraphics[angle=0,scale=0.70]{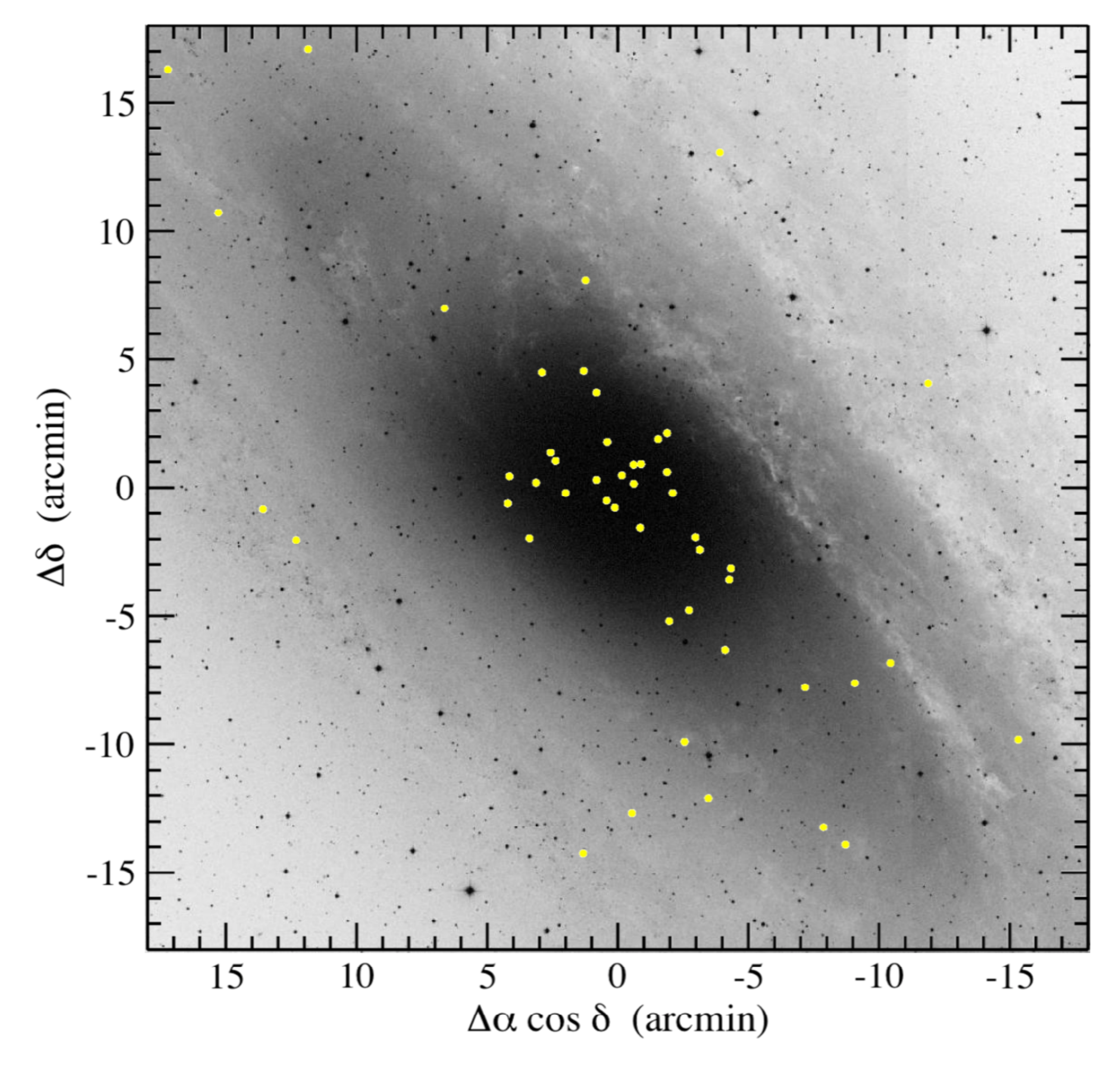}
\caption{The spatial distribution of the 50 novae found within our $36'\times36'$ field of survey~B, superimposed on a DSS2 red image of M31.
\label{fig:4m_spatial}
}
\end{figure*}

\section{The Nova Spatial Distribution}

The spatial distributions of the 203 novae in survey~A and the 50 novae found in survey~B are shown superimposed on (negative) DSS2 red images of M31 in
Figures~\ref{fig:0.9m_spatial} and \ref{fig:4m_spatial}, respectively.
In both cases the surface density
of novae appears to generally follow the background light of M31,
increasing markedly toward the center of the galaxy. 

\subsection{Comparison with the Background light}

\begin{figure}
\includegraphics[angle=0,scale=0.42]{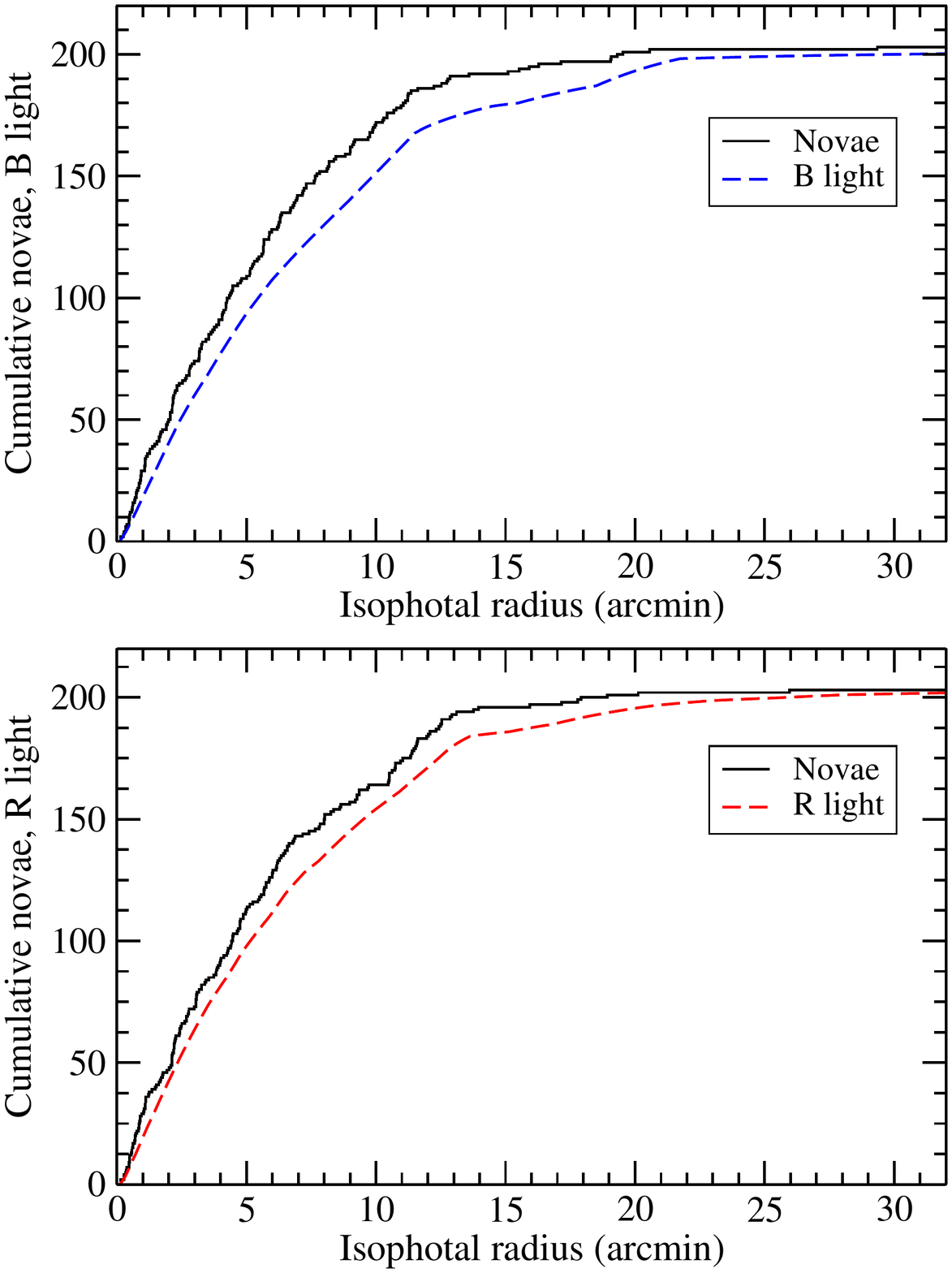}
\caption{The cumulative nova density from survey~A compared with the integrated background light of M31. Top panel: B light from [\citet{1958ApJ...128..465D}; Bottom panel: R light from \citet{1987AJ.....94..306K}. The likelihoods that the nova density match the background light are 0.57\% and 5.5\% for the $B$ and $R$ bands, respectively.}
\label{fig:cumlightBR09m}
\end{figure}

\begin{figure}
\includegraphics[angle=0,scale=0.42]{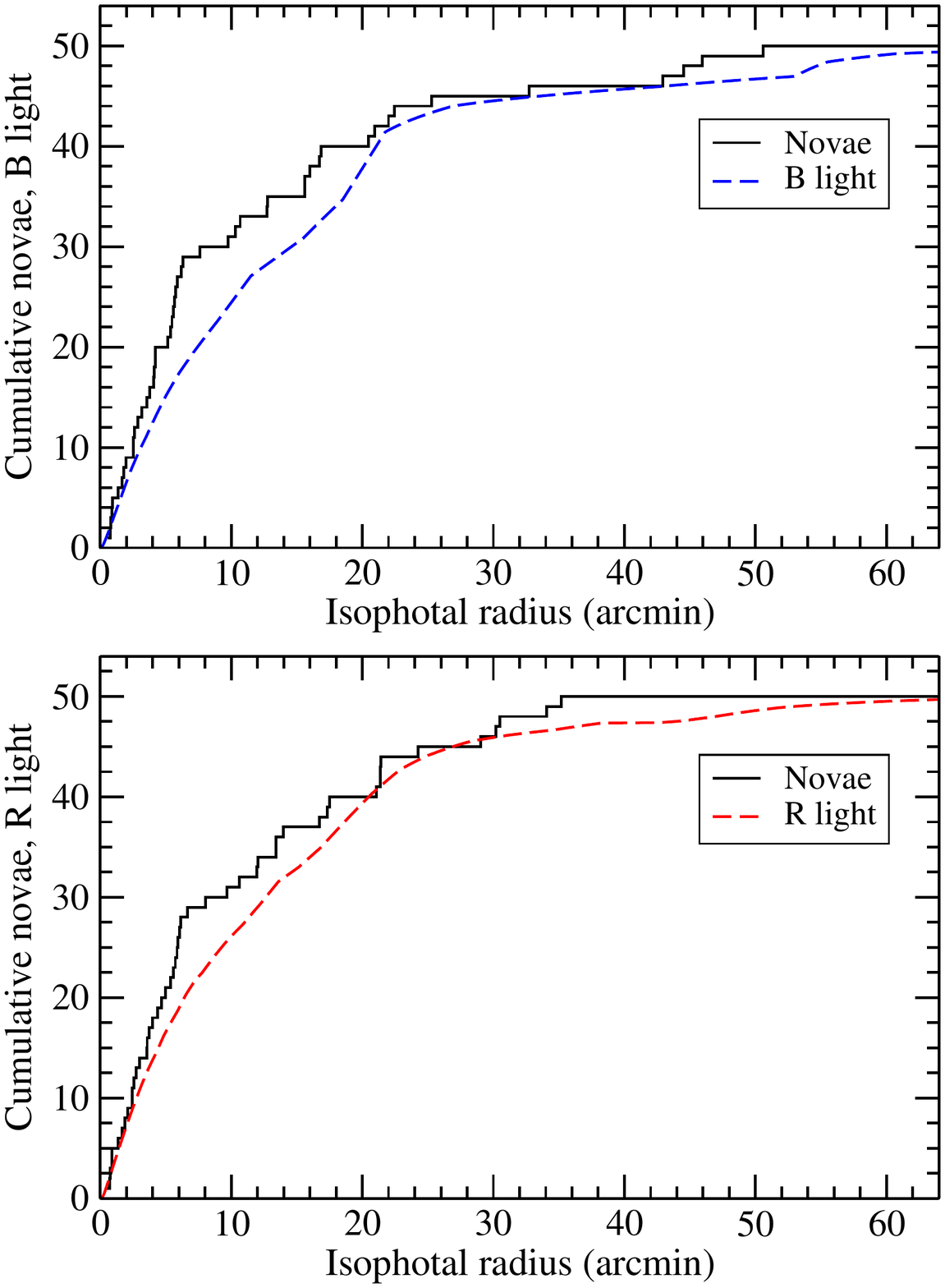}
\caption{The cumulative nova density from survey~B compared with the integrated background light of M31. Top panel: B light from [ref[]; Bottom panel: R light from \citet{1987AJ.....94..306K} The likelihoods that the nova density match the background light are 1.5\% and 10\%, for the $B$ and $R$-bands, respectively.
}
\label{fig:cumlightBR4m}
\end{figure}

A quantitative assessment of how well the nova density follows
the surface brightness of the galaxy can be achieved by comparing their cumulative distributions as a function of galactocentric radius (i.e., the semimajor axis of the elliptical isophote passing through the position of a given nova). These cumulative distributions of novae and the background $B$ and $R$-band light from the surface photometry
of \citet{1958ApJ...128..465D} and \citet{1987AJ.....94..306K} are shown in Figures~\ref{fig:cumlightBR09m} and \ref{fig:cumlightBR4m} for surveys~A and~B, respectively.

As expected, the nova distributions follow the background light of the galaxy more closely at longer wavelengths (i.e., better in $R$ compared to $B$), and better for survey~B with its larger field-of-view ($36'\times36'$) than for the smaller $20'\times20'$ field covered by survey~A. This difference can be attributed to the fact that the portion of M31 covered in survey~A is dominated by the galaxy's bulge population. The bulge light, consisting mostly of lower mass Pop II stars is ``redder" than the disk light, and traces the mass in stars more closely than does the $B$-band light, which is more affected by a smaller number of massive blue  main-sequence stars associated with the galaxy's disk population. In all cases, Kolomogorov-Smirnov (KS) tests show that the probabilities that the distributions are drawn from the same parent distribution are unlikely. The best fit, when the 50 novae from survey~B is compared with the galaxy's $R$ band light, yields a KS statistic of just 0.1, indicating that the nova and light distributions differ with 90\% confidence.

In order to explore
whether the poor fit to the background light results from a nova density distribution that varies with stellar population,
it is useful to compare the nova distribution with the galaxy's bulge and disk light separately,
as well as with different proportions of bulge and disk light.

\subsection{Bulge-Disk Separations}

Optical surface photometry of M31 that can be used to separate the bulge and disk components of the galaxy's
light has been undertaken in several studies going back to
the photographic work of \citet{1958ApJ...128..465D}. In recent years, infrared surface photometry made possible with
2MASS \citep{2003AJ....125..525J} and {\it Spitzer\/} \citep{2004ApJS..154...10F} has become
available, which should better trace the mass of the galaxy. Armed with these data, an extensive set of M31
bulge-disk separations have been performed by \citet{2011ApJ...739...20C} who modeled the infrared light
with a Sersic bulge profile combined with an exponential disk component. These decompositions have been
performed using multiple techniques, including non-linear least-squares and Markov Chain Monte-Carlo
techniques on 1D radial light profiles from major and minor axis cuts, or on azimuthal averages. 
For the fitting of 2D images, they have adopted the GALFIT
galaxy/point source fitting algorithm of \citet{2002AJ....124..266P}. As discussed by
\citet{2011ApJ...739...20C}, the 2D GALFIT models are less affected by non-axisymmetric features like the galaxy's
spiral arms. So, for our purposes, we have adopted their Model~P decomposition of the M31 IRAC ($3.6\mu$) bulge, disk, and total  infrared light. We have explored other models, and found that the results of our analysis are not sensitive to the choice of model used in the bulge-disk decomposition.

\begin{figure*}
\plotone{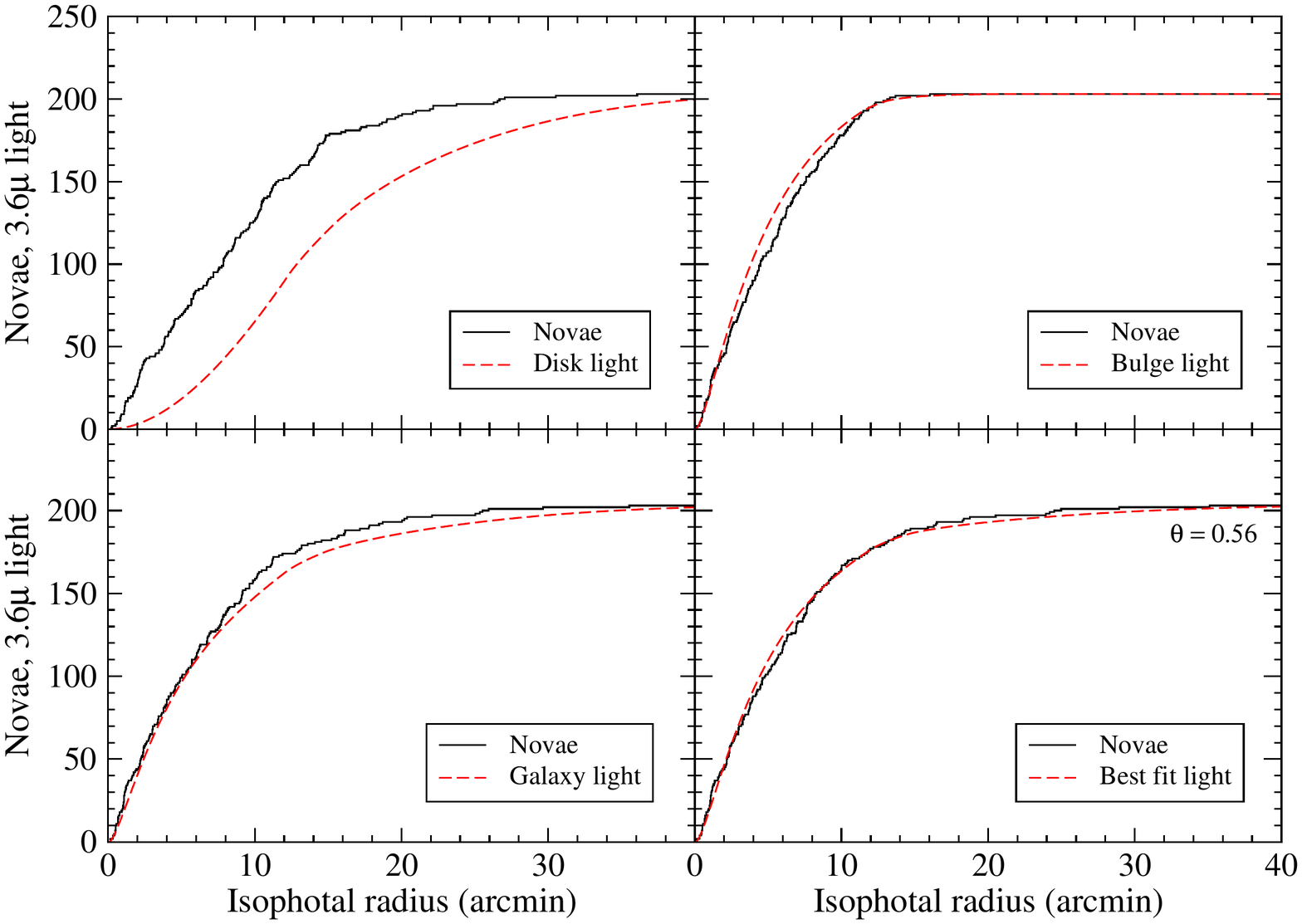}
\caption{The cumulative distributions of novae within our $20'\times20'$ survey region compared with the cumulative distributions of M31 light from Model~P of \citet{2011ApJ...739...20C}.
The nova distribution is compared with the disk, bulge,
total galaxy, and the best fit disk+bulge ($\theta=0.56$) light in the upper left, upper right, lower left and lower right panels, respectively, with corresponding KS values comparing the nova and light distributions of $\sim$0, 0.053, 0.22, and 0.74.
}
\label{fig:0.9m_cumlightP}
\end{figure*}

\begin{figure*}
\plotone{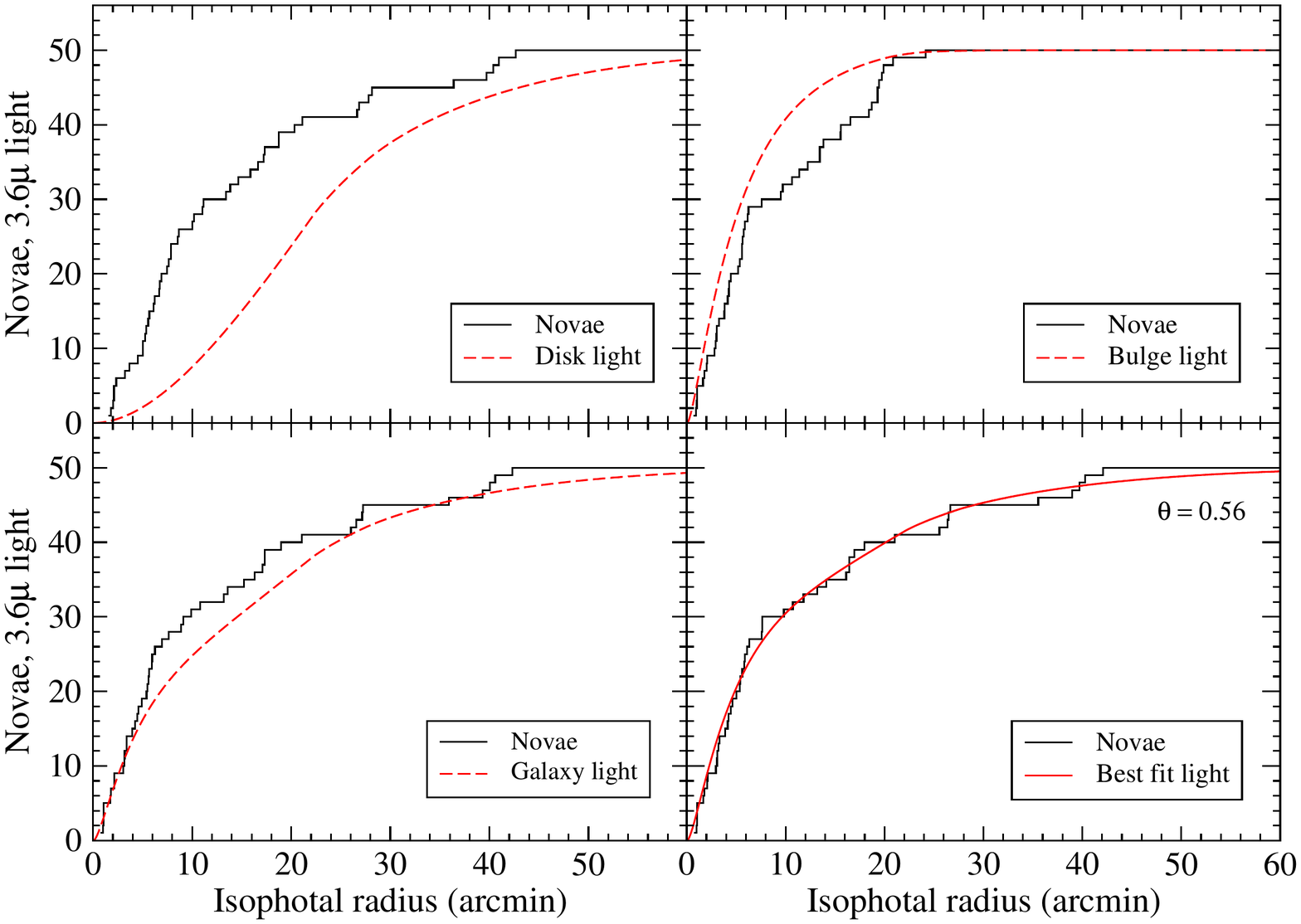}
\caption{The cumulative distributions of novae within our $36'\times36'$ survey region of the 4-m mosaic images compared with the cumulative distribution of M31 light from Model~P of \citet{2011ApJ...739...20C}. As with Figure~\ref{fig:0.9m_cumlightP},
the nova distribution is compared with the disk, bulge,
total galaxy, and the best fit disk+bulge ($\theta=0.56$) light in the upper left, upper right, lower left and lower right panels, respectively, with corresponding KS values comparing the nova and light distributions of $\sim$0, 0.034, 0.26, and 0.90.
}
\label{fig:4m_cumlightP}
\end{figure*}

Figure~\ref{fig:0.9m_cumlightP}
shows the cumulative M31 nova distribution for the $20'\times20'$ region of survey~A as
compared with the cumulative (disk, bulge, and combined) light from model P of \citet{2011ApJ...739...20C}.
As has been found in several previous M31 studies \citep[e.g.,][]{1987ApJ...318..520C,1989AJ.....97.1622C,2001ApJ...563..749S}, when the bulge and disk light are considered separately,
it is clear that the novae follow the galaxy's bulge light considerably better than they do the disk light\footnote{See \citet{1997ApJ...487L..45H} for a different interpretation.}.
In the case of our survey~A in particular, this is not surprising
given that the light in the central $20'\times20'$ surveyed region is dominated by the bulge, which
contains $\sim$70\% of the enclosed light. However, a much better fit is achieved when the overall (bulge+disk) galaxy
light is considered (lower left panel of Figure~\ref{fig:0.9m_cumlightP}). Furthermore, the cumulative nova distribution follows $3.6\mu$ galaxy light considerably better (KS=0.22) than it does the $B$ and $R$-band light (KS=0.0057 and KS=0.055, respectively) as seen in Figures~\ref{fig:cumlightBR09m} and \ref{fig:cumlightBR4m}.

Figure~\ref{fig:4m_cumlightP} shows the cumulative
nova and light distributions for the $36'\times36'$ region of survey~B. As seen in survey~A, the nova distribution follows the bulge light better than the disk light, but the larger coverage of survey~B results in an even better fit of the cumulative nova distribution to the overall $3.6\mu$ galaxy light (KS=0.26). Once again, the nova distribution follows the near infrared $3.6\mu$ light much
better than in the case of the $B$ and $R$-band comparisons.

The apparent affinity of novae for M31's bulge population is consistent with the results of \citet{2011ApJ...734...12S} who found that $\sim$80\% of spectroscopically-confirmed M31 novae belong to the
Fe~II class \citep[e.g., see][]{1992AJ....104..725W}. In studies of the Milky Way, \citet{1998ApJ...506..818D} have shown that Fe~II and He/N novae are associated primarily with the bulge and disk, respectively.

\subsubsection{Optimum Bulge-Disk Nova Densities}

We can determine the optimum ratio of bulge and disk nova densities required to provide the best match to
the overall galaxy light by performing a maximum likelihood analysis that compares the spatial positions of the novae with the bulge and disk
light distributions, as originally described in \citet{1987ApJ...318..520C}. The observed spatial distribution of novae is compared with a
theoretical distribution function where the probability of a nova
occurring at a given location in the galaxy depends both
on the relative disk-to-bulge luminosity at that location and
on an assumed ratio of disk and bulge nova rates, defined as 
$\theta=\rho_\mathrm{disk}/\rho_\mathrm{bulge}$. Specifically, the probability of a
nova appearing at a position $x$ in the galaxy is given by:
\begin{equation}
P(x,\theta) = {L_\mathrm{bulge}(x) + \theta L_\mathrm{disk}(x) \over \sum_{x}[L_\mathrm{bulge}(x) + \theta L_\mathrm{disk}(x)]},
\end{equation}
where the relative luminosity of the disk to the bulge
has been determined from the decomposition
of the overall galaxy light as described earlier.
The value of $\theta$ is
varied until the probability that the observed distribution is drawn
from the theoretical distribution $P(x,\theta)$ is maximized.
In practice, this is accomplished by defining a merit function,
\begin{equation}
Q(\theta)=\sum_{x} {\rm ln}~P(x,\theta),
\end{equation}
\noindent
and finding the value of $\theta$ corresponding to its maximum.

The maximum-likelihood analysis is most accurate when applied to a large field containing significant bulge and disk populations.
In view of the limited spatial coverage of survey~A, which is dominated by M31's bulge population, we have chosen to apply the maximum likelihood analysis to survey~B given its significantly larger field-of-view that includes roughly equal proportions of bulge and disk light (51\% bulge, 49\% disk).

\begin{figure}
\includegraphics[angle=0,scale=0.32]{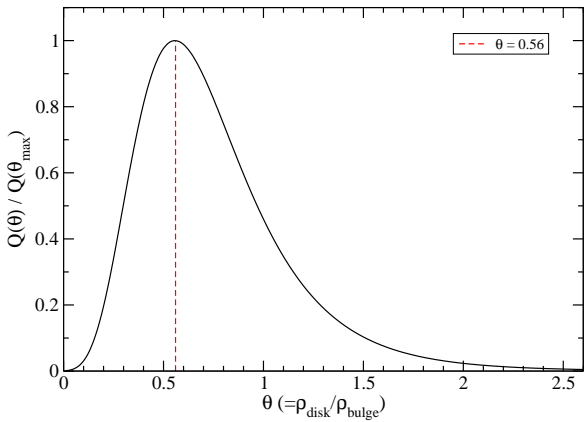}
\caption{
The results of our maximum likelihood analysis plotted as a function of the assumed disk-to-bulge nova density, $\theta$.
The relative bulge and disk surface brightness is based on model P of \citet{2011ApJ...739...20C}.
The analysis has been applied to the wider field 4-m survey (survey~B), because its more extensive coverage ($36'\times36'$)
provides a nearly equal balance between M31's bulge and disk populations. The most probable value of $\theta_\mathrm{max}=0.56^{+0.23}_{-0.17}$.
\label{fig:thetaprob}
}
\end{figure}

The result of our analysis applied to the bulge and disk surface brightness profiles given by Model P of \citet{2011ApJ...739...20C}
is shown in Figure~\ref{fig:thetaprob}.
The optimum fit corresponds to $\theta_\mathrm{max}=0.56^{+0.23}_{-0.17}$ (probable errors) implying a somewhat stronger association of novae with M31's bulge component. However
the resulting distribution is quite broad, suggesting that $\theta_\mathrm{max}$ is relatively poorly constrained. Overall, the analysis is consistent with the conclusion that novae are associated with both the bulge and disk populations of M31, with rates
that are not strongly dependent on stellar population.

The nova density distributions from both surveys~A and~B are compared with the optimum combined light distributions based on $\theta=0.56$ in the lower right panels of Figures~\ref{fig:0.9m_cumlightP} and~\ref{fig:4m_cumlightP}.  For both surveys the fit is remarkably good.

\section{The Global M31 Nova Rate}

In order to determine the global nova rate for M31, our estimate of the rate in the surveyed regions
must be extrapolated to the entire galaxy.
In performing such extrapolations, it has been traditionally assumed that $\theta=1$, i.e., that the luminosity-specific nova rate
is constant across the galaxy's bulge and disk populations. This is the only extrapolation possible without modeling the bulge and disk light separately. Under this assumption, the extrapolation is simply based on the fraction of the galaxy's integrated light in waveband $\lambda$ contained within the surveyed region, $f_S(\lambda)$. To provide the best proxy for the fraction of the mass in stars contained within the surveyed region, longer wavebands ($R$-band through the near 
infrared) are usually preferred.

Table~\ref{tab:NRext} shows the values of $f_S(\lambda)$ and the subsequent global M31 nova rates
based on extrapolations using the light in the bandpasses we have considered (B, R, and the IRAC $3.6\mu$ bands).
With the exception of the $\theta=0.56$ case, where a probable error has been computed,
an error of 10\% in the values of $f_S(\lambda)$ has been assumed when propagating errors from the nova
rates in the surveyed regions to the final extrapolated M31 nova rates.
In the case of the IRAC data, we have used bulge-disk separations given by Model~P from \citet{2011ApJ...739...20C} to enable
extrapolations based on nova densities that vary with stellar population (i.e., values of $\theta$ that differ from unity).

As discussed earlier, extrapolations based on red or infrared photometry are preferred as they are believed to
trace the underlying mass in stars, and, as we have seen, the fits of the cumulative nova density distribution
to the background light improves with increasing wavelength. Thus, we prefer to use the IRAC $3.6\mu$ photometry
to extrapolate the nova rate in our surveyed regions to the entire galaxy.

The question of whether to allow the specific nova rate to vary between the bulge and disk populations is less clear.
On the one hand, the measured luminosity-specific nova rates of galaxies do not appear to vary systematically across a wide range of Hubble types \citep[e.g., see][]{2014ASPC..490...77S}.
Although some late-type galaxies, such as the Large and Small Magellanic Clouds and
the nearly bulge-less spiral M33, appear to have somewhat larger than average luminosity-specific nova rates \citep[e.g., see][]{1994A&A...287..403D}, the value for the early-type
giant elliptical galaxy M87 also appears to be elevated \citep[][]{2016ApJS..227....1S,2017RNAAS...1...11S}. Taken together, these observations suggest that 
the nova rate across M31 may not be particularly sensitive to the galaxy's changing bulge and disk populations.

On the other hand, however, multiple studies of the spatial distribution of novae in
M31 have found that the nova distribution tracks the bulge component of the galaxy
as well, or in some cases better, than the overall background light
\citep[e.g., see][]{1987ApJ...318..520C,2001ApJ...563..749S,2006MNRAS.369..257D,2016PhDT........37K} suggesting that the specific nova rates are in fact affected by stellar population.
Given the considerably improved fits to the background light that are achieved when allowing for differing bulge and disk nova densities, we have chosen to extrapolate the nova rate in our surveyed regions to the entire galaxy
using our optimum $\theta=0.56$ light distribution. Taking the mean\footnote{We have chosen not to apply any weighting. Even though survey~A contains more novae, survey~B covers a significantly larger portion of the galaxy.} of our rates for surveys~A and~B from Table~\ref{tab:NRext} gives a best estimate for the global M31 nova rate of 
$R_\mathrm{M31}=40^{+5}_{-4}$~yr$^{-1}$.
We note that the quoted (formal) uncertainly is based on the propagation of essentially
Poisson errors from our numerical simulations, and therefore does not
fully take into account possible systematic errors in our analysis (e.g., in properly
assessing the effective limiting magnitude of our survey). Thus, these errors likely represent a
lower limit to the true uncertainty in the derived nova rate.

\begin{deluxetable*}{lccccc}
\tablecolumns{6}
\tablecaption{M31 Nova Rate Estimates\label{tab:NRext}}
\tablehead{ &  &  &  & \colhead{$R_\mathrm{M31}$} & \colhead{$\nu_K$} \\
\colhead{Band} & \colhead{$\lambda_{\mathrm{eff}}(\mu)$} & \colhead{$\theta~(=\rho_\mathrm{d}/\rho_\mathrm{b})$} & \colhead{$f_S(\lambda)$} & \colhead{(yr$^{-1}$)} & \colhead{(yr$^{-1}~[10^{10}~L_{\odot,K}]^{-1})$}
}
\startdata
\cutinhead{Survey~A: $R_{\mathrm A}=17.5^{+0.9}_{-0.4}$}
$B$ & 0.44 & 1.0 & $0.34\pm0.02$ & $52^{+5}_{-4}$ & $4.3^{+0.5}_{-0.5}$ \cr
$R$ & 0.64 & 1.0 & $0.37\pm0.03$ & $48^{+4}_{-4}$ & $4.0^{+0.5}_{-0.4}$ \cr
IRAC & 3.6 & 1.0 & $0.38\pm0.03$ & $46^{+4}_{-3}$ & $3.9^{+0.5}_{-0.4}$  \cr
IRAC & 3.6 & $0.56^{+0.23}_{-0.17}$ & $0.48^{+0.07}_{-0.06}$ & $37^{+6}_{-5}$ & $3.0^{+0.5}_{-0.5}$ \cr
\cutinhead{Survey~B: $R_{\mathrm B}=27.0^{+3.8}_{-3.0}$}
$B$ & 0.44 & 1.0 & $0.52\pm0.04$ & $52^{+8}_{-7}$ & $4.4^{+0.8}_{-0.7}$ \cr
$R$ & 0.64 & 1.0 & $0.54\pm0.04$ & $50^{+8}_{-7}$ & $4.2^{+0.7}_{-0.6}$ \cr
IRAC & 3.6 & 1.0 & $0.55\pm0.04$ & $49^{+8}_{-6}$ & $4.1^{+0.7}_{-0.6}$ \cr
IRAC & 3.6 & $0.56^{+0.23}_{-0.17}$ & $0.63^{+0.06}_{-0.05}$ & $43^{+7}_{-6}$ & $3.6^{+0.7}_{-0.6}$
\enddata
\end{deluxetable*}

\subsection{The Luminosity-Specific Nova Rate}

To enable a comparison of the nova rate in M31 with other
galaxies, it is necessary to normalize the rate
to an intrinsic property of the galaxy. Ideally, one would like to
standardize the rate to the total mass in stars. However,
given that the stellar mass cannot be measured directly,
it has been standard practice to normalize extragalactic
nova rates to the infrared (typically the $K$-band) luminosity of the
galaxy, which is thought to trace the underlying stellar 
mass quite well.

In order to compute the integrated $K$-band luminosity of M31
we have chosen to adopt an apparent $K$-band magnitude of $K_\mathrm{M31} = -0.04$ based on the analysis of {\it Spitzer\/} IRAC data presented in \citet{2006ApJ...650L..45B}.
Adopting a distance modulus for M31, $\mu_o(\mathrm{M31}) = 24.38\pm0.05$ \citep{2001ApJ...553...47F}, along with an
absolute $K$-band magnitude for the sun of $M_{K,\odot}=3.27$ \citep{2018ApJS..236...47W}, yields an integrated $K$-band luminosity for M31
of $L_{K}(\mathrm{M31}) = 1.2\times10^{11}~\mathrm{L}_{\odot,K}$. This value is in excellent agreement with a value
of $L_{3.6\mu}(\mathrm{M31}) = 1.17\times10^{11}~\mathrm{L}_{\odot,3.6\mu}$ quoted by \citet{2011ApJ...739...20C}
in their decomposition of M31's IRAC light. The corresponding $K$-band luminosity-specific nova rates $\nu_K$, along with their estimated uncertainties, are given in
the final column of Table~\ref{tab:NRext}.

In the case of our preferred IRAC $\theta=0.56$ extrapolation, we find an average $K$-band luminosity-specific
nova rate based on our two surveys to be $\nu_K=3.3\pm0.4$ novae per year per $10^{10}$ solar luminosities in the $K$ band. This value is slightly higher, but not inconsistent with the mean for galaxies with measured nova rates,
$\nu_K=2.25$~yr$^{-1}~[10^{10}~\mathrm{L}_{K,\odot}]^{-1}$, found by \citet{2014ASPC..490...77S}.

\begin{deluxetable*}{lcccc}
\tablecolumns{5}
\tablecaption{Historical M31 Nova Rate Estimates\label{tab:HistNR}}
\tablehead{\colhead{Authors} & \colhead{Years} & \colhead{Filter} & \colhead{Novae} & \colhead{Rate}
}
\startdata
\citet{1929ApJ....69..103H}	    & $1909-1927$ & $B$	    & 85 & $\sim30$	\cr
\citet{1956AJ.....61...15A}	    & $1953-1954$ & $B$	    & 30 & $26\pm4$	\cr
\citet{1989AJ.....97.1622C}\tablenotemark{a} &  $1955-1986$ & $U,B$ & 142 & $29\pm4$  \cr
\citet{2001ApJ...563..749S}	& $1990-1997$ & \ha\ & 72 & $37^{+12}_{-8}$ \cr
\citet{2006MNRAS.369..257D, 2004MNRAS.353..571D} & $1999-2003$ & $r',i',g'$ & 20 & $65^{+16}_{-15}$  \cr
\citet{2016MNRAS.455..668S} & \dots & \dots & \dots & $\sim106$ \cr
\citet{2016MNRAS.458.2916C}  & \dots  & \dots & \dots & $\sim$97 \cr
This work ($\theta=1$)	     & $1995-2016$ & \ha\  & 253 & $48^{+5}_{-4}$\cr
This work ($\theta=0.56$)	 & $1995-2016$ & \ha\  & 253 & $40^{+5}_{-4}$
\enddata
\tablenotetext{a}{Based on observations from \citet{1964AnAp...27..498R}, \citet{1973Ros}, and \citet{1989AJ.....97...83R}.}
\end{deluxetable*}

\section{Discussion}

The nova rate in M31 has been estimated many times going back to Hubble's early work at Mount
Wilson during the first three decades of the twentieth century.
The results of key studies are summarized in Table~\ref{tab:HistNR}. After beginning with estimates of order 30 per year from the pioneering photographic surveys of \citet{1929ApJ....69..103H}, \citet{1956AJ.....61...15A}, and Rosino \citep[e.g., see][]{1989AJ.....97.1622C}, it is interesting that measurements of the M31 nova rate have crept up in recent years, with a rate of $65^{+16}_{-15}$~yr$^{-1}$ \citep{2006MNRAS.369..257D} becoming the currently accepted value.

Within the past decade, the observation of several faint, but rapidly fading novae in M31 \citep{2011ApJ...735...94K}, along with the discovery of the remarkable recurrent nova M31N~2008-12a \citep{2012ATel.4503....1S, 2014ApJ...786...61T, 2014A&A...563L...9D, 2014A&A...563L...8H} with its recurrence time of only a year, have fueled speculation that a significant population of faint and rapidly evolving (i.e., ``fast") novae might have been missed in most nova surveys. If so, the nova rate in M31, and other galaxies for that matter, could be significantly
higher than had been measured previously. Following up on this possibility, two recent studies attempted to take this putative population of faint and fast novae into account, finding that M31's nova rate could be as high as 100 per year.

In one such study, \citet{2016MNRAS.458.2916C} computed population synthesis models
estimating nova rates in galaxies with differing star formation histories
and morphological types. In the case of M31, they estimated a global nova rate of 97~yr$^{-1}$. In another study, \citet{2016MNRAS.455..668S} attempted to correct the nova rates that had been based on the earlier nova surveys
of \citet{1956AJ.....61...15A} and \citet{2006MNRAS.369..257D} for their potential bias against the discovery
of faint and fast novae. Based on these corrections,
they estimated a global nova rate for
M31 of order 106 per year.

Despite the increased spatial and temporal coverage of M31 and the Galaxy made possible by the contributions of automated surveys like the All-Sky Automated Survey for Supernovae \citep[ASAS-SN,][]{2017PASP..129j4502K} and Palomar's Zwicky Transient Facility \citep[ZTF,][]{2019PASP..131a8002B}, the extensive observations of the Galactic bulge from the
Optical Gravitational Lensing Experiment (OGLE) survey \citep[e.g., see][]{2015ApJS..219...26M},
not to mention the ever more comprehensive contributions made by amateur astronomers, compelling observational evidence for a significant population of faint and fast novae has failed to materialize. And now, the results of the comprehensive 20-year survey presented here, where we find an M31 nova rate of $40^{+5}_{-4}$~yr$^{-1}$, similarly provides no support for a population of faint and fast novae. Even the mean of our population-independent ($\theta=1$) extrapolations, which yields a slightly higher value of $48^{+5}_{-3}$~yr$^{-1}$, which we consider to provide a firm upper limit on the nova rate in M31, does not suggest
the presence of a stealth population of hitherto undiscovered novae. 

To summarize, regardless of how we choose to perform the extrapolation from the surveyed region of M31 to the entire galaxy, the results of the present study suggests an M31 nova rate that is significantly lower than that found by \citet{2006MNRAS.369..257D}, and far below the values estimated by \citet{2016MNRAS.458.2916C} and \citet{2016MNRAS.455..668S}. Finally, as we noted earlier (see Figure~\ref{fig:f1}), over the 7-year period between 2016 and 2020, the average annual number of nova candidates reported in M31 was $\sim$36~yr$^{-1}$. Thus, a corollary of our analysis is that the intensive coverage of M31 appears to now be detecting almost all of the novae that arise in the galaxy.
A caveat is that since not all M31 nova candidates reported annually are spectroscopically confirmed, some objects may actually be Long-Period Variable stars masquerading as novae \citep[e.g., see][]{2008ATel.1851....1S, 2010BlgAJ..14...54T}. If so, the average reported rate may be best considered an upper limit on the true observed nova rate in M31.

\subsection{Implications for the Galactic Nova Rate}

The eruptions of novae are thought to play a significant role in the chemical evolution of the Galaxy \citep[e.g.,][]{2022ApJ...933L..30K,2022MNRAS.509.3258M,2020ApJ...895...70S,2015ApJ...808L..14I,2003PhDT........13L}. In addition to producing a fraction of the $^7$Li and the short-lived isotopes $^{22}$Na and $^{26}$Al, novae are believed to be important in the production of the CNO isotopes, particularly $^{15}$N, where novae may account for a significant fraction of its Galactic abundance \citep{2003MNRAS.342..185R}. An accurate knowledge of the nova rate is therefore essential to a full understanding of Galactic chemical evolution.

The nova rate in the Milky Way, like that in M31, has been estimated many times over the years, both through
extrapolations of the observed nova rate to the entire Galaxy, and through a comparison of the Milky Way to external galaxies with measured nova rates.
Historically, there has been a tension between estimates based upon these two approaches,
with the former having resulted in generally higher rates compared with the latter
\citep[e.g., see Table~10,][for a summary]{2021ApJ...912...19D}.
In particular, recent studies based on extrapolation of the observed nova rate \citep[e.g.,][]{2017ApJ...834..196S,2018MNRAS.476.4162O,2021ApJ...912...19D} have found Galactic rates of $50^{+31}_{-23}$~yr$^{-1}$, $67^{+21}_{-17
}$~yr$^{-1}$, and $43.7^{+19.5}_{-8.7}$~yr$^{-1}$, respectively\footnote{Note that the \citet{2018MNRAS.476.4162O} rate is for the Galactic disk only. Adding the Galactic bulge estimate of $13.8\pm2.6$~yr$^{-1}$ \citep[][]{2015ApJS..219...26M} suggests an overall
Galactic nova rate of $\sim80$~yr$^{-1}$.}, while extragalactic comparisons
\citep[e.g.,][]{1994A&A...286..786D,2000ApJ...530..193S,2014ASPC..490...77S} have
predicted lower Galactic rates of $\approx$20~yr$^{-1}$, $27^{+10}_{-8}$~yr$^{-1}$, and $\sim25$~yr$^{-1}$, respectively.

The discrepancy between the two approaches has diminished recently thanks to the improved spatial and temporal coverage
made possible by all-sky surveys such as ASAS-SN and GAIA \citep{2016A&A...595A...1G}. The analysis of data from these surveys has brought the latest Galactic nova rate estimate down to
below 30~yr$^{-1}$ \citep[][]{2021ApJ...922...25K,2022arXiv220614132K}, which is now in line with the average of those based on extragalactic scalings.

In light of our new and significantly lower estimate of the nova rate in M31, we are in a position to re-determine the Galactic nova rate based on the
ratio of the (infrared) luminosities of our Galaxy and M31.
The absolute $K$-band magnitude of the Galaxy has been estimated by \citet{1996ApJ...473..687M} and by \citet{2001ApJ...556..181D} who found $M_K=-24.06$ and $M_K=-24.02$, respectively. Recalling that
$M_{K,\odot}=3.27$ gives an average estimate of
$L_{K}(\mathrm{G}) = 8.4\times10^{10}~\mathrm{L}_{\odot,K}$ for the $K$-band luminosity of the
Milky Way.
Adopting $L_{K}(\mathrm{M31}) = 1.2\times10^{10}~\mathrm{L}_{\odot,K}$ determined earlier gives a Milky Way to M31 $K$-band
luminosity ratio,
$L_{K}(\mathrm{G})/L_{K}(\mathrm{M31}) = 0.70\pm0.10$, where we have assumed a 10\% error in the luminosities of the individual galaxies. This ratio, coupled with our derived M31 nova rate of $40^{+5}_{-4}$~yr$^{-1}$, yields a global Galactic nova rate of $R_\mathrm{G}=28^{+5}_{-4}$~yr$^{-1}$ for the Milky Way. This value is fully consistent with both earlier extragalactic scaling estimates, and with the recent Galaxy-based estimate of $26\pm5$~yr$^{-1}$ by \citet{2022arXiv220614132K}. The remarkable agreement among these estimates provides renewed confidence in the accuracy and fidelity of both our Galactic and M31 nova rate determinations.

\section{Summary and Conclusions}

The principal conclusions of our extensive survey of novae in M31 can be summarized as follows:

{\bf (1)} During the course of a two-decade long survey spanning the years 1995 through 2016 we have discovered or detected a total of
262 novae in M31, 40 of which are previously unreported. Of these, 253 novae were included as part of two homogeneous surveys, survey~A (203 novae) and survey~B (50 novae). Survey~A was conducted with the KPNO/WIYN~0.9-m reflector and covered the central $20'\times20'$ field of M31, while survey~B
employed the KPNO~4-m reflector and covered a larger $36'\times36'$ field, also centered on the nucleus of M31.

{\bf (2)} After correcting for temporal and spatial incompleteness, nova rates of $R_\mathrm{A}=17.5^{+0.9}_{-0.4}$
and $R_\mathrm{B}=27.0^{+3.8}_{-3.0}$ were found for the areas covered in survey~A and survey~B, respectively.

{\bf (3)} The spatial distributions of the novae discovered in surveys~A and~B were compared with the background light
of M31 in several photometric bands. The nova distributions follow the galaxy's light reasonably well, with
the best fits achieved using the near infrared IRAC $3.6\mu$ light as given by Model~P of \citet{2011ApJ...739...20C}.

{\bf (4)} The nova rates in the surveyed regions were extrapolated to the entire galaxy by using the ratio of M31's total
infrared $3.6\mu$ light to that contained within the surveyed regions. Two different extrapolations were considered: A population-independent extrapolation assuming that the nova density in M31 was the same in M31's bulge and disk (i.e., $\theta~[=\rho_\mathrm{d}/\rho_\mathrm {b}] = 1$), and a population-dependent extrapolation based on a maximum likelihood analysis showing that
the bulge nova density exceeds that of the disk ($\theta~[=\rho_\mathrm{d}/\rho_\mathrm{b}] = 0.56^{+0.23}_{-0.17}$).

Our population-dependent extrapolation, which we have argued is appropriate for M31, leads to a global nova rate of $R_\mathrm{M31}=40^{+5}_{-4}$~yr$^{-1}$. The population-independent extrapolation ($\theta$=1), which assumes a larger disk nova contribution, gives a somewhat higher rate of $R_\mathrm{M31}=48^{+5}_{-3}$~yr$^{-1}$, and can be considered as providing an upper limit to
the nova rate in M31. Both rates are significantly lower than that found by \citet{2006MNRAS.369..257D},
and more in line with earlier estimates of M31's nova rate \citep[e.g., $37^{+12}_{-8}$~yr$^{-1}$,][]{2001ApJ...563..749S}.

{\bf (5)} When normalized to the $K$-band luminosity of M31, we find a luminosity-specific nova rate of
$3.3\pm0.4$ novae per year per $10^{10}$ solar luminosities in the $K$ band, which is typical of other galaxies with measured nova rates.

{\bf (6)} Our new determination of the nova rate in M31, which is based upon the largest sample of novae ever used in a single study of this galaxy, is inconsistent with
recent suggestions \citep[e.g.,][]{2016MNRAS.455..668S,2016MNRAS.458.2916C} that the nova rate in M31 (and the Galaxy) might approach 100~yr$^{-1}$. As a corollary, we find no evidence
for the existence of a significant population of faint and fast novae such as M31N~2008-12a.

{\bf (7)} Scaling our M31 nova rate to the Milky Way using the relative $K$-band luminosities of each galaxy,
we estimate a Galactic nova rate of $R_\mathrm{G}=28^{+5}_{-4}$~yr$^{-1}$.

We conclude on a optimistic note that after more than a century of observations and the discovery of $\sim$500 novae in the Galaxy and $\sim$1300 in M31, the annual rates of novae
in these two galaxies appear to now be well constrained and mutually consistent. Future observations should be directed at better understanding the role the underlying stellar population plays in determining the properties of emergent novae.

\begin{acknowledgments}
We would like to thank the referee, Massimo Della Valle, for several
insightful suggestions that helped improve the presentation of our results.

The work presented here is based in part on observations at
Kitt Peak National Observatory at NSF’s NOIRLab, which is
managed by the Association of Universities for Research in
Astronomy (AURA) under a cooperative agreement with the
National Science Foundation. The authors are honored to be
permitted to conduct astronomical research on Iolkam Du’ag
(Kitt Peak), a mountain with particular significance to the
Tohono O’odham.  This paper is dedicated to the firefighters
who saved the telescopes on Kitt Peak during the Contreras
Fire in June 2022.

We are indebted to the many thousands of teachers and
students who have participated in this project as part of the
RBSE, TLRBSE, and RBSE-U programs, the names of whom are too
many to list.  

This material is based upon work supported by the National
Science Foundation under Grant Nos. ESIE-9619028,
ESIE-0101982, DUE-0618441, DUE-0618849, and DUE-0920293. Any
opinions, findings, and conclusions or recommendations
expressed in this material are those of the authors and do
not necessarily reflect the views of the National Science
Foundation.

D.L. Corbett, M. Rene, and D. Hernandez were supported
through a NASA grant awarded to the Arizona/NASA Space Grant
Consortium. The material contained in this document is based
upon work supported by a National Aeronautics and Space
Administration (NASA) grant or cooperative agreement. Any
opinions, findings, conclusions or recommendations expressed
in this material are those of the author and do not
necessarily reflect the views of NASA.
\end{acknowledgments}

\facilities{WIYN:0.9-m, Mayall, Bok, KPNO:2.1-m, McGraw-Hill}

\software{IRAF \citep{1986SPIE..627..733T}, SAOImageDS9 \citep{2003ASPC..295..489J}, Astropy \citep{astropy:2013, astropy:2018} }

\pagebreak

\bibliography{m31novae}{}

\begin{thebibliography}{}
\expandafter\ifx\csname natexlab\endcsname\relax\def\natexlab#1{#1}\fi
\providecommand{\url}[1]{\href{#1}{#1}}
\providecommand{\dodoi}[1]{doi:~\href{http://doi.org/#1}{\nolinkurl{#1}}}
\providecommand{\doeprint}[1]{\href{http://ascl.net/#1}{\nolinkurl{http://ascl.net/#1}}}
\providecommand{\doarXiv}[1]{\href{https://arxiv.org/abs/#1}{\nolinkurl{https://arxiv.org/abs/#1}}}

\bibitem[{{Arp}(1956)}]{1956AJ.....61...15A}
{Arp}, H.~C. 1956, \aj, 61, 15, \dodoi{10.1086/107284}

\bibitem[{{Astropy Collaboration} {et~al.}(2013){Astropy Collaboration},
  {Robitaille}, {Tollerud}, {Greenfield}, {Droettboom}, {Bray}, {Aldcroft},
  {Davis}, {Ginsburg}, {Price-Whelan}, {Kerzendorf}, {Conley}, {Crighton},
  {Barbary}, {Muna}, {Ferguson}, {Grollier}, {Parikh}, {Nair}, {Unther},
  {Deil}, {Woillez}, {Conseil}, {Kramer}, {Turner}, {Singer}, {Fox}, {Weaver},
  {Zabalza}, {Edwards}, {Azalee Bostroem}, {Burke}, {Casey}, {Crawford},
  {Dencheva}, {Ely}, {Jenness}, {Labrie}, {Lim}, {Pierfederici}, {Pontzen},
  {Ptak}, {Refsdal}, {Servillat}, \& {Streicher}}]{astropy:2013}
{Astropy Collaboration}, {Robitaille}, T.~P., {Tollerud}, E.~J., {et~al.} 2013,
  \aap, 558, A33, \dodoi{10.1051/0004-6361/201322068}

\bibitem[{{Astropy Collaboration} {et~al.}(2018){Astropy Collaboration},
  {Price-Whelan}, {Sip{\H{o}}cz}, {G{\"u}nther}, {Lim}, {Crawford}, {Conseil},
  {Shupe}, {Craig}, {Dencheva}, {Ginsburg}, {Vand erPlas}, {Bradley},
  {P{\'e}rez-Su{\'a}rez}, {de Val-Borro}, {Aldcroft}, {Cruz}, {Robitaille},
  {Tollerud}, {Ardelean}, {Babej}, {Bach}, {Bachetti}, {Bakanov}, {Bamford},
  {Barentsen}, {Barmby}, {Baumbach}, {Berry}, {Biscani}, {Boquien}, {Bostroem},
  {Bouma}, {Brammer}, {Bray}, {Breytenbach}, {Buddelmeijer}, {Burke},
  {Calderone}, {Cano Rodr{\'\i}guez}, {Cara}, {Cardoso}, {Cheedella}, {Copin},
  {Corrales}, {Crichton}, {D'Avella}, {Deil}, {Depagne}, {Dietrich}, {Donath},
  {Droettboom}, {Earl}, {Erben}, {Fabbro}, {Ferreira}, {Finethy}, {Fox},
  {Garrison}, {Gibbons}, {Goldstein}, {Gommers}, {Greco}, {Greenfield},
  {Groener}, {Grollier}, {Hagen}, {Hirst}, {Homeier}, {Horton}, {Hosseinzadeh},
  {Hu}, {Hunkeler}, {Ivezi{\'c}}, {Jain}, {Jenness}, {Kanarek}, {Kendrew},
  {Kern}, {Kerzendorf}, {Khvalko}, {King}, {Kirkby}, {Kulkarni}, {Kumar},
  {Lee}, {Lenz}, {Littlefair}, {Ma}, {Macleod}, {Mastropietro}, {McCully},
  {Montagnac}, {Morris}, {Mueller}, {Mumford}, {Muna}, {Murphy}, {Nelson},
  {Nguyen}, {Ninan}, {N{\"o}the}, {Ogaz}, {Oh}, {Parejko}, {Parley}, {Pascual},
  {Patil}, {Patil}, {Plunkett}, {Prochaska}, {Rastogi}, {Reddy Janga},
  {Sabater}, {Sakurikar}, {Seifert}, {Sherbert}, {Sherwood-Taylor}, {Shih},
  {Sick}, {Silbiger}, {Singanamalla}, {Singer}, {Sladen}, {Sooley},
  {Sornarajah}, {Streicher}, {Teuben}, {Thomas}, {Tremblay}, {Turner},
  {Terr{\'o}n}, {van Kerkwijk}, {de la Vega}, {Watkins}, {Weaver}, {Whitmore},
  {Woillez}, {Zabalza}, \& {Astropy Contributors}}]{astropy:2018}
{Astropy Collaboration}, {Price-Whelan}, A.~M., {Sip{\H{o}}cz}, B.~M., {et~al.}
  2018, \aj, 156, 123, \dodoi{10.3847/1538-3881/aabc4f}

\bibitem[{{Barmby} {et~al.}(2006){Barmby}, {Ashby}, {Bianchi}, {Engelbracht},
  {Gehrz}, {Gordon}, {Hinz}, {Huchra}, {Humphreys}, {Pahre},
  {P{\'e}rez-Gonz{\'a}lez}, {Polomski}, {Rieke}, {Thilker}, {Willner}, \&
  {Woodward}}]{2006ApJ...650L..45B}
{Barmby}, P., {Ashby}, M.~L.~N., {Bianchi}, L., {et~al.} 2006, \apjl, 650, L45,
  \dodoi{10.1086/508626}

\bibitem[{{Bellm} {et~al.}(2019){Bellm}, {Kulkarni}, {Graham}, {Dekany},
  {Smith}, {Riddle}, {Masci}, {Helou}, {Prince}, {Adams}, {Barbarino},
  {Barlow}, {Bauer}, {Beck}, {Belicki}, {Biswas}, {Blagorodnova}, {Bodewits},
  {Bolin}, {Brinnel}, {Brooke}, {Bue}, {Bulla}, {Burruss}, {Cenko}, {Chang},
  {Connolly}, {Coughlin}, {Cromer}, {Cunningham}, {De}, {Delacroix}, {Desai},
  {Duev}, {Eadie}, {Farnham}, {Feeney}, {Feindt}, {Flynn}, {Franckowiak},
  {Frederick}, {Fremling}, {Gal-Yam}, {Gezari}, {Giomi}, {Goldstein},
  {Golkhou}, {Goobar}, {Groom}, {Hacopians}, {Hale}, {Henning}, {Ho}, {Hover},
  {Howell}, {Hung}, {Huppenkothen}, {Imel}, {Ip}, {Ivezi{\'c}}, {Jackson},
  {Jones}, {Juric}, {Kasliwal}, {Kaspi}, {Kaye}, {Kelley}, {Kowalski},
  {Kramer}, {Kupfer}, {Landry}, {Laher}, {Lee}, {Lin}, {Lin}, {Lunnan},
  {Giomi}, {Mahabal}, {Mao}, {Miller}, {Monkewitz}, {Murphy}, {Ngeow},
  {Nordin}, {Nugent}, {Ofek}, {Patterson}, {Penprase}, {Porter}, {Rauch},
  {Rebbapragada}, {Reiley}, {Rigault}, {Rodriguez}, {van Roestel}, {Rusholme},
  {van Santen}, {Schulze}, {Shupe}, {Singer}, {Soumagnac}, {Stein}, {Surace},
  {Sollerman}, {Szkody}, {Taddia}, {Terek}, {Van Sistine}, {van Velzen},
  {Vestrand}, {Walters}, {Ward}, {Ye}, {Yu}, {Yan}, \&
  {Zolkower}}]{2019PASP..131a8002B}
{Bellm}, E.~C., {Kulkarni}, S.~R., {Graham}, M.~J., {et~al.} 2019, \pasp, 131,
  018002, \dodoi{10.1088/1538-3873/aaecbe}

\bibitem[{{Capaccioli} {et~al.}(1989){Capaccioli}, {Della Valle}, {D'Onofrio},
  \& {Rosino}}]{1989AJ.....97.1622C}
{Capaccioli}, M., {Della Valle}, M., {D'Onofrio}, M., \& {Rosino}, L. 1989,
  \aj, 97, 1622, \dodoi{10.1086/115104}

\bibitem[{{Chen} {et~al.}(2016){Chen}, {Woods}, {Yungelson}, {Gilfanov}, \&
  {Han}}]{2016MNRAS.458.2916C}
{Chen}, H.-L., {Woods}, T.~E., {Yungelson}, L.~R., {Gilfanov}, M., \& {Han}, Z.
  2016, \mnras, 458, 2916, \dodoi{10.1093/mnras/stw458}

\bibitem[{{Ciardullo} {et~al.}(1987){Ciardullo}, {Ford}, {Neill}, {Jacoby}, \&
  {Shafter}}]{1987ApJ...318..520C}
{Ciardullo}, R., {Ford}, H.~C., {Neill}, J.~D., {Jacoby}, G.~H., \& {Shafter},
  A.~W. 1987, \apj, 318, 520, \dodoi{10.1086/165388}

\bibitem[{{Ciardullo} {et~al.}(1990){Ciardullo}, {Shafter}, {Ford}, {Neill},
  {Shara}, \& {Tomaney}}]{1990ApJ...356..472C}
{Ciardullo}, R., {Shafter}, A.~W., {Ford}, H.~C., {et~al.} 1990, \apj, 356,
  472, \dodoi{10.1086/168855}

\bibitem[{{Coelho} {et~al.}(2008){Coelho}, {Shafter}, \&
  {Misselt}}]{2008ApJ...686.1261C}
{Coelho}, E.~A., {Shafter}, A.~W., \& {Misselt}, K.~A. 2008, \apj, 686, 1261,
  \dodoi{10.1086/591517}

\bibitem[{{Courteau} {et~al.}(2011){Courteau}, {Widrow}, {McDonald},
  {Guhathakurta}, {Gilbert}, {Zhu}, {Beaton}, \&
  {Majewski}}]{2011ApJ...739...20C}
{Courteau}, S., {Widrow}, L.~M., {McDonald}, M., {et~al.} 2011, \apj, 739, 20,
  \dodoi{10.1088/0004-637X/739/1/20}

\bibitem[{{Darnley} {et~al.}(2014){Darnley}, {Williams}, {Bode}, {Henze},
  {Ness}, {Shafter}, {Hornoch}, \& {Votruba}}]{2014A&A...563L...9D}
{Darnley}, M.~J., {Williams}, S.~C., {Bode}, M.~F., {et~al.} 2014, \aap, 563,
  L9, \dodoi{10.1051/0004-6361/201423411}

\bibitem[{{Darnley} {et~al.}(2004){Darnley}, {Bode}, {Kerins}, {Newsam}, {An},
  {Baillon}, {Novati}, {Carr}, {Cr{\'e}z{\'e}}, {Evans}, {Giraud-H{\'e}raud},
  {Gould}, {Hewett}, {Jetzer}, {Kaplan}, {Paulin-Henriksson}, {Smartt},
  {Stalin}, \& {Tsapras}}]{2004MNRAS.353..571D}
{Darnley}, M.~J., {Bode}, M.~F., {Kerins}, E., {et~al.} 2004, \mnras, 353, 571,
  \dodoi{10.1111/j.1365-2966.2004.08087.x}

\bibitem[{{Darnley} {et~al.}(2006){Darnley}, {Bode}, {Kerins}, {Newsam}, {An},
  {Baillon}, {Belokurov}, {Calchi Novati}, {Carr}, {Cr{\'e}z{\'e}}, {Evans},
  {Giraud-H{\'e}raud}, {Gould}, {Hewett}, {Jetzer}, {Kaplan},
  {Paulin-Henriksson}, {Smartt}, {Tsapras}, \& {Weston}}]{2006MNRAS.369..257D}
---. 2006, \mnras, 369, 257, \dodoi{10.1111/j.1365-2966.2006.10297.x}

\bibitem[{{De} {et~al.}(2021){De}, {Kasliwal}, {Hankins}, {Sokoloski}, {Adams},
  {Ashley}, {Babul}, {Bagdasaryan}, {Delacroix}, {Dekany}, {Greffe}, {Hale},
  {Jencson}, {Karambelkar}, {Lau}, {Mahabal}, {McKenna}, {Moore}, {Ofek},
  {Sharma}, {Smith}, {Soon}, {Soria}, {Srinivasaragavan}, {Tinyanont},
  {Travouillon}, {Tzanidakis}, \& {Yao}}]{2021ApJ...912...19D}
{De}, K., {Kasliwal}, M.~M., {Hankins}, M.~J., {et~al.} 2021, \apj, 912, 19,
  \dodoi{10.3847/1538-4357/abeb75}

\bibitem[{{de Vaucouleurs}(1958)}]{1958ApJ...128..465D}
{de Vaucouleurs}, G. 1958, \apj, 128, 465, \dodoi{10.1086/146564}

\bibitem[{{Della Valle} \& {Izzo}(2020)}]{2020A&ARv..28....3D}
{Della Valle}, M., \& {Izzo}, L. 2020, \aapr, 28, 3,
  \dodoi{10.1007/s00159-020-0124-6}

\bibitem[{{della Valle} \& {Livio}(1994)}]{1994A&A...286..786D}
{della Valle}, M., \& {Livio}, M. 1994, \aap, 286, 786

\bibitem[{{Della Valle} \& {Livio}(1998)}]{1998ApJ...506..818D}
{Della Valle}, M., \& {Livio}, M. 1998, \apj, 506, 818, \dodoi{10.1086/306275}

\bibitem[{{della Valle} {et~al.}(1994){della Valle}, {Rosino}, {Bianchini}, \&
  {Livio}}]{1994A&A...287..403D}
{della Valle}, M., {Rosino}, L., {Bianchini}, A., \& {Livio}, M. 1994, \aap,
  287, 403

\bibitem[{{Drimmel} \& {Spergel}(2001)}]{2001ApJ...556..181D}
{Drimmel}, R., \& {Spergel}, D.~N. 2001, \apj, 556, 181, \dodoi{10.1086/321556}

\bibitem[{{Fazio} {et~al.}(2004){Fazio}, {Hora}, {Allen}, {Ashby}, {Barmby},
  {Deutsch}, {Huang}, {Kleiner}, {Marengo}, {Megeath}, {Melnick}, {Pahre},
  {Patten}, {Polizotti}, {Smith}, {Taylor}, {Wang}, {Willner}, {Hoffmann},
  {Pipher}, {Forrest}, {McMurty}, {McCreight}, {McKelvey}, {McMurray}, {Koch},
  {Moseley}, {Arendt}, {Mentzell}, {Marx}, {Losch}, {Mayman}, {Eichhorn},
  {Krebs}, {Jhabvala}, {Gezari}, {Fixsen}, {Flores}, {Shakoorzadeh}, {Jungo},
  {Hakun}, {Workman}, {Karpati}, {Kichak}, {Whitley}, {Mann}, {Tollestrup},
  {Eisenhardt}, {Stern}, {Gorjian}, {Bhattacharya}, {Carey}, {Nelson},
  {Glaccum}, {Lacy}, {Lowrance}, {Laine}, {Reach}, {Stauffer}, {Surace},
  {Wilson}, {Wright}, {Hoffman}, {Domingo}, \& {Cohen}}]{2004ApJS..154...10F}
{Fazio}, G.~G., {Hora}, J.~L., {Allen}, L.~E., {et~al.} 2004, \apjs, 154, 10,
  \dodoi{10.1086/422843}

\bibitem[{{Franck} {et~al.}(2012){Franck}, {Shafter}, {Hornoch}, \&
  {Misselt}}]{2012ApJ...760...13F}
{Franck}, J.~R., {Shafter}, A.~W., {Hornoch}, K., \& {Misselt}, K.~A. 2012,
  \apj, 760, 13, \dodoi{10.1088/0004-637X/760/1/13}

\bibitem[{{Freedman} {et~al.}(2001){Freedman}, {Madore}, {Gibson}, {Ferrarese},
  {Kelson}, {Sakai}, {Mould}, {Kennicutt}, {Ford}, {Graham}, {Huchra},
  {Hughes}, {Illingworth}, {Macri}, \& {Stetson}}]{2001ApJ...553...47F}
{Freedman}, W.~L., {Madore}, B.~F., {Gibson}, B.~K., {et~al.} 2001, \apj, 553,
  47, \dodoi{10.1086/320638}

\bibitem[{{Gaia Collaboration} {et~al.}(2016){Gaia Collaboration}, {Prusti},
  {de Bruijne}, {Brown}, {Vallenari}, {Babusiaux}, {Bailer-Jones}, {Bastian},
  {Biermann}, {Evans}, {Eyer}, {Jansen}, {Jordi}, {Klioner}, {Lammers},
  {Lindegren}, {Luri}, {Mignard}, {Milligan}, {Panem}, {Poinsignon},
  {Pourbaix}, {Randich}, {Sarri}, {Sartoretti}, {Siddiqui}, {Soubiran},
  {Valette}, {van Leeuwen}, {Walton}, {Aerts}, {Arenou}, {Cropper}, {Drimmel},
  {H{\o}g}, {Katz}, {Lattanzi}, {O'Mullane}, {Grebel}, {Holland}, {Huc},
  {Passot}, {Bramante}, {Cacciari}, {Casta{\~n}eda}, {Chaoul}, {Cheek}, {De
  Angeli}, {Fabricius}, {Guerra}, {Hern{\'a}ndez}, {Jean-Antoine-Piccolo},
  {Masana}, {Messineo}, {Mowlavi}, {Nienartowicz}, {Ord{\'o}{\~n}ez-Blanco},
  {Panuzzo}, {Portell}, {Richards}, {Riello}, {Seabroke}, {Tanga},
  {Th{\'e}venin}, {Torra}, {Els}, {Gracia-Abril}, {Comoretto},
  {Garcia-Reinaldos}, {Lock}, {Mercier}, {Altmann}, {Andrae}, {Astraatmadja},
  {Bellas-Velidis}, {Benson}, {Berthier}, {Blomme}, {Busso}, {Carry},
  {Cellino}, {Clementini}, {Cowell}, {Creevey}, {Cuypers}, {Davidson}, {De
  Ridder}, {de Torres}, {Delchambre}, {Dell'Oro}, {Ducourant}, {Fr{\'e}mat},
  {Garc{\'\i}a-Torres}, {Gosset}, {Halbwachs}, {Hambly}, {Harrison}, {Hauser},
  {Hestroffer}, {Hodgkin}, {Huckle}, {Hutton}, {Jasniewicz}, {Jordan},
  {Kontizas}, {Korn}, {Lanzafame}, {Manteiga}, {Moitinho}, {Muinonen},
  {Osinde}, {Pancino}, {Pauwels}, {Petit}, {Recio-Blanco}, {Robin}, {Sarro},
  {Siopis}, {Smith}, {Smith}, {Sozzetti}, {Thuillot}, {van Reeven}, {Viala},
  {Abbas}, {Abreu Aramburu}, {Accart}, {Aguado}, {Allan}, {Allasia},
  {Altavilla}, {{\'A}lvarez}, {Alves}, {Anderson}, {Andrei}, {Anglada Varela},
  {Antiche}, {Antoja}, {Ant{\'o}n}, {Arcay}, {Atzei}, {Ayache}, {Bach},
  {Baker}, {Balaguer-N{\'u}{\~n}ez}, {Barache}, {Barata}, {Barbier}, {Barblan},
  {Baroni}, {Barrado y Navascu{\'e}s}, {Barros}, {Barstow}, {Becciani},
  {Bellazzini}, {Bellei}, {Bello Garc{\'\i}a}, {Belokurov}, {Bendjoya},
  {Berihuete}, {Bianchi}, {Bienaym{\'e}}, {Billebaud}, {Blagorodnova},
  {Blanco-Cuaresma}, {Boch}, {Bombrun}, {Borrachero}, {Bouquillon}, {Bourda},
  {Bouy}, {Bragaglia}, {Breddels}, {Brouillet}, {Br{\"u}semeister},
  {Bucciarelli}, {Budnik}, {Burgess}, {Burgon}, {Burlacu}, {Busonero}, {Buzzi},
  {Caffau}, {Cambras}, {Campbell}, {Cancelliere}, {Cantat-Gaudin}, {Carlucci},
  {Carrasco}, {Castellani}, {Charlot}, {Charnas}, {Charvet}, {Chassat},
  {Chiavassa}, {Clotet}, {Cocozza}, {Collins}, {Collins}, {Costigan}, {Crifo},
  {Cross}, {Crosta}, {Crowley}, {Dafonte}, {Damerdji}, {Dapergolas}, {David},
  {David}, {De Cat}, {de Felice}, {de Laverny}, {De Luise}, {De March}, {de
  Martino}, {de Souza}, {Debosscher}, {del Pozo}, {Delbo}, {Delgado},
  {Delgado}, {di Marco}, {Di Matteo}, {Diakite}, {Distefano}, {Dolding}, {Dos
  Anjos}, {Drazinos}, {Dur{\'a}n}, {Dzigan}, {Ecale}, {Edvardsson}, {Enke},
  {Erdmann}, {Escolar}, {Espina}, {Evans}, {Eynard Bontemps}, {Fabre},
  {Fabrizio}, {Faigler}, {Falc{\~a}o}, {Farr{\`a}s Casas}, {Faye}, {Federici},
  {Fedorets}, {Fern{\'a}ndez-Hern{\'a}ndez}, {Fernique}, {Fienga}, {Figueras},
  {Filippi}, {Findeisen}, {Fonti}, {Fouesneau}, {Fraile}, {Fraser}, {Fuchs},
  {Furnell}, {Gai}, {Galleti}, {Galluccio}, {Garabato}, {Garc{\'\i}a-Sedano},
  {Gar{\'e}}, {Garofalo}, {Garralda}, {Gavras}, {Gerssen}, {Geyer}, {Gilmore},
  {Girona}, {Giuffrida}, {Gomes}, {Gonz{\'a}lez-Marcos},
  {Gonz{\'a}lez-N{\'u}{\~n}ez}, {Gonz{\'a}lez-Vidal}, {Granvik}, {Guerrier},
  {Guillout}, {Guiraud}, {G{\'u}rpide}, {Guti{\'e}rrez-S{\'a}nchez}, {Guy},
  {Haigron}, {Hatzidimitriou}, {Haywood}, {Heiter}, {Helmi}, {Hobbs},
  {Hofmann}, {Holl}, {Holland}, {Hunt}, {Hypki}, {Icardi}, {Irwin}, {Jevardat
  de Fombelle}, {Jofr{\'e}}, {Jonker}, {Jorissen}, {Julbe}, {Karampelas},
  {Kochoska}, {Kohley}, {Kolenberg}, {Kontizas}, {Koposov}, {Kordopatis},
  {Koubsky}, {Kowalczyk}, {Krone-Martins}, {Kudryashova}, {Kull}, {Bachchan},
  {Lacoste-Seris}, {Lanza}, {Lavigne}, {Le Poncin-Lafitte}, {Lebreton},
  {Lebzelter}, {Leccia}, {Leclerc}, {Lecoeur-Taibi}, {Lemaitre}, {Lenhardt},
  {Leroux}, {Liao}, {Licata}, {Lindstr{\o}m}, {Lister}, {Livanou}, {Lobel},
  {L{\"o}ffler}, {L{\'o}pez}, {Lopez-Lozano}, {Lorenz}, {Loureiro},
  {MacDonald}, {Magalh{\~a}es Fernandes}, {Managau}, {Mann}, {Mantelet},
  {Marchal}, {Marchant}, {Marconi}, {Marie}, {Marinoni}, {Marrese},
  {Marschalk{\'o}}, {Marshall}, {Mart{\'\i}n-Fleitas}, {Martino}, {Mary},
  {Matijevi{\v{c}}}, {Mazeh}, {McMillan}, {Messina}, {Mestre}, {Michalik},
  {Millar}, {Miranda}, {Molina}, {Molinaro}, {Molinaro}, {Moln{\'a}r},
  {Moniez}, {Montegriffo}, {Monteiro}, {Mor}, {Mora}, {Morbidelli}, {Morel},
  {Morgenthaler}, {Morley}, {Morris}, {Mulone}, {Muraveva}, {Musella},
  {Narbonne}, {Nelemans}, {Nicastro}, {Noval}, {Ord{\'e}novic},
  {Ordieres-Mer{\'e}}, {Osborne}, {Pagani}, {Pagano}, {Pailler}, {Palacin},
  {Palaversa}, {Parsons}, {Paulsen}, {Pecoraro}, {Pedrosa}, {Pentik{\"a}inen},
  {Pereira}, {Pichon}, {Piersimoni}, {Pineau}, {Plachy}, {Plum}, {Poujoulet},
  {Pr{\v{s}}a}, {Pulone}, {Ragaini}, {Rago}, {Rambaux}, {Ramos-Lerate},
  {Ranalli}, {Rauw}, {Read}, {Regibo}, {Renk}, {Reyl{\'e}}, {Ribeiro},
  {Rimoldini}, {Ripepi}, {Riva}, {Rixon}, {Roelens}, {Romero-G{\'o}mez},
  {Rowell}, {Royer}, {Rudolph}, {Ruiz-Dern}, {Sadowski}, {Sagrist{\`a}
  Sell{\'e}s}, {Sahlmann}, {Salgado}, {Salguero}, {Sarasso}, {Savietto},
  {Schnorhk}, {Schultheis}, {Sciacca}, {Segol}, {Segovia}, {Segransan},
  {Serpell}, {Shih}, {Smareglia}, {Smart}, {Smith}, {Solano}, {Solitro},
  {Sordo}, {Soria Nieto}, {Souchay}, {Spagna}, {Spoto}, {Stampa}, {Steele},
  {Steidelm{\"u}ller}, {Stephenson}, {Stoev}, {Suess}, {S{\"u}veges}, {Surdej},
  {Szabados}, {Szegedi-Elek}, {Tapiador}, {Taris}, {Tauran}, {Taylor},
  {Teixeira}, {Terrett}, {Tingley}, {Trager}, {Turon}, {Ulla}, {Utrilla},
  {Valentini}, {van Elteren}, {Van Hemelryck}, {van Leeuwen}, {Varadi},
  {Vecchiato}, {Veljanoski}, {Via}, {Vicente}, {Vogt}, {Voss}, {Votruba},
  {Voutsinas}, {Walmsley}, {Weiler}, {Weingrill}, {Werner}, {Wevers},
  {Whitehead}, {Wyrzykowski}, {Yoldas}, {{\v{Z}}erjal}, {Zucker}, {Zurbach},
  {Zwitter}, {Alecu}, {Allen}, {Allende Prieto}, {Amorim},
  {Anglada-Escud{\'e}}, {Arsenijevic}, {Azaz}, {Balm}, {Beck}, {Bernstein},
  {Bigot}, {Bijaoui}, {Blasco}, {Bonfigli}, {Bono}, {Boudreault}, {Bressan},
  {Brown}, {Brunet}, {Bunclark}, {Buonanno}, {Butkevich}, {Carret}, {Carrion},
  {Chemin}, {Ch{\'e}reau}, {Corcione}, {Darmigny}, {de Boer}, {de Teodoro}, {de
  Zeeuw}, {Delle Luche}, {Domingues}, {Dubath}, {Fodor}, {Fr{\'e}zouls},
  {Fries}, {Fustes}, {Fyfe}, {Gallardo}, {Gallegos}, {Gardiol}, {Gebran},
  {Gomboc}, {G{\'o}mez}, {Grux}, {Gueguen}, {Heyrovsky}, {Hoar}, {Iannicola},
  {Isasi Parache}, {Janotto}, {Joliet}, {Jonckheere}, {Keil}, {Kim},
  {Klagyivik}, {Klar}, {Knude}, {Kochukhov}, {Kolka}, {Kos}, {Kutka}, {Lainey},
  {LeBouquin}, {Liu}, {Loreggia}, {Makarov}, {Marseille}, {Martayan},
  {Martinez-Rubi}, {Massart}, {Meynadier}, {Mignot}, {Munari}, {Nguyen},
  {Nordlander}, {Ocvirk}, {O'Flaherty}, {Olias Sanz}, {Ortiz}, {Osorio},
  {Oszkiewicz}, {Ouzounis}, {Palmer}, {Park}, {Pasquato}, {Peltzer}, {Peralta},
  {P{\'e}turaud}, {Pieniluoma}, {Pigozzi}, {Poels}, {Prat}, {Prod'homme},
  {Raison}, {Rebordao}, {Risquez}, {Rocca-Volmerange}, {Rosen}, {Ruiz-Fuertes},
  {Russo}, {Sembay}, {Serraller Vizcaino}, {Short}, {Siebert}, {Silva},
  {Sinachopoulos}, {Slezak}, {Soffel}, {Sosnowska}, {Strai{\v{z}}ys}, {ter
  Linden}, {Terrell}, {Theil}, {Tiede}, {Troisi}, {Tsalmantza}, {Tur},
  {Vaccari}, {Vachier}, {Valles}, {Van Hamme}, {Veltz}, {Virtanen}, {Wallut},
  {Wichmann}, {Wilkinson}, {Ziaeepour}, \& {Zschocke}}]{2016A&A...595A...1G}
{Gaia Collaboration}, {Prusti}, T., {de Bruijne}, J.~H.~J., {et~al.} 2016,
  \aap, 595, A1, \dodoi{10.1051/0004-6361/201629272}

\bibitem[{{G{\"u}th} {et~al.}(2010){G{\"u}th}, {Shafter}, \&
  {Misselt}}]{2010ApJ...720.1155G}
{G{\"u}th}, T., {Shafter}, A.~W., \& {Misselt}, K.~A. 2010, \apj, 720, 1155,
  \dodoi{10.1088/0004-637X/720/2/1155}

\bibitem[{{Hatano} {et~al.}(1997){Hatano}, {Branch}, {Fisher}, \&
  {Starrfield}}]{1997ApJ...487L..45H}
{Hatano}, K., {Branch}, D., {Fisher}, A., \& {Starrfield}, S. 1997, \apjl, 487,
  L45, \dodoi{10.1086/310862}

\bibitem[{{Henze} {et~al.}(2014){Henze}, {Ness}, {Darnley}, {Bode}, {Williams},
  {Shafter}, {Kato}, \& {Hachisu}}]{2014A&A...563L...8H}
{Henze}, M., {Ness}, J.~U., {Darnley}, M.~J., {et~al.} 2014, \aap, 563, L8,
  \dodoi{10.1051/0004-6361/201423410}

\bibitem[{{Hubble}(1929)}]{1929ApJ....69..103H}
{Hubble}, E.~P. 1929, \apj, 69, 103, \dodoi{10.1086/143167}

\bibitem[{{Izzo} {et~al.}(2015){Izzo}, {Della Valle}, {Mason}, {Matteucci},
  {Romano}, {Pasquini}, {Vanzi}, {Jordan}, {Fernandez}, {Bluhm}, {Brahm},
  {Espinoza}, \& {Williams}}]{2015ApJ...808L..14I}
{Izzo}, L., {Della Valle}, M., {Mason}, E., {et~al.} 2015, \apjl, 808, L14,
  \dodoi{10.1088/2041-8205/808/1/L14}

\bibitem[{{Jarrett} {et~al.}(2003){Jarrett}, {Chester}, {Cutri}, {Schneider},
  \& {Huchra}}]{2003AJ....125..525J}
{Jarrett}, T.~H., {Chester}, T., {Cutri}, R., {Schneider}, S.~E., \& {Huchra},
  J.~P. 2003, \aj, 125, 525, \dodoi{10.1086/345794}

\bibitem[{{Joye} \& {Mandel}(2003)}]{2003ASPC..295..489J}
{Joye}, W.~A., \& {Mandel}, E. 2003, in Astronomical Society of the Pacific
  Conference Series, Vol. 295, Astronomical Data Analysis Software and Systems
  XII, ed. H.~E. {Payne}, R.~I. {Jedrzejewski}, \& R.~N. {Hook}, 489

\bibitem[{{Kasliwal} {et~al.}(2011){Kasliwal}, {Cenko}, {Kulkarni}, {Ofek},
  {Quimby}, \& {Rau}}]{2011ApJ...735...94K}
{Kasliwal}, M.~M., {Cenko}, S.~B., {Kulkarni}, S.~R., {et~al.} 2011, \apj, 735,
  94, \dodoi{10.1088/0004-637X/735/2/94}

\bibitem[{{Kato} {et~al.}(2014){Kato}, {Saio}, {Hachisu}, \&
  {Nomoto}}]{2014ApJ...793..136K}
{Kato}, M., {Saio}, H., {Hachisu}, I., \& {Nomoto}, K. 2014, \apj, 793, 136,
  \dodoi{10.1088/0004-637X/793/2/136}

\bibitem[{{Kaur}(2016)}]{2016PhDT........37K}
{Kaur}, A. 2016, PhD thesis, Clemson University, South Carolina

\bibitem[{{Kawash} {et~al.}(2021){Kawash}, {Chomiuk}, {Rodriguez}, {Strader},
  {Sokolovsky}, {Aydi}, {Kochanek}, {Stanek}, {Mukai}, {De}, {Shappee},
  {Holoien}, {Prieto}, \& {Thompson}}]{2021ApJ...922...25K}
{Kawash}, A., {Chomiuk}, L., {Rodriguez}, J.~A., {et~al.} 2021, \apj, 922, 25,
  \dodoi{10.3847/1538-4357/ac1f1a}

\bibitem[{{Kawash} {et~al.}(2022){Kawash}, {Chomiuk}, {Strader}, {Sokolovsky},
  {Aydi}, {Kochanek}, {Stanek}, {Kostrzewa-Rutkowska}, {Hodgkin}, {Mukai},
  {Shappee}, {Jayasinghe}, {Rizzo Smith}, {Holoien}, {Prieto}, \&
  {Thompson}}]{2022arXiv220614132K}
{Kawash}, A., {Chomiuk}, L., {Strader}, J., {et~al.} 2022, arXiv e-prints,
  arXiv:2206.14132.
\newblock \doarXiv{2206.14132}

\bibitem[{{Kemp} {et~al.}(2022){Kemp}, {Karakas}, {Casey}, {C{\^o}t{\'e}},
  {Izzard}, \& {Osborn}}]{2022ApJ...933L..30K}
{Kemp}, A.~J., {Karakas}, A.~I., {Casey}, A.~R., {et~al.} 2022, \apjl, 933,
  L30, \dodoi{10.3847/2041-8213/ac7c72}

\bibitem[{{Kent}(1987)}]{1987AJ.....94..306K}
{Kent}, S.~M. 1987, \aj, 94, 306, \dodoi{10.1086/114472}

\bibitem[{{Kochanek} {et~al.}(2017){Kochanek}, {Shappee}, {Stanek}, {Holoien},
  {Thompson}, {Prieto}, {Dong}, {Shields}, {Will}, {Britt}, {Perzanowski}, \&
  {Pojma{\'n}ski}}]{2017PASP..129j4502K}
{Kochanek}, C.~S., {Shappee}, B.~J., {Stanek}, K.~Z., {et~al.} 2017, \pasp,
  129, 104502, \dodoi{10.1088/1538-3873/aa80d9}

\bibitem[{{Lyke}(2003)}]{2003PhDT........13L}
{Lyke}, J.~E. 2003, PhD thesis, University of Minnesota, Twin Cities

\bibitem[{{Malhotra} {et~al.}(1996){Malhotra}, {Spergel}, {Rhoads}, \&
  {Li}}]{1996ApJ...473..687M}
{Malhotra}, S., {Spergel}, D.~N., {Rhoads}, J.~E., \& {Li}, J. 1996, \apj, 473,
  687, \dodoi{10.1086/178181}

\bibitem[{{Massey} {et~al.}(2007){Massey}, {McNeill}, {Olsen}, {Hodge},
  {Blaha}, {Jacoby}, {Smith}, \& {Strong}}]{2007AJ....134.2474M}
{Massey}, P., {McNeill}, R.~T., {Olsen}, K.~A.~G., {et~al.} 2007, \aj, 134,
  2474, \dodoi{10.1086/523658}

\bibitem[{{Massey} {et~al.}(2006){Massey}, {Olsen}, {Hodge}, {Strong},
  {Jacoby}, {Schlingman}, \& {Smith}}]{2006AJ....131.2478M}
{Massey}, P., {Olsen}, K.~A.~G., {Hodge}, P.~W., {et~al.} 2006, \aj, 131, 2478,
  \dodoi{10.1086/503256}

\bibitem[{{Molaro} {et~al.}(2022){Molaro}, {Izzo}, {D'Odorico}, {Aydi},
  {Bonifacio}, {Cescutti}, {Harvey}, {Hernanz}, {Selvelli}, \& {della
  Valle}}]{2022MNRAS.509.3258M}
{Molaro}, P., {Izzo}, L., {D'Odorico}, V., {et~al.} 2022, \mnras, 509, 3258,
  \dodoi{10.1093/mnras/stab3106}

\bibitem[{{Monet} {et~al.}(2003){Monet}, {Levine}, {Canzian}, {Ables}, {Bird},
  {Dahn}, {Guetter}, {Harris}, {Henden}, {Leggett}, {Levison}, {Luginbuhl},
  {Martini}, {Monet}, {Munn}, {Pier}, {Rhodes}, {Riepe}, {Sell}, {Stone},
  {Vrba}, {Walker}, {Westerhout}, {Brucato}, {Reid}, {Schoening}, {Hartley},
  {Read}, \& {Tritton}}]{2003AJ....125..984M}
{Monet}, D.~G., {Levine}, S.~E., {Canzian}, B., {et~al.} 2003, \aj, 125, 984,
  \dodoi{10.1086/345888}

\bibitem[{{Mr{\'o}z} {et~al.}(2015){Mr{\'o}z}, {Udalski}, {Poleski},
  {Soszy{\'n}ski}, {Szyma{\'n}ski}, {Pietrzy{\'n}ski}, {Wyrzykowski},
  {Ulaczyk}, {Koz{\l}owski}, {Pietrukowicz}, \&
  {Skowron}}]{2015ApJS..219...26M}
{Mr{\'o}z}, P., {Udalski}, A., {Poleski}, R., {et~al.} 2015, \apjs, 219, 26,
  \dodoi{10.1088/0067-0049/219/2/26}

\bibitem[{{Neill} \& {Shara}(2004)}]{2004AJ....127..816N}
{Neill}, J.~D., \& {Shara}, M.~M. 2004, \aj, 127, 816, \dodoi{10.1086/381484}

\bibitem[{{Nomoto}(1982)}]{1982ApJ...253..798N}
{Nomoto}, K. 1982, \apj, 253, 798, \dodoi{10.1086/159682}

\bibitem[{{{\"O}zd{\"o}nmez} {et~al.}(2018){{\"O}zd{\"o}nmez}, {Ege},
  {G{\"u}ver}, \& {Ak}}]{2018MNRAS.476.4162O}
{{\"O}zd{\"o}nmez}, A., {Ege}, E., {G{\"u}ver}, T., \& {Ak}, T. 2018, \mnras,
  476, 4162, \dodoi{10.1093/mnras/sty432}

\bibitem[{{Peng} {et~al.}(2002){Peng}, {Ho}, {Impey}, \&
  {Rix}}]{2002AJ....124..266P}
{Peng}, C.~Y., {Ho}, L.~C., {Impey}, C.~D., \& {Rix}, H.-W. 2002, \aj, 124,
  266, \dodoi{10.1086/340952}

\bibitem[{Rector {et~al.}(2019)Rector, Puckett, Wooten, Vogt, Coble, \&
  Pilachowski}]{10.1088/2514-3433/ab2b42ch7}
Rector, T.~A., Puckett, A.~W., Wooten, M.~M., {et~al.} 2019, in Astronomy
  Education, Volume 1, 2514-3433 (IOP Publishing), 7--1 to 7--10,
  \dodoi{10.1088/2514-3433/ab2b42ch7}

\bibitem[{{Ritchey}(1917)}]{1917PASP...29..210R}
{Ritchey}, G.~W. 1917, \pasp, 29, 210, \dodoi{10.1086/122638}

\bibitem[{{Romano} \& {Matteucci}(2003)}]{2003MNRAS.342..185R}
{Romano}, D., \& {Matteucci}, F. 2003, \mnras, 342, 185,
  \dodoi{10.1046/j.1365-8711.2003.06526.x}

\bibitem[{{Rosino}(1964)}]{1964AnAp...27..498R}
{Rosino}, L. 1964, Annales d'Astrophysique, 27, 498

\bibitem[{{Rosino}(1973)}]{1973Ros}
---. 1973, \aaps, 9, 347

\bibitem[{{Rosino} {et~al.}(1989){Rosino}, {Capaccioli}, {D'Onofrio}, \& {della
  Valle}}]{1989AJ.....97...83R}
{Rosino}, L., {Capaccioli}, M., {D'Onofrio}, M., \& {della Valle}, M. 1989,
  \aj, 97, 83, \dodoi{10.1086/114959}

\bibitem[{{Shafter}(2017)}]{2017ApJ...834..196S}
{Shafter}, A.~W. 2017, \apj, 834, 196, \dodoi{10.3847/1538-4357/834/2/196}

\bibitem[{{Shafter}(2019)}]{2019enhp.book.....S}
---. 2019, {Extragalactic Novae; A historical perspective},
  \dodoi{10.1088/2514-3433/ab2c63}

\bibitem[{{Shafter} {et~al.}(2000){Shafter}, {Ciardullo}, \&
  {Pritchet}}]{2000ApJ...530..193S}
{Shafter}, A.~W., {Ciardullo}, R., \& {Pritchet}, C.~J. 2000, \apj, 530, 193,
  \dodoi{10.1086/308349}

\bibitem[{{Shafter} {et~al.}(2014){Shafter}, {Curtin}, {Pritchet}, {Bode}, \&
  {Darnley}}]{2014ASPC..490...77S}
{Shafter}, A.~W., {Curtin}, C., {Pritchet}, C.~J., {Bode}, M.~F., \& {Darnley},
  M.~J. 2014, in Astronomical Society of the Pacific Conference Series, Vol.
  490, Stellar Novae: Past and Future Decades, ed. P.~A. {Woudt} \& V.~A.~R.~M.
  {Ribeiro}, 77.
\newblock \doarXiv{1307.2296}

\bibitem[{{Shafter} {et~al.}(2012){Shafter}, {Hornoch}, {Ciardullo}, {Darnley},
  \& {Bode}}]{2012ATel.4503....1S}
{Shafter}, A.~W., {Hornoch}, K., {Ciardullo}, J. V.~R., {Darnley}, M.~J., \&
  {Bode}, M.~F. 2012, The Astronomer's Telegram, 4503, 1

\bibitem[{{Shafter} \& {Irby}(2001)}]{2001ApJ...563..749S}
{Shafter}, A.~W., \& {Irby}, B.~K. 2001, \apj, 563, 749, \dodoi{10.1086/324044}

\bibitem[{{Shafter} {et~al.}(2017){Shafter}, {Kundu}, \&
  {Henze}}]{2017RNAAS...1...11S}
{Shafter}, A.~W., {Kundu}, A., \& {Henze}, M. 2017, Research Notes of the
  American Astronomical Society, 1, 11, \dodoi{10.3847/2515-5172/aa9847}

\bibitem[{{Shafter} {et~al.}(2009){Shafter}, {Rau}, {Quimby}, {Kasliwal},
  {Bode}, {Darnley}, \& {Misselt}}]{2009ApJ...690.1148S}
{Shafter}, A.~W., {Rau}, A., {Quimby}, R.~M., {et~al.} 2009, \apj, 690, 1148,
  \dodoi{10.1088/0004-637X/690/2/1148}

\bibitem[{{Shafter} {et~al.}(2008){Shafter}, {Ciardullo}, {Burwitz}, {Henze},
  {Pietsch}, {Milne}, {Misselt}, {Williams}, {Hartmann}, {Updike}, {Bode}, \&
  {Darnley}}]{2008ATel.1851....1S}
{Shafter}, A.~W., {Ciardullo}, R., {Burwitz}, V., {et~al.} 2008, The
  Astronomer's Telegram, 1851, 1

\bibitem[{{Shafter} {et~al.}(2011){Shafter}, {Darnley}, {Hornoch},
  {Filippenko}, {Bode}, {Ciardullo}, {Misselt}, {Hounsell}, {Chornock}, \&
  {Matheson}}]{2011ApJ...734...12S}
{Shafter}, A.~W., {Darnley}, M.~J., {Hornoch}, K., {et~al.} 2011, \apj, 734,
  12, \dodoi{10.1088/0004-637X/734/1/12}

\bibitem[{{Shafter} {et~al.}(2021){Shafter}, {Hornoch}, {Ben{\'a}{\v{c}}ek},
  {Gal{\'a}d}, {Jan{\'\i}k}, {Jury{\v{s}}ek}, {Kotkov{\'a}}, {Kurf{\"u}rst},
  {Ku{\v{c}}{\'a}kov{\'a}}, {Ku{\v{s}}nir{\'a}k}, {Li{\v{s}}ka}, {Paunzen},
  {Skarka}, {{\v{S}}koda}, {Wolf}, {Zasche}, \& {Zejda}}]{2021ApJ...923..239S}
{Shafter}, A.~W., {Hornoch}, K., {Ben{\'a}{\v{c}}ek}, J., {et~al.} 2021, \apj,
  923, 239, \dodoi{10.3847/1538-4357/ac2c79}

\bibitem[{{Shara} {et~al.}(2016){Shara}, {Doyle}, {Lauer}, {Zurek}, {Neill},
  {Madrid}, {Miko{\l}ajewska}, {Welch}, \& {Baltz}}]{2016ApJS..227....1S}
{Shara}, M.~M., {Doyle}, T.~F., {Lauer}, T.~R., {et~al.} 2016, \apjs, 227, 1,
  \dodoi{10.3847/0067-0049/227/1/1}

\bibitem[{{Soraisam} {et~al.}(2016){Soraisam}, {Gilfanov}, {Wolf}, \&
  {Bildsten}}]{2016MNRAS.455..668S}
{Soraisam}, M.~D., {Gilfanov}, M., {Wolf}, W.~M., \& {Bildsten}, L. 2016,
  \mnras, 455, 668, \dodoi{10.1093/mnras/stv2359}

\bibitem[{{Soraisam} {et~al.}(2017){Soraisam}, {Gilfanov}, {Kupfer}, {Masci},
  {Shafter}, {Prince}, {Kulkarni}, {Ofek}, \& {Bellm}}]{2017A&A...599A..48S}
{Soraisam}, M.~D., {Gilfanov}, M., {Kupfer}, T., {et~al.} 2017, \aap, 599, A48,
  \dodoi{10.1051/0004-6361/201629368}

\bibitem[{{Starrfield} {et~al.}(2020){Starrfield}, {Bose}, {Iliadis}, {Hix},
  {Woodward}, \& {Wagner}}]{2020ApJ...895...70S}
{Starrfield}, S., {Bose}, M., {Iliadis}, C., {et~al.} 2020, \apj, 895, 70,
  \dodoi{10.3847/1538-4357/ab8d23}

\bibitem[{{Starrfield} {et~al.}(2016){Starrfield}, {Iliadis}, \&
  {Hix}}]{2016PASP..128e1001S}
{Starrfield}, S., {Iliadis}, C., \& {Hix}, W.~R. 2016, \pasp, 128, 051001,
  \dodoi{10.1088/1538-3873/128/963/051001}

\bibitem[{{Taneva} {et~al.}(2010){Taneva}, {Valcheva}, {Ovcharov}, \&
  {Nedialkov}}]{2010BlgAJ..14...54T}
{Taneva}, N., {Valcheva}, A., {Ovcharov}, E., \& {Nedialkov}, P. 2010,
  Bulgarian Astronomical Journal, 14, 54

\bibitem[{{Tang} {et~al.}(2014){Tang}, {Bildsten}, {Wolf}, {Li}, {Kong}, {Cao},
  {Cenko}, {De Cia}, {Kasliwal}, {Kulkarni}, {Laher}, {Masci}, {Nugent},
  {Perley}, {Prince}, \& {Surace}}]{2014ApJ...786...61T}
{Tang}, S., {Bildsten}, L., {Wolf}, W.~M., {et~al.} 2014, \apj, 786, 61,
  \dodoi{10.1088/0004-637X/786/1/61}

\bibitem[{{Tody}(1986)}]{1986SPIE..627..733T}
{Tody}, D. 1986, in Society of Photo-Optical Instrumentation Engineers (SPIE)
  Conference Series, Vol. 627, Instrumentation in astronomy VI, ed. D.~L.
  {Crawford}, 733, \dodoi{10.1117/12.968154}

\bibitem[{{Townsley} \& {Bildsten}(2005)}]{2005ApJ...628..395T}
{Townsley}, D.~M., \& {Bildsten}, L. 2005, \apj, 628, 395,
  \dodoi{10.1086/430594}

\bibitem[{{Warner}(1995)}]{1995cvs..book.....W}
{Warner}, B. 1995, {Cataclysmic variable stars}, Vol.~28

\bibitem[{{Williams}(1992)}]{1992AJ....104..725W}
{Williams}, R.~E. 1992, \aj, 104, 725, \dodoi{10.1086/116268}

\bibitem[{{Williams} \& {Shafter}(2004)}]{2004ApJ...612..867W}
{Williams}, S.~J., \& {Shafter}, A.~W. 2004, \apj, 612, 867,
  \dodoi{10.1086/422833}

\bibitem[{{Willmer}(2018)}]{2018ApJS..236...47W}
{Willmer}, C. N.~A. 2018, \apjs, 236, 47, \dodoi{10.3847/1538-4365/aabfdf}

\bibitem[{Wooten {et~al.}(2018)Wooten, Coble, Puckett, \&
  Rector}]{PhysRevPhysEducRes.14.010151}
Wooten, M.~M., Coble, K., Puckett, A.~W., \& Rector, T. 2018, Phys. Rev. Phys.
  Educ. Res., 14, 010151, \dodoi{10.1103/PhysRevPhysEducRes.14.010151}

\end{thebibliography}
\bibliographystyle{aasjournal}

\end{document}